\documentclass{article}

\usepackage{tikz}
\usetikzlibrary{arrows,matrix,graphs,shapes,snakes,automata,backgrounds,petri,positioning,calc,decorations.pathmorphing}
\usepackage[latin1]{inputenc}
\usepackage{verbatim}
\usepackage[leqno]{amsmath}
\usepackage{amssymb}
\usepackage{latexsym}
\usepackage{amsfonts}
\usepackage{proof}
\usepackage{array}
\usepackage{calc}
\usepackage{mathdots}
\usepackage{stmaryrd}
\usepackage{harvard}
\usepackage[english]{babel}
\usepackage{extarrows}
\usepackage{rotating}
\usepackage{hyperref}
\usepackage{linguex}
\usepackage{mathbbol}
\usepackage{varioref}
\usepackage{floatpag}

\floatpagestyle{empty}
\newcommand{\shfocus}{\pm}
\newcommand{\leftimpl}{\|}
\newcommand{\rightimpl}{\|}
\newcommand{\leftexpl}{\|}
\newcommand{\rightexpl}{\|}

\newtheorem{theorem}{Theorem}[section]
\newtheorem{lemma}[theorem]{Lemma}
\newtheorem{proposition}[theorem]{Proposition}
\newtheorem{definition}[theorem]{Definition}
\newtheorem{corollary}[theorem]{Corollary}
\newcommand{\lolli}{\multimap}
\newcommand{\bs}{\backslash}
\newcommand{\bo}{[}
\newcommand{\bc}{]}
\newcommand{\ra}{\rightarrow}
\newcommand{\editout}[1]{}
\newcommand{\vdashpos}{\vdash_p}
\newcommand{\vdashneg}{\vdash_n}
\newcommand{\sbst}{\mathbb{s}}
\newcommand{\qed}{\hfill \ensuremath{\Box}}
\newcommand{\str}{\sigma\rightarrow\sigma}

\newcommand{\concat}{%
  \mathord{\raisebox{1ex}{\scalebox{.7}{$\frown$}}}%
}

\tikzset{invclip/.style={clip,insert path={{[reset cm]
      (-1383.99999pt,-1383.99999pt) rectangle (1383.99999pt,1383.99999pt)
    }}}}
\tikzstyle{reverseclip}=[insert path={(current page.north east) --
  (current page.south east) --
  (current page.south west) --
  (current page.north west) --
  (current page.north east)}
]

\title{Hybrid Type-Logical Grammars,\\ First-Order Linear Logic and
  the Descriptive Inadequacy of Lambda Grammars\thanks{This work has
    benefitted from the generous
support of the French agency Agence Nationale de la Recherche as part of the project Polymnie
(ANR-12-CORD-0004).}}
\author{Richard Moot}

\begin{document}
\maketitle

\section{Introduction}

Hybrid type-logical grammars \cite{kl12gap,kl13emp,kl13coord} are a relatively new framework in
computational linguistics, which combines insights from the Lambek
calculus \cite{lambek} and lambda grammars
\cite{oehrle,muskens01lfg,muskens03lambda} --- lambda grammars are also called, depending on the authors,
abstract categorial grammars \cite{groote01acg} and linear grammars \cite{pollard11linear}, though with somewhat
different notational conventions\footnote{I prefer the term lambda
  grammars, since I think it most clearly describes the
  system. Though the term abstract categorial grammars appears to be
  more common, and I use it from time to time in this article, I will argue in Section~\ref{inadeq} that abstract
  categorial grammars/lambda grammars are \emph{unlike} all other
  versions of categorial grammars in important ways.}. The resulting
combined system solves some know problems of both the Lambek calculus
and of lambda grammars and the additional expressiveness of hybrid
type-logical grammars permits the treatment of linguistic phenomena
such as gapping
which have no satisfactory solution in either subsystem. 

The goal of this paper is to prove that hybrid
type-logical grammars are a fragment of first-order linear logic. This
embedding result has several important consequences: it not only
provides a simple new proof theory for the calculus, thereby clarifying
the proof-theoretic foundations of hybrid type-logical grammars, but, since the
translation is simple and direct, it also provides several new parsing
strategies for hybrid type-logical grammars. Second, NP-completeness of
hybrid type-logical grammars follows immediately.

The main embedding result also sheds new light on problems with
lambda grammars, which are a subsystem of hybrid
type-logical grammars and hence a special case of the translation into
first-order linear logic.
Abstract categorial grammars are attractive both because of
their simplicity --- they use the simply typed lambda calculus, one of the most widely used
tools in formal semantics, to compute surface structure (strings)
as well as to compute logical form (meanings) --- and because of the fact that they
provide a natural account of quantifier scope and extraction; for
both, the analysis is superior to the Lambek calculus analysis. So it
is easy to get the impression that lambda grammars are an unequivocal improvement
over Lambek grammars.

In reality, the picture is much more nuanced: while lambda grammars
have some often discussed advantages over Lambek grammars, there are
several cases --- notably coordination, but we will see in Section~\ref{inadeq}
that this is true for any analysis where the Lambek calculus uses
non-atomic arguments --- where the Lambek grammar analysis
is clearly superior. Many key examples illustrating the elegance
of categorial grammars with respect to the syntax-semantics interface fail to have a satisfactory treatment in abstract
categorial grammars.

However, whether or not lambda grammars are an improvement over the
Lambek calculus is ultimately not the most important question. Since
there is a large number of formal systems which improve upon the
Lambek calculus, it makes much more sense to compare lambda grammars
to these extensions, which include, among many others, Hybrid
Type-Logical Grammars, the Displacement calculus \cite{mvf11displacement} and multimodal
type-logical grammars \cite{mmli,M95}. These extended Lambek calculi all keep the things
that worked in the Lambek calculus but improve on the analysis in ways
which allow the treatment of more complex phenomena in syntax and
especially in the syntax-semantics interface. Compared to these systems, the inadequacies of lambda
grammars are evident: even for the things lambda grammars do right (quantifier scope and
extraction), there are phenomena, such as reflexives and gapping, which
are handled by the same mechanisms as quantifier scope and extraction
in alternative theories, yet which cannot be adequately handled by
lambda grammars.  The abstract categorial grammar treatment suffers
from problems of overgeneration and problems at the syntax-semantics
interface unlike any other categorial grammar. I will discuss some
possible solutions for lambda grammars, but it is clear that a major
redesign of the theory is necessary. The most painless solution seems
to be a move either to hybrid type-logical grammars or directly to
first-order linear logic: both are simple, conservative extensions
which solve the many problems of lambda grammars while staying close
to the spirit of lambda grammars.

This paper is structured as follows. Section~\ref{sec:mill1} will
introduce first-order linear logic and Section~\ref{sec:stl} will
provide some background about the simply typed lambda
calculus. These two introductory sections can be skimmed by people familiar
with first-order linear logic and the simply typed lambda calculus respectively. Section~\ref{sec:hybrid} will introduce hybrid type-logical
grammars and in Section~\ref{sec:equiv} we will give a translation of
hybrid type-logical grammars into first-order linear logic and prove
its correctness. Section~\ref{sec:comparison} will then compare
the Lambek calculus and several of its extensions through their
translations in first-order linear logic. This comparison points to a
number of potential problems for lambda grammars. We will
discuss these problems, as well as some potential solutions in Section~\ref{inadeq}.
Finally, the last section will contain some concluding remarks.

\section{First-order Linear Logic}
\label{sec:mill1}

Linear logic was introduced by \citeasnoun{Girard} as a logic
which restricts the structural rules which apply freely in classical logic. 
The multiplicative, intuitionistic fragment of first-order linear
logic (which in the following, I will call either MILL1 or simply
first-order linear logic), can be seen as a resource-conscious version of first-order
intuitionistic logic. Linear implication, written $A \multimap B$, is
a variant of intuitionistic implication $A\Rightarrow B$ with the
additional constraint that the $A$ argument formula is used
\emph{exactly once}. So, looking at linear logic from the context of grammatical analysis, we would assign an intransitive verb the
formula $np\multimap s$, indicating it is a formula which
combines with a single $np$ (noun phrase) to form an $s$ (a sentence).

Linear logic is a commutative logic. In the context of language
modelling, this means our languages are closed under permutations of
the input string, which does not make for a good linguistic principle
(at least not a good \emph{universal} one and a principle which is at
least debatable even in languages which allow relatively free word order). We need some way to restrict or
control commutativity. The Lambek calculus \cite{lambek} has the simplest such restriction: we drop
the structural rule of commutativity altogether. This means linear implication
$A \multimap B$ splits into two implications: $A\backslash B$, which
looks for an $A$ to its left to form a $B$, and $B/A$, which looks for
an $A$ to its right to form a $B$. In the Lambek calculus, we would
therefore refine the assignment to
intransitive verbs from $np\multimap s$ to $np\backslash s$, indicating the intransitive verb
is looking for the subject to its left.

In first-order linear logic, we can choose a more versatile solution,
namely using first-order variables to encode word order. We assign
atomic formulas a pair of string positions: $np$ becomes $np(0,1)$,
meaning it is a noun phrase spanning position 0 (its leftmost
position) to 1 (its rightmost position). Using pairs of (integer)
variables to represent strings is standard in parsing algorithms.
The addition of quantifiers makes things more interesting. For example, we can assign  the formula $\forall x. n(3,x)
\multimap np(2,x)$ to
a determiner ``the'' which spans positions $2,3$. This means it is looking for a noun which starts
at its right (that is the leftmost position of this noun is the
rightmost position of the determiner, 3) but ends at any position $x$
to produce a noun phrase which starts at position 2 (the leftmost
position of the determiner) and ends at position $x$ (the rightmost
position of the noun).  Combined with a noun $n(3,4)$, this would
allow us to instantiate $x$ to 4 and produce $np(2,4)$. In other
words, the formula given to the determiner indicates it is looking for
a noun to its right in order to produce a noun phrase, using a form of
``concatenation by instantiation of variables'' which should be familiar to
anyone who has done some logic programming or who has a basic
familiarity with parsing in general \cite{PS87,dedpar}. Similarly, we can assign an
intransitive verb at position 1,2 the formula $\forall y. np(y,1)
\multimap s(y,2)$ to indicate it is looking for a noun phrase to its
left to form a sentence, as the Lambek calculus formula $np\backslash
s$ for intransitive verbs does --- this correspondence between
first-order linear logic and the Lambek calculus is fully general and
discussed fully in \cite{mill1} and briefly in the next section.

\subsection{MILL1}
\label{sec:nounif}

After this informal introduction to first-order linear logic, it is
time to be a bit more precise. We will not need function symbols in
the current paper, so
\emph{terms} are either variables denoted $x, y, z, \ldots $ (a
countably infinite number) or constants, for which I
will normally use integers $0, 1, \ldots$, giving an $m$-word string
$m+1$ string positions, from $0$ to $m$. The atomic formulas are of the form $a(t_1,
\ldots, t_m)$ with $t_i$ terms, $a$ a predicate symbol (we only need a
finite, typically smal, number of predicate symbols, often only the
following four: $n$ for noun,
$np$ for noun phrase, $s$ for sentence, $pp$ for predicate phrase) and $m$ its
arity. Our language does \emph{not} contain the identity relation symbol ``=''. Given this set of atomic formulas $\mathcal{A}$ and the set of variables $\mathcal{V}$, the set of formulas is
defined as follows\footnote{We need neither the multiplicative
  conjunction $\otimes$ nor the existential quantifier $\exists$ in
  this paper, though adding them to the logic poses no problems. The
  natural deduction rules for $\exists$ and $\otimes$ are slightly
  more complicated than those for $\forall$ and $\multimap$ but the
  basic proof net building blocks don't change, see
  for example \cite{quant,mill1,moot13lambek}.}.

$$
\mathcal{F} ::= \mathcal{A} \; | \; \mathcal{F} \multimap \mathcal{V} \; | \; \forall \mathcal{V}. \mathcal{F} %\; | \; \exists V. F
$$

We treat formulas as syntactically equivalent up to renaming of bound
variables, so substituting $\forall y. A [x:=y]$ (where $A$ does not
contain $y$ before this substitution is made) for $\forall x. A$
inside a formula $B$ will produce an equivalent formula, for
example $\forall x. a(x) \equiv \forall y. a(y)$.

Table~\ref{tab:millnd} shows the natural deduction rules for
first-order linear logic. The variable $x$ in the $\forall E$ and
$\forall I$ rules is called the \emph{eigenvariable} of the
rule. The $\forall I$ rule has the condition that the variable $y$
which is replaced by the eigenvariable
 does not occur in undischarged hypotheses of the proof and that $x$
 does not occur in $A$ before the substitution is made\footnote{It is
   sometimes more convenient to use the following $\forall I$
   rule $$\infer[\forall I^*]{\forall x. A}{A}
 $$ \noindent with the condition there are no free occurrence of $x$
 in open hypotheses. The rule of Table~\ref{tab:millnd} is more
 convient in the following section when we use meta-variables, where
 it becomes ``replace all occurrences of a (meta-)variable by $x$,
 then quantify over $x$''.}.
Throughout this paper, we will use the standard convention in
first-order (linear) logic \cite{quant,empires,bpt} that every occurrence of
a quantifier $\forall$, $\exists$ in a sequent uses
a distinct variable and in addition that no variable occurs both free
and bound in a sequent.

\begin{table}
$$
\begin{array}{ccc}
%\infer[\otimes E_i]{C}{A \otimes B & \infer*{C}{[A]^i[B]^i}}
%&&
%\infer[\otimes I]{A\otimes B}{A & B} \\
%\\
\infer[\multimap E]{B}{A & A \multimap B} &&
\infer[\multimap I]{A \multimap B}{\infer*{B}{[A]^i}} \\
\\
%\infer[\exists E_i^*]{C}{\exists x .A & \infer*{C}{[A]^i}} &&
%\infer[\exists I]{\exists x.A}{A[x:=t]} \\
%\\
\infer[\forall E]{A[x:=t]}{\forall x. A} & &
\infer[\forall I^*]{\forall x. A}{A[y:=x]} \\
\end{array}
$$
\caption{Natural deduction rules for first-order linear logic}
\label{tab:millnd}
\end{table}

As shown in \cite{mill1}, we can translate Lambek calculus sequents
and formulas into first-order linear logic as follows.
\begin{gather*}
A_1, \ldots, A_n \vdash B = \\
\| A_1 \|^{0,1}, \ldots \| A_n \|^{n-1,n} \vdash \| B \|^{0,n}
\end{gather*}

\begin{align*}
\| a \|^{x,y} & =  a(x,y) \\
\| A/ B \|^{x,y} &= \forall z. \| B \|^{y,z} \multimap \| A \|^{x,z} \\
\| B\backslash A \|^{y,z} &= \forall x \| B \|^{x,y} \multimap \| A \|^{x,z}
\end{align*}

The integers 0 to $n$ represent the positions of the formulas in the
sequent and the translations for complex formulas introduce
universally quantified variables. The translation for $A/B$ states
that if we have a formula $A/B$ at positions $x,y$ then for any $z$ if
we find a formula $B$ at positions $y,z$ (that is, to the immediate
right of our $A/B$ formula) then we have an $A$ at positions $x,z$,
starting at the left position of the $A/B$ formula and ending at the
right position of the $B$ argument. In
other words, a formula $A/B$ is something which combines with a $B$
to its right to form an $A$, just like its Lambek calculus counterpart.

Using this translation, we can see that the first-order linear logic
formulas used for the determiner and the intransitive verb in the
previous section correspond to the translations of $np/n$ at position
$2,3$ and $np\backslash s$ at position $1,2$ respectively.

To give a simple example of a first-order linear logic proof, we shown
a derivation of ``every student ran'', corresponding to the Lambek
calculus sequent.
$$
(s/(np\backslash s))/n, n, np\backslash s \vdash s
$$

We first translate the sequent into first-order linear logic.
$$
\| (s/(np\backslash s))/n \|^{0,1}, \| n\|^{1,2}, \| np\backslash s
\|^{2,3} \vdash  \| s \|^{0,3}
$$
Then translate the formulas as follows.
$$
\forall y. [ n(1,y) \multimap \forall z. [ \forall x. [np(x,y) \multimap
s(x,z)] \multimap s(0,z) ] ], n(1,2), \forall v. [ np(v,2) \multimap
s(v,3) \vdash s(0,3) ]
$$

We can then show that ``every student ran'' is a grammatical sentence
under these formula assignments as follows.

$$
\infer[\multimap E]{s(0,3)}{\infer[\forall E]{\forall x. [np(x,2) \multimap
s(x,3)] \multimap s(0,3)}{
\infer[\multimap E]{\forall z. [ \forall x. [np(x,2) \multimap
s(x,z)] \multimap s(0,z) ]}{
\infer[\forall E]{n(1,2) \multimap \forall z. [ \forall x. [np(x,2) \multimap
s(x,z)] \multimap s(0,z) ]}{\forall y. [ n(1,y) \multimap \forall z. [ \forall x. [np(x,y) \multimap
s(x,z)] \multimap s(0,z) ] ]} & n(1,2)}} & \forall v. [np(v,2)
\multimap s(v,3)]}
$$

The application of the final $\multimap E$ rule is valid, since $\forall x. [np(x,2) \multimap
s(x,3)] \equiv \forall v. [np(v,2)
\multimap s(v,3)]$.
%On the right branch of the proof, we rename the variable $v$ to $x$ to
%make the $\forall E$ rule valid. This is a valid application of the
%$\forall I$ rule since all premisses are closed formulas.

%Just like first-order logic, we have the following basic fact.

\begin{definition}[Universal closure]\label{def:cl} If $A$ is a formula we denote the set of free
  variables of $A$ by $\textit{FV}(A)$. 

For an antecedent
  $\Gamma=A_1,\ldots,A_n$, $\textit{FV}(\Gamma) = \textit{FV}(A_1)
  \cup \dots \cup \textit{FV}(A_n)$.

The \emph{universal closure} of a formula $A$ with $\textit{FV}(A) = \{x_1,\ldots,x_n\}$, denoted
$\textit{Cl}(A)$, is the formula $\forall x_1 \ldots \forall x_n . A$.

The \emph{universal closure} of a formula $A$ \emph{modulo antecedent}
$\Gamma$,  written $\textit{Cl}_{\Gamma}(A)$, is defined by universally
quantifying over the free variables in $A$ which do not occur in
$\Gamma$. If $\textit{FV}(A) \setminus \textit{FV}(\Gamma) = \{
x_1,\ldots, x_n\}$, then $\textit{Cl}_{\Gamma}(A) = \forall x_1
\ldots \forall x_n . A$.
\end{definition}

\begin{proposition}\label{prop:cl} $\Gamma \vdash A$ iff $\Gamma \vdash \textit{Cl}_{\Gamma}(A)$.
\end{proposition}

\paragraph{Proof} If the closure modulo $\Gamma$ prefixes $n$ universal
quantifiers to $A$, we can go from $\Gamma \vdash A$ to $\Gamma \vdash
\textit{Cl}_{\Gamma}(A)$ by using the $\forall I$ rule $n$ times (the
quantified variables added for the closure have been chosen to respect the condition on the rule) and
in the opposite direction by using the $\forall E$ rule $n$ times. 
\qed

%\begin{proposition} The following statements $\Gamma \vdash C$ are
%  derivable in MILL1 and the statements $\Gamma \nvdash C$ are
%  underivable (assuming no free occurrences of $x$ in $A$).
%\begin{align} A \multimap \forall x. B \vdash \forall x. [A \multimap
 % B]  \\
%
%\end{align}
%\end{proposition}

\subsection{MILL1 with focusing and unification}
\label{sec:unif}

The $\forall E$ rule, as formulated in the previous section, has the
disadvantage that it requires us to choose a term $t$ with which to
replace $x$ and that making the right choice for $t$ requires some
insight into how the resulting formula will be used in the rest of the proof. In
the example of the preceding section we need to make two such
``educated guesses'': we instantiate $y$ to 2 to allow the
elimination rule with minor premiss $n(1,2)$ and we instantiate $z$ to 3
to produce the desired conclusion $s(0,3)$.

The
standard solution to automate this process in first-order logic
theorem proving is to change the $\forall E$ rule:
instead of directly replacing the quantified variable by the ``right''
choice, we replace it by a meta-variable (I
will use the Prolog-like notation $A$, $B$, $\ldots$ for these
variables, or, when confusion with the notation $A$ and $B$ for arbitrary
formulas is possible $C$, $D$, $E$, $\ldots$, $V$, $W$, $X$,
$\ldots$).
These meta-variables will represent our current
knowledge about the term with which we will replace a given quantified
variable. The MGU we compute for the endsequent will correspond to
the most general instantiations of these variables in the given proof (that is, all other
instantiations can be obtained from this final MGU by means of additional substitutions).

The $\multimap E$ rule \emph{unifies} the $B$ formulas of the
argument and minor premiss of the rule (so the two occurrences of $B$
need only be unifiable instead of identical). Remember that the unification
of two atomic formulas $a(x_1,\ldots,x_m)$ and $b(y_1,\ldots,y_n)$ is
only defined when $a=b$ and $m=n$ and that unification tries to find
the most general instantiation of all free variables such that $x_i =
y_i$ (for all $1 \leq i \leq n = m$) and fails if no such
instantiation exists. The  presence of an explicit quantifier presents a
complication, but only a minor one: bound variables are treated just
like constants which, in addition, must respect the variable condition. 

More precisely, the unification of %a positive and a negative formula
two formulas is defined as
follows. %Negative occurrences of the universal quantifier come from
%the translation function, whereas positive occurrences of the universal
%quantifier can only come from the $\forall I$ rule.
%\begin{align*}
%\textit{unify}(a(x_1,\ldots,x_n)^-, a(y_1,\ldots,y_n)^+) &=
%\textit{unify}(x_i,y_i)\ \textrm{for all}\ 1 \leq i \leq n \\
%\textit{unify}((A\multimap B)^-, (C\multimap D)^+) &= \textit{unify}(C^-,A^+),\textit{unify}(B^-,D^+)\\
%\textit{unify}((\forall x.A)^-, (\forall y.B)^+) & = \textit{unify}(A[x:=y]^-,B^+) 
%\end{align*}
\begin{align*}
\textit{unify}(a(x_1,\ldots,x_n), a(y_1,\ldots,y_n)) &=
\textit{unify}(x_i,y_i)\ \textrm{for all}\ 1 \leq i \leq n \\
\textit{unify}(A_1\multimap B_1, A_2\multimap B_2) &= \textit{unify}(A_2,
A_1),\textit{unify}(B_1, B_2)\\
\textit{unify}(\forall x.A, \forall y.B) & = \textit{unify}(A, B[y:=x]) 
\end{align*}

The $\forall$ case assumes there are no free occurrences of $x$ in $B$
before substitution. It is defined in such a way that it is
independent of the actual variable names used for the quantifier (as
mentioned, we use a different variable for each occurrence of a
quantifier) and bound occurrences of $x_i$ and $y_i$ are treated as
constants in the $\textit{unify}(x_i,y_i)$ clause, subject to the
following condition:
%Unification \emph{fails} if 
if  we compute a substitution $D:=x$ for a
formula $A$ and $x$ is not free
for $D$ in $A$ then unification \emph{fails}. In other words, the substitution cannot introduce new
bound variables, so for example $\forall y. a(D,y)$ and $\forall
z. a(z,z)$ fail to unify, since $D$ is not free for $y$ in $\forall
y. a(D,y)$, and therefore we  cannot legally substitute $y$ for $D$
since it would result in an ``accidental capture'', creating a new
bound occurrence of $y$.\footnote{In such cases, substitution
  succeeds but does nothing and subsequent unification fails, since
  the formulas are not alphabetic variants after substitution.}

%We translate a sequent $A_1,\ldots,A_n \vdash B$ by removing all negative
%$occurrences of $\forall$ and replacing the corresponding eigenvariable
%$occurrences by fresh meta-variables. 

As second problem with natural deduction proof search is that we can
have subproofs like the following.
$$
\begin{array}{cc}
\infer[\forall E]{a(y)}{\infer[\forall I]{\forall x. a(x)}{a(y)}} &
\infer[\multimap E]{B}{\infer[\multimap I_i]{A\multimap B}{\infer*{B}{[A]^i}} & A}
\end{array}
$$
In both cases, we introduce a connective and then immediately
eliminate it. A natural deduction proof is called \emph{normal} if is
does not contain any subproof of the forms shown above. One of the classic results for natural deduction is
\emph{normalization} which states that we can eliminate such detours
\cite{glt,bpt}. In the case of linear logic, removing such detours is
even guaranteed to decrease the size of the proof.

We use a form of
\emph{focalized} natural deduction \cite{focus,fnd}, which is a
syntactic variant of natural deduction guaranteed to generate only
normal natural deduction proofs. We use two turnstiles, the
negative $\vdashneg$ and the positive $\vdashpos$ (for the reader
familiar with focused
proofs, $\Gamma \vdash C \Downarrow$  corresponds to $\Gamma \vdashneg
C$ and $\Gamma \vdash C \Uparrow$ to $\Gamma \vdashpos C$).

We will call a sequent $\Gamma \vdashpos C$ a \emph{positive} sequent
(and $C$ a positive formula) and a sequent $\Gamma \vdashneg C$ a
\emph{negative} sequent (and $C$ a negative formula).

\begin{table}
\begin{center}
\textbf{Lexicon}
\end{center}
\vspace{-1\baselineskip}

$$
A \vdashneg A
$$

\vspace{.3\baselineskip}
\begin{center}
\textbf{Axiom/Hypothesis}
\end{center}
\vspace{-1\baselineskip}

$$
A \vdashneg A
$$

\vspace{.3\baselineskip}
\begin{center}
\textbf{Shift Focus}
\end{center}
\vspace{-1\baselineskip}

$$
\infer[\shfocus]{\Gamma \vdashpos A}{\Gamma \vdashneg A}
$$

\vspace{.3\baselineskip}
\begin{center}
\textbf{Logical Rules}
\end{center}
\vspace{-.5\baselineskip}

$$
\infer[\multimap E]{\sbst(\Gamma),\sbst(\Delta) \vdashneg \sbst(A)}{\Gamma \vdashneg B \multimap
  A & \Delta \vdashpos B}
$$

$$
\infer[\multimap I]{\Gamma \vdashpos B \multimap A}{\Gamma, B \vdashpos A}
$$

$$
\infer[\forall E]{\Gamma \vdashneg A[x:=D]}{\Gamma\vdashneg \forall
  x. A}
$$

$$
\infer[\forall I^*]{\Gamma \vdashpos \forall x.A}{\Gamma \vdashpos A[y:=x]}
$$
\caption{Focused first-order linear logic with unification}
\label{tab:rulesimpl}
\end{table}

Table~\ref{tab:rulesimpl} shows the rules of first-order linear logic
in this format. For the lexicon rule, we require that the formula $A$ is closed. The formula $A$ of the hypothesis rule \emph{can} contain free variables.\editout{
 hypothesis rule contains no constants
but can contain free
variables, though free variables are constrained to occur once on each side of
the turnstile\footnote{That is, we cannot have a hypothesis like
  $a(x,x)\vdash a(x,x)$. As we will see, this restriction corresponds
  to a type of linearity of the variables and, though non-standard, it
  will simplify the proofs which follow. Note that this restriction is
compatible with the standard analysis of  \textit{wh}-movement in
categorial grammars: an $np$ trace will start as $np(x,y)\vdash
np(x,y)$ and get further instantiated by unification later in the
proof.}.}

For the $\forall I$ rule, $y$ is either a variable or a meta-variable
which has no free occurrences in any undischarged hypothesis.

For the $\multimap E$ rule, %the formula $B$ occurs once in its
%positive and once in its negative incarnation and
$\sbst$ is the most general unifier of $\langle \Gamma, B\rangle$ and
$\langle \Delta, B\rangle$. That is, we unify the two occurrences of $B$ in
their respective contexts, using unification for complex formulas as
defined above. 
%This unification can require application of the
%$\forall I$ rule to ensure the unification of a negative formula
%containing a quantifier with a positive formula which does not, as is
%clear from the way unification of quantified formulas is defined --- we%
%will present an example below.
 %(we treat quantifiers as ``transparent'' and
%bound variables as constants for the purpose of this unification, so,
%for example, the negative formula $\forall x. a(x,A)$ and the positive
%formula $a(B,2)$ unify instantiating $A$ to 2 and $B$ to $x$, but this
%same positive formula would \emph{fail} to unify with $a(2,B)$ since
%$x$ and $2$ are different constants).
%two occurrences of $B$ may have distinct instantiations of variables
%before unification. 
The resulting most general unifier is then applied
to the two contexts and to $A$ (replacing, if necessary, any variables shared
between $A$ and $B$ in the formula $A$). 
%The $\multimap E$ rule has
%the additional constraint the the conclusion of the rule must satisfy
%the variable convention and therefore that no variable occurs both
%bound and free in the sequent.

%The $\forall I$ rule allows us to replace any free variable not occurring
%in $\Gamma$ by the fresh variable $x$ (note that we are allowed to use
%a variable not occurring in $A$).

We can see from the rules that axioms start negative and stay negative
as long as they are the major premiss of a $\multimap E$ rule or the
premiss of a $\forall E$ rule. We must
switch to positive sequents to use the introduction rules or to use
the sequent as the minor premiss of a $\multimap E$
rule. 

The ``detour'' subproofs we have seen
above cannot receive 
%by labeling all formulas as positive and negative using the
%labeling required by the rule applications. In both cases, this will
%lead to
a consistent labeling: the formula $A\multimap B$ is the
conclusion of a $\multimap I$ rule and must therefore be on the
right-hand side
of a positive sequent, however, it is also the major premiss of a
$\multimap E$ rule and must therefore be on the right-hand side of a negative
sequent (it is easily verified there is no way to transform a positive
sequent into a negative sequent, however the point is that the
\emph{original} detour receives an inconsistent labeling).%: the only way for a positive sequent to become
%negative is to be the minor premiss of a $\multimap E$ rule, but the
%minor premiss disappears and cannot be recovered).

$$
\begin{array}{cc}
\infer[\forall E]{\vdashneg a(y)}{\infer[???]{\vdashneg \forall x. a(x)}{\infer[\forall I]{\vdashpos \forall x. a(x)}{\vdashpos a(y)}}} &
\infer[\multimap E]{\vdashneg B}{\infer[???]{\vdashneg A \multimap
    B}{\infer[\multimap I_i]{\vdashpos A\multimap B}{\infer*{\vdashpos
        B}{[\vdashneg A]^i}}} & \vdashpos A}
\end{array}
$$

\begin{definition}\label{def:track}
A \emph{principal branch} is a sequence of negative sequents which starts at a
hypothesis, then follows all elimination rules from (major) premiss
to conclusion ending at a focus shift rule (this corresponds to the
normal notion of principal branch from e.g.\ \cite{glt}; a sequence of
negative sequents can only pass through the major premiss of a
$\multimap E$ rule and through the single premiss of a $\forall E$ rule).

A \emph{track} is a path
of negative sequents followed by a focus shift followed by a path of
positive sequents. A track ends either
in the conclusion of the proof or in the minor premiss of a
$\multimap E$ rule. 

The \emph{main track} of a proof is the track which ends in its conclusion
 (these definitions corresponds to the
standard notion of  track and main track in normal proofs, see e.g.\ \cite{bpt}).
\end{definition}

This suggests a relation between focused proofs and normal natural
deduction proofs, which is made explicit in the following two propositions.

\begin{proposition}\label{prop:focus} For every natural deduction proof of $\Gamma\vdash B$, there is a
  focused natural deduction proof with unification of $\Gamma \vdashpos B$.
\end{proposition}

\paragraph{Proof} We first transform the natural deduction
proof of $\Gamma\vdash B$ into a normal natural deduction proof, then
proceed by induction on the length of the proof and show that we can
create both a proof of $\Gamma \vdashpos B$ and a substitution
$\sbst$. We proceed by induction on the depth of the proof.

If $d=1$, we have an  axiom or hypothesis rule, which we translate as
follows.

$$
\infer[\shfocus]{A\vdashpos A}{A\vdashneg A}
$$

If $d > 1$ we proceed by case analysis on the last rule.

The only case which requires some attention is the $\multimap E$
case. Given that the proof is normal, we have a normal (sub)proof
which ends in a $\multimap E$ rule. We are therefore on the principal
branch of this subproof and we know that a principal branch starts with
an axiom/lexicon rule then passes only $\forall E$ rules and $\multimap E$ rules through
their major premiss. Hence, the last rule producing the major premiss
in the original proof must either have been an axiom/lexicon rule or
an elimination rule for $\multimap$ or $\forall$.

Now induction hypothesis gives us a proof $\delta_1$ of  $\Gamma
\vdashpos B\multimap A$ and a proof $\delta_2$ of $\Delta \vdashpos
B$. However, given that the last rule of the proof which produces
$\delta_1$ was either axiom/lexicon, the $\forall E$ rule or the
$\multimap E$ rule --- all of which have negative sequents as their
conclusion --- the last rule of
$\delta_1$ must have been the focus shift rule. Removing this focus
shift rule produces a valid proof $\delta_{1'}$ of $\Gamma \vdashneg B\multimap A$,
which we can combine with the proof $\delta_2$ of $\Delta\vdashpos B$ as follows.
$$
\infer[\shfocus]{\Gamma, \Delta \vdashpos A}{\infer{\Gamma, \Delta
    \vdashneg A}{\infer*[\delta_{1'}]{\Gamma\vdashneg B \multimap A}{} & \infer*[\delta_2]{\Delta \vdashpos B}{}}}
$$

Note that this is again a proof which ends with a focus shift rule.

Since the original proof uses the stricter notion of \emph{identity}
(instead of unifiability) for the $B$ formulas, we need not change the
substitution we have computed so far and therefore leave $\Gamma$,
$\Delta$ and $A$ unchanged.

For the $\forall E$ rule, induction hypothesis gives us a proof
$\delta$ of $\Gamma \vdashpos \forall x. A$, by reasoning similar to
the case for $\multimap E$, we know the last rule of $\delta$ was a
focus shift rule, which we can remove, then extend the proof as follows.
$$
\infer[\shfocus]{\Gamma\vdashpos A[x:=D]}{\infer[\forall
  E]{\Gamma\vdashneg A[x:=D]}{\infer*[\delta]{\Gamma \vdashneg \forall x. A}{}}}
$$

Adding the substitution $D := t$ (where $t$ is the term used for the
in the original $\forall E$ rule) to the unifier.

The cases for $\forall I$ and $\multimap I$ are trivial, since we can
extend the proof with the same rule.
\qed

\begin{proposition} For every focused natural deduction proof, there
  is a natural deduction proof. %Moreover, if the focused proof
%  restricts the focus shift rule to atomic formulas, this proof is eta-long.
\end{proposition}

\paragraph{Proof} If we remove the focus shift rule and replace both
$\vdashneg$ and $\vdashpos$ by $\vdash$ then we only need to give
specific instantiations for the $\forall E$ rules. The most general
unifier $\sbst$ computed for the complete proof gives us such values for each
(negatively) quantified variable (if wanted, remaining meta-variables can be
replaced by free variables). \qed

The following is a standard property of normal natural deduction
proofs (and therefore of focused natural deduction proofs).

\begin{proposition} Focused proofs satisfy the \emph{subformula}
  property. That is, any formula occurring in a proof of  $\Gamma
  \vdashpos B$ (or $\Gamma \vdashneg B$) is a subformula either of
  $\Gamma$ or of $B$.
\end{proposition}

The following proposition is easily verified by induction on $A$ and
using the correspondence between natural deduction proofs and $\lambda$-terms.

\begin{proposition} We can restrict the focus shift rule to atomic
  formulas $A$. When we do so, we only produce long normal form proofs
  (which correspond to beta normal eta long
  lambda terms).
\end{proposition}

\editout{
\begin{proposition}\label{propa} We can assume, without loss of generality, that a
  conclusion of the form $\forall x. (B \multimap A)$ has been obtained
  by a consecutive use of the $\multimap I$ and $\forall I$ rules.
\end{proposition}

\paragraph{Proof} This is immediate from the setup of the focalised
system, given that we can restrict the use of the focus shift rule to
atomic formulas. So we are in the following situation

$$
\infer[\forall I]{\Gamma \vdashpos \forall x. (B\multimap A)}{\Gamma
  \vdashpos B \multimap A}
$$

\noindent and, since $B\multimap A$ is a complex formula, we cannot
change to a negative sequent. Hence, the only available rule given the
polarity of the sequent is
$\multimap I$.

An easy alternative proof can be given using the proof nets of the
following section, which is the basic observation that combinations of
positive links can be treated as a unit (ie.\ as a single, combined rule). \qed
}

The proof from the previous section looks as follows in the 
unification-based version of first-order linear logic, though we use a
form with implicit antecedents to economize on horizontal space and
to make comparison with the proof of the previous section
easier. This proof produces the most general unifier $Y=2$, $Z=3$,
corresponding to the explicit instantiations for $y$ and $z$ at the $\forall E$ rules
in the previous proof. 
%Compared to the previous proof, we implicitly apply the two
%$\forall E$ rules at the same time (even though the second occurrence
%is not the main connective of the formula). Unification produces
%$Y=2$ and $Z=3$, which corresponds to the substitutions of the
%previous proof.

\medskip
\noindent\scalebox{.83}{$$
\infer[\shfocus]{\vdashpos s(0,3)}{\infer[\multimap E]{\vdashneg s(0,3)}{
\infer[\forall E]{\vdashneg \forall x. [np(x,2)\multimap s(x,Z)] \multimap s(0,Z)}{\infer[\multimap E]{\vdashneg \forall z. [ \forall x. [np(x,2) \multimap
s(x,z)] \multimap s(0,z) ]}{
\infer[\forall E]{\vdashneg n(1,Y) \multimap \forall z. [ \forall x. [np(x,Y) \multimap
s(x,z)] \multimap s(0,z) ]}{\vdashneg \forall y. [ n(1,y) \multimap
\forall z. [ \forall x. [np(x,y) \multimap
s(x,z)] \multimap s(0,z)] ] }   & \infer[\shfocus]{\vdashpos n(1,2)}{\vdashneg
  n(1,2)}}} & \infer[\shfocus]{\vdashpos \forall v. [np(v,2)
\multimap s(v,3)]}{\vdashpos \forall v. [np(v,2)
\multimap s(v,3)] }}}
$$}
\medskip

\editout{
\medskip
\noindent\scalebox{.83}{$$
\infer[\multimap E]{s(0,3)}{
\infer[\multimap E]{\forall x. [np(x,2) \multimap
s(x,Z)] \multimap s(0,Z) }{
\infer[\shfocus]{n(1,Y) \multimap \forall x. [np(x,Y) \multimap
s(x,Z)] \multimap s(0,Z) }{\forall y. [ n(1,y) \multimap \forall z. [ \forall x. [np(x,y) \multimap
s(x,z)] \multimap s(0,z) ] ]} & n(1,2)} & \infer[\forall I]{\forall x. [np(x,2)
\multimap s(x,3)]}{\infer[\shfocus]{np(V,2)\multimap s(V,3) }{ \forall v. [ np(v,2)\multimap s(v,3) ]}}}
$$}
\medskip
}

Restricting
focus shift ($\shfocus$) to atomic formulas, produces the following
proof in long normal form.
% since the left
%branch of the proof ends with a formula $\forall y.A$, but it is not
%in eta-long form. We can expand it into an eta-long proof as shown below
%and verify this proof does satisfy the form of
%Proposition~\ref{propa}. 
Remark that our hypothesis in this proof
is not $np(V,2)$ but $np(U,W)$ which unifies with $np(V,2)$ at the
 $\multimap E$ rule immediately below it.

\noindent\scalebox{.75}{$$
\infer[\shfocus]{\vdashpos s(0,3)}{
   \infer[\multimap E]{\vdashneg s(0,3)}{
      \infer[\forall E]{\vdashneg \forall x. [np(x,2) \multimap
        s(x,Z) ] \multimap s(0,Z)}{
         \infer[\multimap E]{ \vdashneg \forall z. [\forall x. [np(x,2) \multimap s(x,z)] \multimap s(0,z) ]}{
            \infer[\forall E]{\vdashneg n(1,Y) \multimap \forall z. [ \forall x. [np(x,Y)  \multimap s(x,z)] \multimap s(0,z)] }{\vdashneg 
        \forall y. [ n(1,y) \multimap \forall z. [ \forall x. [np(x,y) \multimap
        s(x,z)] \multimap s(0,z) ] ]}
%\forall y. [ n(1,y) \multimap \forall z. [ (np(X,y) \multimap s(X,z)) \multimap s(0,z) ] ]
       & \infer[\shfocus]{\vdashpos n(1,2)}{\vdashneg n(1,2)}
   }} & %& 
\!\!\!\!\!\!\!\!\!\!\!\!\infer[\forall I]{\vdashpos \forall w. [np(w,2)\multimap s(w,3)]}{
   \infer[\multimap I_1]{\vdashpos np(V,2)\multimap s(V,3)}{
        \infer[\shfocus]{\vdashpos s(V,3)}{\infer[\multimap E]{\vdashneg s(V,3)}{
              \infer[\shfocus]{\vdashpos np(U,W)}{\infer[\textit{Hyp}_1]{\vdashneg np(U,W)}{}}
           & \infer[\forall E]{\vdashneg np(V,2)\multimap
             s(V,3)}{\vdashneg \forall v. np(v,2) \multimap s(v,3)}
         }}
     }
   }
}}
$$}

\subsection{Proof Nets}
\label{sec:pn}

Proof nets are an elegant alternative to natural deduction and an
important research topic in their own right; for reasons of space we 
provide only an informal introduction --- the reader interested in more
detail is referred to \cite{llintro} for an
introduction and to  \cite{multiplicatives,empires}  for detailed proofs in the context of
linear logic and to \cite{pnlambek,diss,mr12lcg} for introductions in
the context of categorial grammars and the Lambek calculus. Though
proof nets shine especially for the $\exists$ and $\otimes$ rules (where the
natural deduction formulation requires commutative conversions to
decide proof equivalence), they are a useful alternative in the
$\forall$ and $\multimap$ case as well since they provide an easy
combinatorial way to do proof search and therefore make arguments
about non-derivability of statements and serve to count the number of readings.

\citeasnoun{quant} shows that the proof nets of multiplicative linear
logic \cite{Girard,multiplicatives} have a simple extension to the
first-order case. Essentially, a proof net is a graph labeled with (polarized
occurrences of) the
(sub)formulas of  a sequent $\Gamma \vdash C$, subject to some conditions we will
discuss below. Obviously, not all graphs labeled with formulas correspond to
derivable statements. However, we can characterize the proof nets
among the larger class of proof structures (graphs labeled with
formulas which, contrary to proof nets, do \emph{not} necessarily correspond
to proofs) by means of simple graph-theoretic properties.

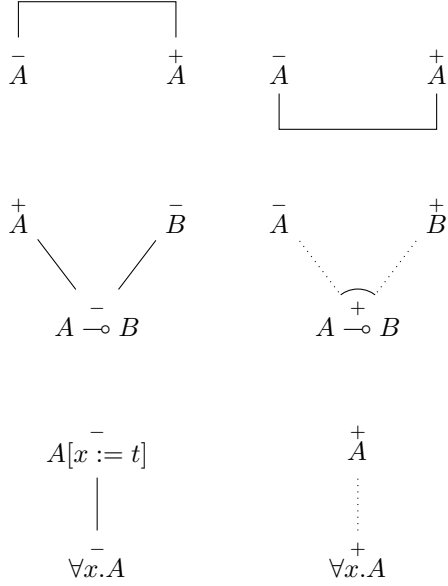
\begin{figure}
\begin{center}
%\begin{tikzpicture}
%\end{tikzpicture}
%\end{center}
%\vspace{\baselineskip}
%\begin{center}
\begin{tikzpicture}
\node (forallnc) {$\overset{-}{\forall x. A}$};
\node (forallnp) [above=2em of forallnc] {$\overset{-}{A[x:=t]}$};
\draw (forallnc) -- (forallnp);
\node (forallpc) [right=7em of forallnc] {$\overset{+}{\forall x. A}$};
\node (forallpp) [above=2em of forallpc] {$\overset{+}{A_{\rule{0pt}{1.55ex}}}$};
\draw [dotted] (forallpc) -- (forallpp);
%
%\node (existsnc) [right=7em of forallpc] {$\overset{-}{\exists x. A}$};
%\node (existsnp) [above=2em of existsnc] {$\overset{-}{A_{\rule{0pt}{1.55ex}}}$};
%\draw [dotted] (existsnc) -- (existsnp);
%
%\node (otimesnc) [above=7em of existsnc] {$\overset{-}{A\otimes B}$};
%\node (tmponl) [left=0.66em of otimesnc] {};
%\node (aotimesnc) [above=2.5em of tmponl] {$\overset{-}{A}$};
%\node (tmponr) [right=0.66em of otimesnc] {};
%\node (botimesnc) [above=2.5em of tmponr] {$\overset{-}{B}$};
%\begin{scope}
%\begin{pgfinterruptboundingbox}
%\path [clip] (otimesnc.center) circle (2.5ex) [reverseclip];
%\end{pgfinterruptboundingbox}
%\draw [dotted] (otimesnc.center) -- (botimesnc);
%\draw [dotted] (otimesnc.center) -- (aotimesnc);
%\end{scope}
%\begin{scope}
%\path [clip] (aotimesnc) -- (otimesnc.center) -- (botimesnc);
%\draw (otimesnc.center) circle (2.5ex);
%\end{scope}
%
%\node (otimespc) [right=7em of otimesnc] {$\overset{+}{A\otimes B}$};
%\node (tmpopl) [left=0.66em of otimespc] {};
%\node (aotimespc) [above=2.5em of tmpopl] {$\overset{+}{A}$};
%\node (tmpopr) [right=0.66em of otimespc] {};
%\node (botimespc) [above=2.5em of tmpopr] {$\overset{+}{B}$};
%\draw (otimespc) -- (aotimespc);
%\draw (otimespc) -- (botimespc);
%
%\node (existspc) [below=7em of otimespc] {$\overset{+}{\exists x. A}$};
%\node (existspp) [above=2em of existspc] {$\overset{+}{A[x:=t]}$};
%\draw  (existspc) -- (existspp);
%
\node (lollinc) [above=7em of forallnc] {$\overset{-}{A\multimap B}$};
\node (tmplnl) [left=0.66em of lollinc] {};
\node (alollin) [above=2.5em of tmplnl] {$\overset{+}{A}$};
\draw (lollinc) -- (alollin);
\node (tmplnr) [right=0.66em of lollinc] {};
\node (blollin) [above=2.5em of tmplnr] {$\overset{-}{B}$};
\draw (lollinc) -- (blollin);
\node (lollipc) [above=7em of forallpc] {$\overset{+}{A\multimap B}$};
\node (tmplpl) [left=0.66em of lollipc] {};
\node (alollip) [above=2.5em of tmplpl] {$\overset{-}{A}$};
\node (tmplpr) [right=0.66em of lollipc] {};
\node (blollip) [above=2.5em of tmplpr] {$\overset{+}{B}$};
\begin{scope}
\begin{pgfinterruptboundingbox}
\path [clip] (lollipc.center) circle (2.5ex) [reverseclip];
\end{pgfinterruptboundingbox}
\draw [dotted] (lollipc.center) -- (blollip);
\draw [dotted] (lollipc.center) -- (alollip);
\end{scope}
\begin{scope}
\path [clip] (alollip) -- (lollipc.center) -- (blollip);
\draw (lollipc.center) circle (2.5ex);
\end{scope}
% Axiom/Cut
\node (anx) [above=3.5em of alollin] {$\overset{-}{A}$};
\node (anp) [above=3.5em of blollin] {$\overset{+}{A}$};
\node (anxa) [above=1em of anx] {};
\node (anpa) [above=1em of anp] {};
\draw (anx) -- (anxa.center) -- (anpa.center) -- (anp);
\node (cnx) [above=3.5em of alollip] {$\overset{-}{A}$};
\node (cnp) [above=3.5em of blollip] {$\overset{+}{A}$};
\node (cnxb) [below=1em of cnx] {};
\node (cnpb) [below=1em of cnp] {};
\draw (cnx) -- (cnxb.center) -- (cnpb.center) -- (cnp);
\end{tikzpicture}
\end{center}
\caption{Links for proof structures in the $\forall$, $\multimap$
  fragment of first-order linear logic.}
\label{fig:links}
\end{figure}

The basic building blocks of proof structures are \emph{links}, as
shown in 
Figure~\ref{fig:links}. We will call the formulas displayed below the link their
\emph{conclusion} and the formulas displayed above it their
premisses. The axiom link (top left) has no premisses and two
conclusions, the cut link has no conclusions and two premisses, the
binary logical links have two premisses ($A$ and $B$) and one
conclusion $A\multimap B$ and the unary logical links have one premiss
$A$ and one conclusion $\forall x.A$. We will call $x$ the
eigenvariable of the link and require that all links use distinct
variables.% --- this restriction is standard in linear logic \cite{quant,empires}.
%Throughout this paper, we will use the standard convention in
%irst-order linear logic \cite{quant,empires} that every occurrence of
%a quantifier $\forall$, $\exists$ in a sequent uses
%a distinct variable.

Given a statement $A_1, \ldots, A_n \vdash C$
we can unfold the formulas using the logical links of the figure, using the
negative links for the $A_i$ and the positive link for $C$. Since
there is only one type of link for each combination of
connective/polarity, we unfold our formulas
deterministically\footnote{For the negative $\forall$ this is not
  immediately obvious, since we need to choose a suitable term $t$. We
  will discuss this case below but we will essentially use
  meta-variables and
  unification just like we did for natural deduction in Section~\ref{sec:unif}.}, until
we end up at the atomic formulas and have produced a ``formula
forest'', a sequence of formula decomposition trees labeled with some additional
information (polarity labels and dashed lines), which is sometimes called a \emph{proof frame}.

We turn this proof
frame into a \emph{proof structure} by connecting atomic formulas of
opposite polarity in such a way there is a perfect matching between
the positive and negative atoms. This step can already fail, for
example if the number of positive and negative occurrences of an
atomic formula differ but also because of incompatible atomic
formulas
like $a(0,1)$ and $a(x,1)$, with $x$ the eigenvariable of a
$\forall^+$ link. More generally, it can be the case that there is no
coherent substitution which allows us to perform a complete matching
of the atomic formulas using axiom links. These restrictions on the
instantiations of variables are a powerful tool for proof search \cite{moot07filter,moot13lambek}.

Proof structures are essentially
graphs where some of the links are drawn with dashed lines; the binary
dashed lines are paired, as indicated by the connecting arc. We will
call the dashed logical links ($\forall^+$ and $\multimap^+$) the
\emph{positive} links and the solid logical links ($\forall^-$ and $\multimap^-$)
the \emph{negative} links. The terms positive and negative
links only apply to the logical links; the axiom and cut link
are neither positive nor negative. A proof structure containing only
negative logical links is just a graph labeled with polarized formulas.

Figure~\ref{fig:psex} shows the proof net which corresponds to the
natural deduction proof of Section~\ref{sec:nounif}. To save space, we
have noted only the main connective at each link; the full formula can
be obtained unambiguously from the context. We have also been free
in the way we ordered the premisses of the $\multimap$ links, which
allows us to give a planar presentation of the axiom links, much like
Lambek calculus proof nets. However, there is no planarity requirement
in the proof net calculus; the first-order variables offer more
flexibility than simple planarity. For the $\forall^-$ links, we have
annotated the substitutions next to the link. If we use a
unification-based presentation, as we did for natural deduction in
Section~\ref{sec:unif}, we can ``read off'' these substitutions from the most
general unifier computed for the axioms (as opposed to natural
deduction, the axioms and not the $\multimap E$ rule, which
corresponds to the $\multimap^-$ link, are responsible for the
unification of variables).

\begin{figure}
\begin{center}
\begin{tikzpicture}
\node (b1) at (6,0) {$\overset{-}{\forall y}$};
\node (bv1) at (6,1.2) {$\overset{-}{\multimap}$};
\node (npos) at (7,2.4) {$\overset{+}{n(1,2)}$};
\node (fn1) at (5,2.4) {$\overset{-}{\forall z}$};
\node (fi1) at (5,3.6) {$\overset{-}{\multimap}$};
\node (s1) at (4,4.8) {$\overset{-}{s(0,3)}$};
\node (fx) at (6,4.8) {$\overset{+}{\forall x}$};
\node (pi) at (6,6) {$\overset{+}{\multimap}$};
\node (s2) at (5,7.2) {$\overset{+}{s(x,3)}$};
\node (np1) at (7,7.2) {$\overset{-}{np(x,2)}$};
\node (n2) at (9,0) {$\overset{-}{n(1,2)}$};
\node (b3) at (12,0) {$\overset{-}{\forall v}$};
\node (f3) at (12,1.2) {$\overset{-}{\multimap}$};
\node (np2) at (11,2.4) {$\overset{+}{np(x,2)}$};
\node (s3) at (13,2.4) {$\overset{-}{s(x,3)}$};
\node (g) at (14,0) {$\overset{+}{s(0,3)}$};
%
%\path[->] (sisi) edge node[right] {$\lambda y^{\sigma}$} (sbis);
%\path[draw] (b1) node[right] {$y:=2$} (bv1);
\draw (b1) -- (bv1);
\node (y) at (6.6,0.6) {$y:=2$};
\draw (bv1) -- (npos);
\draw (bv1) -- (fn1);
\draw (fn1) -- (fi1);
\node (z) at (5.6,3.0) {$z:=3$};
\draw (fi1) -- (s1);
\draw (fi1) -- (fx);
\draw[dashed] (fx) -- (pi);
\draw[dashed] (pi) -- (s2);
\draw[dashed] (pi) -- (np1);
\draw (b3) -- (f3);
\node (v) at (12.6,0.6) {$v:=x$};
\draw (f3) -- (np2);
\draw (f3) -- (s3);
% axioms
\draw (npos) -- (7,3.2) -- (9,3.2) -- (n2);
\draw (np1) -- (7,8.0) -- (11,8.0) -- (np2);
\draw (s2) -- (5,8.6) -- (13,8.6) -- (s3);
\draw (s1) -- (4,9.2) -- (14,9.2) -- (g);
\end{tikzpicture}
\end{center}
\caption{Proof net corresponding to the natural deduction proof of
  Section~\ref{sec:nounif}}
\label{fig:psex}
\end{figure}

A proof structure is a \emph{proof net} if the statement $A_1,\ldots, A_n
\vdash C$ is derivable, that is, given the proof of
Section~\ref{sec:nounif}, we know the proof structure of
Figure~\ref{fig:psex} is a proof net. However, this definition is
not very useful, since it depends on finding a proof in some other
proof system; we would like to use the proof structure itself
to directly decide whether or not the statement is derivable. However, it is possible to distinguish the proof
nets from the other proof structures by simple graph-theoretic
properties. To do so, we first introduce some auxiliary notions, which
turn the graph-like proof structures into standard graphs. Since
axiom, cut and the
negative links already produce normal graphs ($\multimap^-$
corresponds to two edges, all other links to a single edge in the graph), we only need a way to
remove the positive links.

\begin{definition}
A \emph{switching} is a choice for each positive link as follows.
\begin{itemize}
\item For each $\multimap^+$ link, we choose one its
premisses ($A$ or $B$).
\item For each $\forall^+$ link, we choose \emph{either} its premiss $A$ \emph{or} any of
the formulas in the proof structure containing a free occurrence of
the eigenvariable of the link.
\end{itemize}

A given a switching $s$, a \emph{correction graph} is a proof structure
where we replace all dashed links by a link from the conclusion of the
link to the formula chosen by the switching $s$.
\end{definition}

\begin{theorem} \cite{quant} A proof structure is a \emph{proof net} iff all its
  correction graphs are acyclic and connected.
\end{theorem}

Defined like this, it would seem that deciding whether or not a proof
structure is a proof net is rather complicated: there are potentially
many correction graphs --- we have two independent possibilities for each $\multimap^+$
link and generally at least two subformulas containing the
eigenvariable of each $\forall^+$ link, giving $2^n$ correction graphs
for $n$ positive links --- and we need verify all of them. Fortunately,
there are very efficient alternatives: linear time in the
quantifier-free case \cite{murong,pnlinear} and at most squared time, though possibly better,
in the case with quantifiers \cite{moot13lambek}.

Going back to the example shown in Figure~\ref{fig:psex}, we can see
that there are two positive links and twelve correction graphs: there
are six free occurrences of $x$ --- four in atomic formulas and two
additional occurrences in the 
conclusions ($\multimap^+$ and $\multimap^-$) which combine these
atomic formulas into $np(x,2) \multimap s(x,2)$ --- times the two independent possibilities for switching
$\multimap^+$ left or right. We can verify that all twelve
possibilities produce acyclic, connected graphs. Removing the positive
links splits the graph into three connected components: the single
node labeled $\multimap^+$ (representing $(np(x,2) \multimap
s(x,2))^+$), a component containing the intransitive verb ending at
the axioms to $s(x,3)^+$ and $np(x,2)^-$ and a final component
containing the rest of the graph, ending at the conclusion of the
$\forall^+$ link (which has been disconnected from its premiss). Now, any
switching for the $\multimap^+$ link will connect its isolated
conclusion node
to the component containing  $s(x,3)^+$ and $np(x,2)^-$  (via one or
the other of these nodes), leaving two connected components. Finally,
all free occurrences of the variable $x$ occur in this newly created
component, therefore any choice for a switching of the $\forall^+$
link will join these disconnected components into a single, connected component.
Since each choice connected two disjoint components, we have not
generated any cycles.

We can also show that this is the only possible proof structure for
the given logical statement: there is only one choice for the $n$
formulas, one choice for the $np$ formulas though two choices for the
$s$ formulas. However, the alternative proof structure would link $s(0,z)$ to
$s(x,z)$ (for some value of $z$), which fails because $x$, being the
eigenvariable of a $\forall^+$ link, cannot be instantiated to 0.

As a second example, let's show how we can use correction graphs to
show underivability.
Though it is clear that the switching for the universal quantifier
must refer to free occurrences of its eigenvariable somewhere (as do
its counterparts in natural deduction and sequent calculus), it is not
so easy to find a small example in the $\forall,\multimap$ fragment
where this condition is necessary to show underivability, since
finding a global instantiation of the variables is already a powerful
constraint on proof structures. However, the existential quantifier
and the universal quantifier differ only in the labeling of
formulas for the links and we need the formula labeling only for
determining the free variables.

\begin{figure}
\begin{center}
\begin{tikzpicture}
\node (b1) at (1.5cm,0cm) {$\overset{-}{\forall x. a(x)  \multimap b}$};
\node (lax)  at (0cm,1.5cm) {$\overset{+}{\forall x. a(x)}$};
\node (lb) at (3cm,1.5cm)  {$\overset{-}{b}$};
\node (g) at (6cm,0cm) {$\overset{+}{\exists y. [ a(y) \multimap b ] }$};
\node (g2) at (6cm,1.5cm) {$\overset{+}{a(x) \multimap b}$};
\draw (g) -- (g2);
\node  (ga) at (7.5cm,3cm) {$\overset{-}{a(x)}$};
\node  (gb) at (4.5cm,3cm) {$\overset{+}{b}$};
\node (la) at (0cm,3cm) {$\overset{+}{a(x)}$};
\draw (ga) -- (7.5cm,4.5cm) -- (0cm,4.5cm) -- (la);
\draw (gb) -- (4.5cm,4cm) -- (3cm,4cm) -- (lb);
\draw[dashed] (lax) -- (la);
\draw (b1) -- (lax);
\draw (b1) -- (lb);
\draw[dashed] (g2) -- (ga);
\draw[dashed] (g2) -- (gb);
\end{tikzpicture}
\end{center}
\caption{Proof structure which is not a proof net}
\label{fig:noproof}
\end{figure}
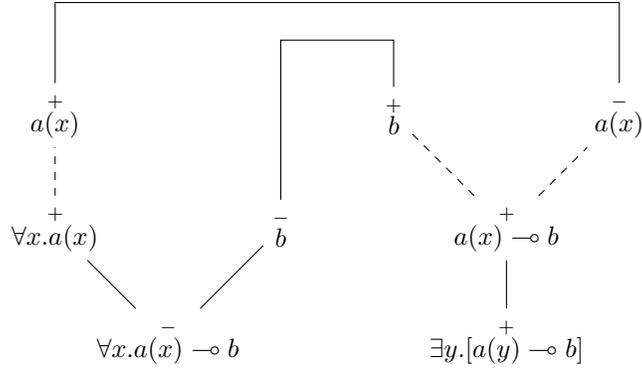

A proof structure of the underivable sequent $(\forall
x. a(x))  \multimap b \nvdash \exists y. [ a(y) \multimap b ] $ is
shown  in Figure~\ref{fig:noproof}. It is easy to verify this is the
unique proof structure corresponding to this sequent. This sequent is used for
computing the prenex normal form of a formula in classical logic
(replacing $\multimap$ by $\Rightarrow$), but it is invalid in
intuitionistic logic and linear logic since it depends on the
structural rule of right contraction.

\begin{figure}
\begin{center}
\begin{tikzpicture}
\node (b1) at (1.5cm,0cm) {$\overset{-}{\forall x. a(x)  \multimap b}$};
\node (lax)  at (0cm,1.5cm) {$\overset{+}{\forall x. a(x)}$};
\node (lb) at (3cm,1.5cm)  {$\overset{-}{b}$};
\node (g) at (6cm,0cm) {$\overset{+}{\exists y. [ a(y) \multimap b ] }$};
\node (g2) at (6cm,1.5cm) {$\overset{+}{a(x) \multimap b}$};
\draw (g) -- (g2);
\node  (ga) at (7.5cm,3cm) {$\overset{-}{a(x)}$};
\node  (gb) at (4.5cm,3cm) {$\overset{+}{b}$};
\node (la) at (0cm,3cm) {$\overset{+}{a(x)}$};
\draw (ga) -- (7.5cm,4.5cm) -- (0cm,4.5cm) -- (la);
\draw (gb) -- (4.5cm,4cm) -- (3cm,4cm) -- (lb);
%\draw[dashed] (lax) -- (la);
\draw (b1) -- (lax);
\draw (b1) -- (lb);
\draw (g2) -- (gb);
\draw plot [smooth] coordinates {(lax.north) (2.5cm,4.0cm) (5.0cm,4.0cm) (g2.north)};  
\end{tikzpicture}
\end{center}
\caption{A cyclic and disconnected correction graph for the proof
  structure of Figure~\ref{fig:noproof}}
\label{fig:cycle}
\end{figure}
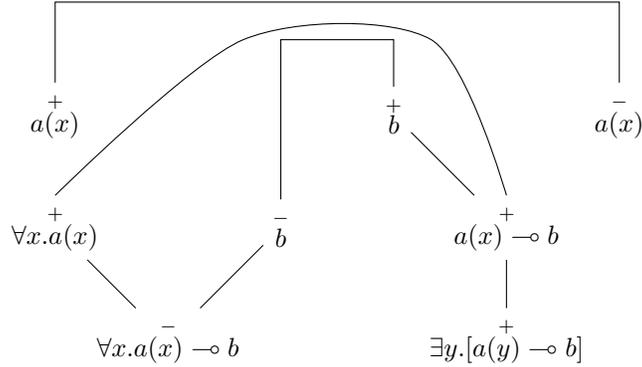

In order to show the sequent is invalid in linear logic, it suffices
to find a switching such that the corresponding correction graph
either contains a cycle or is disconnected. Figure~\ref{fig:cycle} shows a
correction graph for the proof structure of Figure~\ref{fig:noproof}
which is both cyclic and disconnected: the axiom $a(x) \vdash a(x)$
is not connected to the rest of the structure and the connection
between $\forall x. a(x)$ and $a(x) \multimap b$ produces a cycle,
since there is a second path to these two formulas through the axiom
$b\vdash b$.

This concludes our brief introduction to proof nets for first-order
linear logic. We refer the reader to Appendix~A of \cite{glt} for
discussion about the relation between proof nets and natural deduction.

\section{Basic Properties of the Simply Typed Lambda Calculus}
\label{sec:stl}

Before introducing hybrid type-logical grammars, we will first review
some basic properties of the simply typed lambda calculus which will
prove useful in what follows. This section is not intended as a
general introduction to the simply typed lambda calculus: we will
assume the reader has at least some basic knowledge such as can be
found in Chapter~3 of \cite{glt} or other textbooks and some knowledge
about substitution and most general unifiers. For more detail, and for
proofs of the lemmas and propositions of this section, the reader is
referred to \cite{hindley}.

A remark on notation: we will use $\rightarrow$ exclusively as a type
constructor (also when we know we are using it to type a linear lambda term) and
$\multimap$  exclusively as a logical connective.

\begin{definition} A lambda term $M$ is a \emph{linear} lambda term
  iff 
\begin{enumerate}
\item for every subterm $\lambda x. N$ of $M$,  $x$ has exactly one
  occurrence in $N$ (in other words, each abstraction binds exactly one variable occurrence),
\item all free variables of $M$ occur exactly once.
\end{enumerate}
\end{definition}

Table~\ref{tab:typing} lists the Curry-style typing rules for the
linear lambda calculus. For the $\ra E$ rule, $\Gamma$ and $\Delta$
cannot share term variables; for the $\ra I$ rule, $\Gamma$ cannot contain
$x$ (ie.\ $\Gamma, x^{\alpha}$ must be a valid context).

\begin{table}
$$
\begin{array}{c}
x^{\alpha} \vdash x:\alpha \\
\\
\infer[\ra E]{\Gamma,\Delta\vdash (M\,N):\beta}{\Gamma \vdash
  M:\alpha\ra \beta & \Delta \vdash N:\alpha}\\
\\
\infer[\ra I]{\Gamma\vdash \lambda x.M:\alpha\ra\beta}{\Gamma,x^{\alpha} \vdash M:\beta}
\end{array}
$$
\caption{Curry-style typing rules for the linear lambda calculus}
\label{tab:typing}
\end{table}

\begin{proposition}\label{prop:linear} For linear lambda terms, we have the following:
\begin{enumerate}
\item\label{prop:fv} When $M$ is a linear lambda term and $\Gamma \vdash
  M:\alpha$ a deduction of $M$, then the variables occurring in
  $\Gamma$ are exactly the free variables of $M$.
\item If $M$, $N$ are linear lambda terms which do not share free
  variables then $(M\,N)$ is a linear lambda term.
\item If $M$ is a linear lambda term with a free occurrence of $x$
  then $\lambda x.M$ is a linear lambda term.
\item If $M$ is a linear lambda term and
  $M\twoheadrightarrow_{\beta\eta} N$ then $N$ is a linear lambda
  term.
\end{enumerate}
\end{proposition}

% Not sure if we need this
\editout{
Inversion Lemma, case (iii). Proposition 1.2.3, case (iii) from
Barendregt e.a.

\begin{proposition} If $\Gamma\vdash \lambda x.M:A$, then $A \equiv
  B\ra C$ and $\Gamma, x:B \vdash M:C$.
\end{proposition}
}

%Substitution tells us we can replace a variable of a certain type by a
%complex term of that sam

\begin{lemma}[Substitution]\label{lem:subst} If $\Gamma, x:\alpha \vdash M:\beta$,
  $\Delta\vdash N:\alpha$ and $\Gamma$ and $\Delta$ are compatible
  (ie.\ there are no conflicting variable assignments and therefore
  $\Gamma,\Delta$ is a valid context),
  then $\Gamma,\Delta \vdash M[x:=N]:\beta$.
\end{lemma}

%\paragraph{Proof}

The following two results are rather standard, we can find them in
\cite{hindley} as Lemmas~2C1 and 2C2.

\begin{lemma}[Subject Reduction]\label{lemma:sr} Let $M \twoheadrightarrow_{\beta\eta}
  N$, then $\Gamma \vdash M:\alpha \Rightarrow \Gamma \vdash N:\alpha$
\end{lemma}

\begin{lemma}[Subject Expansion]\label{lemma:se} Let $M \twoheadrightarrow_{\beta\eta}
  N$ with $M$ a linear lambda term, then $\Gamma \vdash N:\alpha \Rightarrow \Gamma \vdash M:\alpha$
\end{lemma}

\subsection{Principal types}
\label{sec:pt}

The main notions from Chapter~3 of \cite{hindley} are the following.

\begin{definition}[Principal type] A \emph{principal type} of a term
  $M$ is a type $\alpha$ such that
\begin{enumerate}
\item for some context $\Gamma$ we have
  $\Gamma \vdash M:\alpha$
\item if $\Gamma' \vdash M:\beta$, then there is a substitution
  $\sbst$ such that $\sbst(\alpha) = \beta$.
\end{enumerate}
\end{definition}

\begin{definition}[Principal pair] A \emph{principal pair} for a term
  $M$ is a pair $\langle \Gamma, \alpha\rangle$ such that $\Gamma
  \vdash M:\alpha$ and for all $\beta$ such that $\Gamma \vdash
  M:\beta$ there is a substitution $\sbst$ with $\sbst(\alpha) = \beta$
\end{definition}

\begin{definition}[Principal deduction] A \emph{principal deduction}
  for a term $M$ is a derivation $\delta$ of a statement $\Gamma
  \vdash M:\alpha$ such that every other derivation with term $M$ is
  an instance of $\delta$ (ie.\ obtained by globally applying a
  substitution $\sbst$ to all types in the proof).
\end{definition}

From the definitions above, it is clear that if $\delta$ is a
principal deduction for $\Gamma \vdash M:\alpha$ then $\Gamma,\alpha$
is a principal pair and $\alpha$ a principal type of $M$.

%For a closed
%term, $\Gamma$ is empty (by
%Proposition~\ref{prop:linear}.\ref{prop:fv}). However, 
If $M$ contains
free variables $x_1,\ldots x_n$ we can compute the principal
type $\alpha_1 \ra \ldots (\alpha_n \ra \beta)$ of the closed term $\vdash \lambda
x_1,\ldots x_n.M$ which is the same as the principal type for
$x_1^{\alpha_1},\ldots,x_n^{\alpha_n}\vdash M^{\beta}$.

%Note also that computing the principal type $\alpha$ 

\subsection{The principal type algorithm}

The principal type algorithm of \citeasnoun{hindley} is defined as
follows. It is slightly more general and computes principal
\emph{deductions}. It takes as input a lambda term $M$ and outputs either its
principal type $\alpha$ or fails in case $M$ is untypable. We closely
follow Hindley's presentation, keeping his numbering but restricting
ourselves to linear lambda terms; we omit his
correctness proof of the algorithm.

We proceed by induction on the construction of $M$.

{
\renewcommand{\theenumi}{\Roman{enumi}}
\renewcommand{\theenumii}{\alph{enumii}}
\renewcommand{\theenumiii}{\arabic{enumiii}}

\begin{enumerate}
\item If $M$ is a variable, say $x$, then we take an unused type variable
  $\alpha$ and return $x^{\alpha} \vdash x:\alpha$ as principal
  deduction.
\item If $M$ is of the form $\lambda x. N$ and $x$ occurs in $N$ then
  we look at the principal deduction $\delta$ of $N$ by induction hypothesis
  : if we fail to compute a principal deduction for $N$ then there is
  no principal deduction for $\lambda x. N$ either. If such a
  deduction $\delta$ does exist, then we can extend it as follows.

$$
\infer[\multimap I]{\Gamma \vdash \lambda
  x.N:\alpha\ra\beta}{\infer*[\delta]{x^{\alpha},\Gamma\vdash N:\beta}{}}
$$

\item $M$ is of the form $\lambda x. N$ and $x$ does not occur in $N$;
  this case cannot occur since it violates the condition on linear
  lambda terms (we must bind exactly one occurrence of $x$ in $N$), so we fail.

\item\label{pt:appl} $M$ is of the form $(N\, P)$. If the algorithm fails for either
  $N$ or $P$, then $M$ is untypable and we fail. If not, induction
  hypothesis gives us a principal proof $\delta_1$ for $\Gamma \vdash
  N:\gamma'$ and a principal proof $\delta_2$ for $\Delta\vdash
  P:\gamma$.  If necessary, we rename type variables if $\Gamma$ and $\Delta$ such
  that $\Gamma$ and $\Delta$ have no type variables in common. Since
  $M$ is linear, $N$ and $P$ cannot share term variables.

\begin{enumerate}
 \item If $\gamma'$ is of the form $\alpha \rightarrow \beta$
% Since $M$ is
%  eta-long, $N$ is eta-long and therefore of the form $\alpha\ra\beta$
%  (Lemma 3.22 from \citeasnoun{kanazawa11pg}, removing the need for
%  case~\ref{pt:appl}b from \citeasnoun{hindley}, which will be a
%  simplification later).
  then we compute the most general unifier $\sbst$ of $\langle \Gamma,
  \alpha\rangle$ and $\langle \Delta,\gamma\rangle$. If this fails the
  term is untypable; if not we combine the proofs as
  follows.

$$
\infer[\ra E]{\sbst(\Gamma),\sbst(\Delta) \vdash (N\,
  P):\sbst(\beta)}{\infer*[\sbst(\delta_1)]{\sbst(\Gamma)\vdash N:\sbst(\alpha)\ra\sbst(\beta)}{}
  & \infer*[\sbst(\delta_2)]{\sbst(\Delta)\vdash P:\sbst(\gamma)}{}}
$$

 \item If $\gamma'$ is a type variable, then we compute the most
   general unifier $\sbst$ of $\langle \Gamma,\gamma'\rangle$ and $\langle
   \Delta,\gamma\ra\beta\rangle$ (with $\beta$ a fresh type
   variable). If this succeeds and the term is typable, we can produce
   its principal proof as follows.

$$
\infer[\ra E]{\sbst(\Gamma),\sbst(\Delta) \vdash (N\,
  P):\sbst(\beta)}{\infer*[\sbst(\delta_1)]{\sbst(\Gamma)\vdash N:\sbst(\gamma)\ra\sbst(\beta)}{}
  & \infer*[\sbst(\delta_2)]{\sbst(\Delta)\vdash P:\sbst(\gamma)}{}}
$$

\end{enumerate}
\end{enumerate}
}

The main utility of principal types in the current paper
is given by the coherence theorem.

\begin{theorem}[Coherence]\label{thm:coherence} Suppose $\Gamma \vdash N:\alpha$ and let
  $\alpha$ be a principal type of $N$ then $\forall
  P \, \forall \Gamma' \subseteq \Gamma\  \ \Gamma' \vdash P:\alpha
  \ \Longrightarrow\  P \equiv_{\beta\eta} N$
\end{theorem}

The coherence theorem states that a principal type determines a
lambda term uniquely (up to $\beta\eta$ equivalence). Since we work in
a linear system, where weakening is not allowed, we only need the
special case $\Gamma' = \Gamma$. This special case of
Theorem~\ref{thm:coherence} is the following: if $\Gamma\vdash N:\alpha$ with $\alpha$ a principal type of
$N$ then for any $P$ such that $\Gamma \vdash P:\alpha$ we have that
$P \equiv_{\beta\eta} N$.

In brief, the principal type algorithm allows us to
compute the principal type of a given typable lambda term, whereas the
coherence theorem allows us to reconstruct a lambda term (up to
$\beta\eta$ equivalence) from a principal type.

\begin{definition} We say a sequent $\Gamma\vdash C$ is \emph{balanced} if all
  atomic types occurring in the sequent occur exactly twice.
\end{definition}

The following lemmas are easy consequences of 1) the Curry-Howard
isomorphism between linear lambda terms and Intuitionistic Linear
Logic (ILL), which allows us to interpret the linear type constructor
``$\ra$'' as the logical connective ``$\multimap$'' 2) the correspondence between (normal) natural deduction proofs
and (cut-free) proof
nets and 3) the fact that renaming the conclusions of the axiom links
in a proof net gives another proof net.

\begin{lemma}\label{lemma:balanced} If $M$ is a linear
  lambda term with free variables $x_1,\ldots,x_n$ then the principal
  type $\alpha_1 \ra \ldots (\alpha_1 \ra \beta)$ of $\lambda x_1 \ldots
  \lambda x_n.M$ is balanced. Hence the principal type of
  $x_1^{\alpha_1},\ldots x_n^{\alpha_n} \vdash M^{\beta}$ is balanced.
\end{lemma}

\paragraph{Proof} Compute the natural deduction proof of $M$ and
convert it to a ILL proof net. By subject reduction
(Lemma~\ref{lemma:sr}), normalization/cut elimination keeps the type
$\alpha$ invariant. Let $P$ be the cut-free proof net which corresponds to
the natural deduction proof of $M$ and which has the same type as $M$.
We obtain a balanced proof net by using a different atomic formula for
all axiom links. From this proof net, we can obtain all other types of
$M$ by renaming the axiom links (allowing for non-atomic axiom links), hence it is a principal type and it
is balanced by construction. \qed

\begin{lemma}If $M$ is a beta-normal lambda term with free variables
  $x_1,\ldots,x_n$ and if $\lambda x_1,\ldots,\lambda x_n.M$ has a 
  balanced typing then $M$ is linear.
\end{lemma}

\paragraph{Proof} If $\lambda x_1,\ldots,\lambda x_n.M$ has a 
  balanced typing, then from this typing we can construct a unique cut-free ILL
  proof net of $\lambda x_1,\ldots,\lambda x_n.M$. Since it is an ILL
  proof net, this lambda term must be linear and therefore $M$ as well.
\qed

\subsection{Examples}
\label{sec:exc}

To illustrate the principal type algorithm, we give two examples in
this section.

As a first example, we compute the principal proof of $\textbf{C}
\equiv \lambda f . \lambda x .
\lambda  y . ((f\, y)\, x)$ 
%by first eta-expanding it to
%$\lambda f . \lambda x
%\lambda y ((\lambda v.\lambda w.(f\, v)\,w)\, y)\, x)$ and then
%computing the principal type
as follows. % $(\beta \ra \alpha \ra \gamma) \ra \alpha \ra \beta \ra \gamma$.

$$
\infer[\ra I]{\vdash \lambda f. \lambda x.\lambda y.  ((f\,y)\,x):(\beta\ra\alpha\ra\gamma)\ra\alpha\ra\beta\ra\gamma}{
\infer[\ra I]{f^{\beta\ra\alpha\ra\gamma}\vdash \lambda x.\lambda y.  ((f\,y)\,x):\alpha\ra\beta\ra\gamma}{
\infer[\ra I]{f^{\beta\ra\alpha\ra\gamma},x^{\alpha}\vdash\lambda y.  ((f\,y)\,x):\beta\ra\gamma}{
\infer[\ra E]{f^{\beta\ra\alpha\ra\gamma},y^{\beta},x^{\alpha}\vdash
  ((f\,y)\,x):\gamma}{\infer[\ra
  E]{f^{\beta\ra\gamma_1},y^{\beta}\vdash
    (f\,y):\gamma_1}{f^{\gamma_0} \vdash f:\gamma_0 & y^{\beta} \vdash
    y:\beta} & x^{\alpha}\vdash x:\alpha}}}}
$$
%\bigskip

The substitutions $\gamma_0 := \beta\ra\gamma_1$ (for the topmost $\ra
E$ rule) and $\gamma_1 :=
\alpha\ra \gamma$ (for the bottom $\ra E$ rule) have been left implicit
in the proof.

As a second example, the principal proof of $l^{2\ra 1} \vdash \lambda
O. \lambda S . \lambda z. (S\,(l\, (O\, z)))$ is the following.

$$
\infer[\ra I]{l^{2\ra 1}\vdash \lambda O. \lambda S. \lambda
  z. (S\,(l\,(O\, z))):(\beta\ra 2)\ra(1\ra\alpha)\ra\beta\ra\alpha}{
\infer[\ra I]{l^{2\ra 1},O^{\beta\ra
    2}\vdash \lambda S. \lambda z. (S\,(l\,(O\, z))):(1\ra\alpha)\ra\beta\ra\alpha}{
\infer[\ra I]{S^{1\ra\alpha},l^{2\ra 1},O^{\beta\ra
    2}\vdash \lambda z. (S\,(l\,(O\, z))):\beta\ra\alpha}{
\infer[\ra E]{S^{1\ra\alpha},l^{2\ra 1},O^{\beta\ra
    2},z^{\beta}\vdash (S\,(l\,(O\, z))):\alpha}{S^{\alpha_2} \vdash S:\alpha_2 &
\infer[\ra E]{l^{2\ra 1},O^{\beta\ra 2},z^{\beta}\vdash (l\,(O\, z)):1}{l^{2\ra 1}\vdash l:2\ra 1 &\infer[\ra E]{O^{\beta\ra\alpha_1},z^{\beta} \vdash (O\, z):\alpha_1}{O^{\alpha_0}\vdash O:\alpha_0 & z^{\beta}\vdash z:\beta}}}}}}
$$

The substitutions $\alpha_0 := \beta\ra\alpha_1$, $\alpha_1 := 2$,
$\alpha_2 := 1\ra\alpha$ (of the three $\ra E$ rules, from top to
bottom) have again been left implicit.

\section{Hybrid Type-Logical Grammars}
\label{sec:hybrid}

Hybrid type-logical grammars have been introduced in \cite{kl12gap} as
an extension of lambda grammars which combines insights from the
Lambek calculus into lambda grammars. Depending on authors,
lambda grammars \cite{muskens03lambda} are also called abstract
categorial grammars \cite{groote01acg} or linear
grammars \cite{pollard11linear}.

Formulas of hybrid type-logical grammars are defined as follows, where
$\mathcal{F}_2$ are the formulas of hybrid type-logical grammars and
$\mathcal{F}_1$ the formulas of Lambek grammars. $\mathcal{A}$ denotes
the atomic formulas of the Lambek calculus --- we will call these
formulas \emph{simple} atomic formulas, since their denotations are strings --- $\mathcal{B}$
signifies complex atomic formulas, whose denotations are not simple
strings, but string \emph{tuples}. %The definition below essentially
%follows K&L 2014 (though with the distinction of two types of atomic formulas)
\begin{align*}
\mathcal{F}_0 &::= \mathcal{A} \\
\mathcal{F}_1 &::= \mathcal{F}_0 \,\ | \,\ \mathcal{F}_1 / \mathcal{F}_1 \,\
| \,\ \mathcal{F}_1 \backslash \mathcal{F}_1 \\
\mathcal{F}_2 & ::= \mathcal{B} \,\ | \,\ \mathcal{F}_1 \,\ | \,\ \mathcal{F}_2|\mathcal{F}_2
\end{align*}

As is clear from the
recursive definition of formulas above, hybrid type-logical grammars
are a sort of layered or fibred logic. Such logics have been studied
before as extensions of the Lambek calculus by replacing the atomic
formulas in $\mathcal{F}_0$ by feature logic formulas \cite{bj,dm}.

Lambek grammars are obtained by not allowing 
connectives or complex atoms in $\mathcal{F}_2$. From hybrid
type-logical grammars, we obtain
lambda grammars by not allowing connectives in
$\mathcal{F}_1$. Inversely, we can see hybrid type-logical grammars as
lambda grammars where simple atomic formulas have been replaced by
Lambek formulas.

Before presenting the rules of hybrid type-logical grammars, we'll
introduce some notational conventions: $A$ and $B$ range over arbitrary formulas; $C$, $D$ and $E$ denote type variables or type constants; %(in the
%Church version of the system there are
%no type variables and only one type constant $\sigma$).
$n$ and $n-1$
denote type constants corresponding to string positions; $\alpha$ and
$\beta$ denote arbitrary types. Types are written as superscripts to
the terms; $x$, $y$ and $z$ denote term variables; $M$ and $N$ denote arbitrary terms.

Table~\ref{tab:htlg} shows the rules of Hybrid Type-Logical
Grammars. The rules are presented in such a way that they compute principal types in
addition to the terms. We obtain the Church-typed version ---
equivalent to the calculus presented in \cite{kl12gap} --- by
replacing all type variables and constants by the type constant
$\sigma$. For the principal types, we use the Curry-typed version,
though for readability, we often write the types of subterms as superscripts as well.

\editout{

$$
\infer[/E]{\lambda z. w_1(\ldots(w_{n+m}\, z)) : A}{\lambda x.
  w_1(\ldots(w_n\, x)) : A/B & \lambda y. w_{n+1}(\ldots (w_{n+m}\,
  y)) :B }
$$

$$
\infer[\backslash E]{\lambda z. w_1(\ldots(w_{n+m}\, z)) : A}{\lambda x.
  w_1(\ldots(w_n\, x)) :B & \lambda y. w_{n+1}(\ldots (w_{n+m}\, y)):B\backslash A}
$$

$$
\infer[/I]{\lambda z. w_1( \ldots (w_{n-1}\, z):A/B}{\infer*{\lambda z. w_1( \ldots w_{n-1}(w_n\, z)):A}{[w_n:B]}}
$$

$$
\infer[\backslash I]{\lambda z. w_2(\ldots (w_n\, z):B\backslash A}{\infer*{\lambda z. w_1( w_2\ldots (w_n\, z)):A}{[w_1:B]}}
$$

$$
\infer[|E]{t\, u:A}{t:A|B & u:B}
$$

$$
\infer[|I]{\lambda x.t:A|B}{\infer*{t:A}{[x:B]}}
$$

The elimination rules are just function composition for type
$\sigma\ra\sigma$. Crucially, the introduction rules have no
representation in the lambda calculus (without term equations).
}

\begin{table}
\begin{center}
\textbf{Lexicon}
\end{center}
\vspace{-.5\baselineskip}

$$
x^{n\ra n-1}: A \vdash M^{\alpha}:A
$$

\begin{center}
\textbf{Axiom/Hypothesis}
\end{center}
\vspace{-.5\baselineskip}

$$
\infer{x^{\alpha}:A \vdash M^{\alpha}:A}{}
$$

\editout{
$$
\infer[/E]{\Gamma,\Delta \vdash \lambda z. w_1(\ldots(w_{n+m}\, z)) :
  A}{\Gamma\vdash \lambda x.
  w_1(\ldots(w_n\, x)) : A/B & \Delta\vdash\lambda y. w_{n+1}(\ldots (w_{n+m}\,
  y)) :B }
$$

$$
\infer[\backslash E]{\Gamma,\Delta\vdash \lambda z. w_1(\ldots(w_{n+m}\, z)) : A}{\Gamma\vdash\lambda x.
  w_1(\ldots(w_n\, x)) :B & \Delta\vdash\lambda y. w_{n+1}(\ldots (w_{n+m}\, y)):B\backslash A}
$$

$$
\infer[/I]{\Gamma\vdash\lambda z. w_1( \ldots (w_{n-1}\, z)):A/B}{\Gamma,w_n:B\vdash\lambda z. w_1( \ldots w_{n-1}(w_n\, z)):A}
$$

$$
\infer[\backslash I]{\Gamma\vdash\lambda z. w_2(\ldots (w_n\, z)):B\backslash A}{\Gamma,w_1:B\vdash\lambda z. w_1( w_2\ldots (w_n\, z)):A}
$$
}

\editout{
$$
\infer[/E]{\Gamma,\Delta \vdash (\lambda z^E. M\, (N\, z))^{E\ra C}: A}{\Gamma\vdash M^{D\ra C} : A/B & \Delta\vdash N^{E\ra D} :B }
$$

$$
\infer[\backslash E]{\Gamma,\Delta\vdash (\lambda z^E. M\, (N\, z))^{E\ra
  C} :
  A}{\Gamma\vdash M^{D\ra C} :B & \Delta\vdash N^{E\ra D}:B\backslash A}
$$

Strictly speaking the rightmost premiss of both rules, which is shown
to be of type $E\ra D$ should be of type $E\ra F$, with the conclusion
of the rule applying the substitution $F := D$ to $\Delta, E$.}

\begin{center}
\textbf{Logical rules -- Lambek}
\end{center}
\vspace{-.5\baselineskip}

$$
\infer[/E]{\sbst(\Gamma),\sbst(\Delta) \vdash (\lambda z^{\sbst(E)}. M\, (N\, z))^{\sbst(E)\ra \sbst(C)}: A}{\Gamma\vdash M^{F\ra C} : A/B & \Delta\vdash N^{E\ra D} :B }
$$

$$
\infer[\backslash E]{\sbst(\Gamma),\sbst(\Delta)\vdash (\lambda z^{\sbst(E)}. M\, (N\, z))^{\sbst(E)\ra
  \sbst(C)} :
  A}{\Gamma\vdash M^{F\ra C} :B & \Delta\vdash N^{E\ra D}:B\backslash A}
$$

$$
\infer[/I]{\sbst(\Gamma)\vdash ((\lambda x^{\sbst(D)\ra\sbst(C)}. M)\, (\lambda z^{\sbst(F)}.z))^{\sbst(C)\ra
  \sbst(E)}:A/B}{\Gamma,x^{D\ra C}:B\vdash M^{D\ra E}:A}
$$

$$
\infer[\backslash I]{\sbst(\Gamma)\vdash ((\lambda x^{\sbst(C)\ra\sbst(D)}. M)\, (\lambda z^{\sbst(F)}.z))^{\sbst(E)\ra \sbst(C)}:B\backslash
  A}{\Gamma,x^{C\ra D}:B\vdash M^{E\ra D}:A}
$$

\editout{
$$
\infer[/I]{\Gamma\vdash(\lambda z^C. (M[z])^E)^{C\ra
  E}:A/B}{\Gamma,x^{D\ra C}:B\vdash (\lambda z^D. (M[(x\, z)])^E)^{D\ra E}:A}
$$

$$
\infer[\backslash I]{\Gamma\vdash (\lambda z^E. M^C)^{E\ra C}:B\backslash
  A}{\Gamma,x^{C\ra D}:B\vdash (\lambda z^E. (x\, M^{C})^D)^{E\ra D}:A}
$$
}

\begin{center}
\textbf{Logical rules -- lambda grammars}
\end{center}
\vspace{-.5\baselineskip}

$$
\infer[|E]{\sbst(\Gamma),\sbst(\Delta)\vdash (M\, N)^{\sbst(\alpha)} :A}{\Gamma\vdash M^{\beta\ra \alpha}:A|B & \Delta\vdash N^{\gamma}:B}
$$

\editout{
$$
\infer[|E]{\Gamma,\Delta\vdash (M\, N)^{\alpha} :A}{\Gamma\vdash M^{\beta\ra \alpha}:A|B & \Delta\vdash N^{\beta}:B}
$$

As with the other elimination rules, the right premiss of the rule
should have type $\gamma$ and the conclusion of the rule should
compute the most general unifier of $\langle \Gamma; \beta\rangle$ and
$\langle \Delta;\gamma \rangle$ and apply this substitution to the
conclusion of the rule.}

$$
\infer[|I]{\Gamma\vdash (\lambda x^{\beta}.M^{\alpha})^{\beta\ra\alpha}:A|B}{\Gamma,x^{\beta}:B\vdash M^{\alpha}:A}
$$

\caption{Logical rules for hybrid type-logical grammars}
\label{tab:htlg}
\end{table}

%$*$ no free occurrences of $D$ in $\Gamma$, $D$ a type variable

The subsystem containing only the rules for $|$ is simply
lambda grammar. The subsystem containing only the rules for $/$ and
$\bs$ is a notational variant of the Lambek calculus.

% start comments on rules

For the Lexicon rule, $\langle x^{n\ra n-1}, \alpha\rangle$ is a principal pair for $M$
or, equivalently, with $\lambda x. M$ a $\beta$-normal $\eta$-long linear lambda term
and $(n\ra n-1)\ra\alpha$ its principal type). For the
Axiom/Hypothesis rule, $M$ is the eta-expansion of $x:A$.

For the Lambek calculus elimination rule $/E$ and $\bs E$, 
$\sbst$ is the most general unifier of $\langle \Gamma;
F\rangle$ and $\langle \Delta; D\rangle$ (this generally just replaces
$F$ by $D$ but takes care of the cases where $C = D$ or $E = F$ as
well). The concatenation operation of the Lambek calculus corresponds
to function composition on terms and to unification of string
positions on types (much like we have seen in Section~\ref{sec:mill1}).

For the Lambek calculus introduction rules $/I$ and $\bs I$, $\sbst$ is the most general unifier of $\langle \Gamma;
C\ra D\rangle$ (resp.\ $\langle \Gamma;
D\ra C\rangle$) and $\langle \emptyset; F\ra F\rangle$ (ie.\ we simply identify $C$
and $D$ and replace $x$ by the identity function on string positions
--- the empty string).

In the $|E$ rule, $\sbst$ is the most general unifier of $\langle
\Gamma;\beta\rangle$ and $\langle \Delta;\gamma\rangle$.

For convenience, we will often tacitly apply the following rule.

$$
\infer[=_{\beta\eta}]{\Gamma \vdash N^{\alpha}:A}{\Gamma \vdash M^{\alpha}:A &
  M=_{\beta\eta} N}
$$

Though the above rule is not strictly necessary, we use it to simplify
the lambda terms we compute, performing
on-the-fly $\beta$-normalization (ie.\ we replace $M$ by its
beta-normal, or beta-normal-eta-long, form $N$). Since we have both
subject reduction and subject expansion, $M$ and $N$ are guaranteed to
have the same type $\alpha$.

% end

Apart from the types, the system presented in Table~\ref{tab:htlg} is a
notational variant of hybrid type-logical grammars as presented by
Kubota and Levine \citeyear{kl12gap,kl13emp}. We have replaced strings
as basic types by string \emph{positions}
with Church type $\sigma\ra\sigma$. This is a standard strategy in
abstract categorial grammars, akin to the difference lists in Prolog,
which allows us to do without an explicit concatenation operation:
concatenation is simply treated as function composition, as can be
seen from the term assignments for the $/E$ and $\bs E$ rules. The
introduction rules $/I$ and $\bs I$ are presented somewhat differently
than the Kubota and Levine version, who present rules requiring (in
our notation) premisses with term assignments $M \equiv_{\beta\eta} \lambda
z. N[(x\, z)]$ and $M \equiv_{\beta\eta}
(x\, N)$ respectively. The present
formulation has the advantage that it is more robust in the sense that
it does not require us to test that $M$ is $\beta\eta$ equivalent to
the given terms. Though it may appear a bit strange that the $/I$ and
$\bs I$ rules require the
identity of the type variable $D$ between $x$ and
$M$, it is clear that this follows from the intended interpretation,
which requires the string variable $x$ to occur at the beginning
(resp.\ end) of the string denoted by $M$, and this solution seems preferable to
interleaving normalization and pattern matching in our rules.

The types, at least for the $|$ rules, are exactly those computed using the
principal type algorithm of \citeasnoun{hindley} discussed in Section~\ref{sec:pt}. We will see how the types for the
Lambek connectives and the lexicon rule correspond to principal type
computations in the next section.

\subsection{Justification of the principal types for the new rules}
\label{sec:pthybrid}

For $/E$ and $\bs E$, their principal types are justified as
follows; $\sbst_1$ is the most general unifier of $\langle z^G;
G\rangle$ and $\langle \Delta; E\rangle$ --- since $G$ is a type
variable not occurring elsewhere, we can assume without loss of
generality that $\sbst_1$ just replaces $G$ with $E$ --- and $\sbst_2$ is the most
general unifier of $\langle
\sbst_1(\Delta),\sbst_1(z^G);\sbst_1(D)\rangle$ and $\langle \Gamma; F\rangle$.  The important type unification is of $D$ and
$F$ (the unification of $E$ and $G$ affects only a discharged
axiom). %; that is we identify the rightmost position of the left premiss
%of the rule with the leftmost position of the right premiss of the rule.
%We have written $\Gamma'$ for $\Gamma[F:=D]$ and $C'$ for $C[F:=D]$
%(this second case is relevant only if $\Gamma$ is empty).

At the level of the types, the two rules are the same: both correspond
to concatenation.

\editout{
$$
\infer[\ra I]{\lambda z. t\, (u\, z):E\ra C}{
     \infer[\ra E]{t\, (u\, z):C}{
        \infer[\ra E]{u\, z:D}{
                  z:E 
               & u:E\ra D
         }
      &
          t:D\ra C 
     }
}
$$
}

$$
\infer[\ra I]{\sbst_2(\Gamma),\sbst_2(\sbst_1(\Delta))\vdash \lambda z. M\, (N\, z):\sbst_2(\sbst_1(E))\ra \sbst_2(\sbst_1(C))}{
     \infer[\ra E]{\sbst_2(\Gamma),\sbst_2(\sbst_1(\Delta)), \sbst_2(\sbst_1(z^G)) \vdash M\, (N\, z):\sbst_2(\sbst_1(C))}{
        \infer[\ra E]{
\sbst_1(\Delta), \sbst_1(z^G) \vdash N\, z:\sbst_1(D)}{
                  \infer{z^G \vdash z:G}{}
               & \Delta \vdash N:E\ra D
         }
      &
          \Gamma\vdash M:F\ra C 
     }
}
$$

Taking $\sbst = \sbst_1 \cup \sbst_2$, which is possible since
$\Gamma$, $\Delta$ and $z$ are disjoint, gives us the following
proof. 

$$
\infer[\ra I]{\sbst(\Gamma),\sbst(\Delta)\vdash \lambda z. M\, (N\, z):\sbst(E)\ra \sbst(C)}{
     \infer[\ra E]{\sbst(\Gamma),\sbst(\Delta), \sbst(z^G) \vdash M\, (N\, z):\sbst(C)}{
        \infer[\ra E]{
\sbst(\Delta), \sbst(z^G) \vdash N\, z:\sbst(D)}{
                  \infer{z^G \vdash z:G}{}
               & \Delta \vdash N:E\ra D
         }
      &
          \Gamma\vdash M:F\ra C 
     }
}
$$

Since $\sbst_1$ only replaced $G$ by $E$ and $G$ no longer
appears in the conclusion of the proof (the corresponding hypothesis
$z$ has been withdrawn) we can treat $\sbst$ as the most general unifier of
$\langle \Delta;D\rangle$ and $\langle \Gamma;F\rangle$.

\editout{
$$
\infer[\ra I]{\Gamma',\Delta\vdash \lambda z. M\, (N\, z):E\ra C'}{
     \infer[\ra E]{\Gamma',\Delta, z^E \vdash M\, (N\, z):C'}{
        \infer[\ra E]{
\Delta, z^E \vdash N\, z:D}{
                  \infer{z^G \vdash z:G}{}
               & \Delta \vdash N:E\ra D
         }
      &
          \Gamma\vdash M:F\ra C 
     }
}
$$

$$
\infer[\ra I]{\Gamma,\Delta\vdash \lambda z. M\, (N\, z):E\ra C}{
     \infer[\ra E]{\Gamma,\Delta, z^E \vdash M\, (N\, z):C}{
        \infer[\ra E]{
\Delta, z^E \vdash N\, z:D}{
                  \infer{z^E \vdash z:E}{}
               & \Delta \vdash N:E\ra D
         }
      &
          \Gamma\vdash M:D\ra C 
     }
}
$$
}

We compute the principal type for the $/I$ rule as follows.

$$
\infer[\ra E]{\Gamma \vdash (\lambda x.M) \lambda z.z:C\ra E}{\infer[\ra I]{\Gamma \vdash
    \lambda x. M:(D\ra C)\ra D\ra
    E}{\Gamma, x^{D\ra C}\vdash M:D\ra E} & \vdash \infer[\ra
  I]{\lambda z.z: F\ra F}{\infer{z^F \vdash z:F}{}}}
$$

And symmetrically for $\bs I$.

$$
\infer[\ra E]{\Gamma \vdash (\lambda x. M)\lambda z.z:E\ra C}{
   \infer[\ra I]{\Gamma \vdash \lambda x.M:(C\ra D)\ra E\ra D}{
      \Gamma, x^{C\ra D} \vdash M:E\ra D
   }
& \infer[\ra I]{\vdash \lambda z.z:F\ra F}{\infer{z^F\vdash z:F}{}}
}
$$

From the point of view of the principal type computation, we identify
the $C$ and $D$ variables, essentially replacing $x$ by the empty string.

\begin{lemma}\label{lemma:pt} The proof rules for Hybrid type-logical
  grammars of Table~\ref{tab:htlg} compute principal
  types for the lambda terms corresponding to their proofs.
\end{lemma}

\paragraph{Proof} We essentially use the same algorithm as
\citeasnoun{hindley}, which is somewhat simplified by the restriction
to linear lambda terms
which are eta-long. 

The principal types for $/ E$, $\bs E$, $/ I$ and $\bs I$
rules are justified as shown above.

The lexicon rule is justified by
the Substitution Lemma (Lemma~\ref{lem:subst}): given a principal type $\alpha$ for a lexical entry, we replace a
hypothesis of the form $\alpha \vdash \alpha$ by a hypothesis of the
form $n\ra n-1 \vdash \alpha$, where we know this second sequent has a
linear proof. \qed

\editout{
\begin{lemma}\label{lemma:balanced} If the premisses of a rule in hybrid type-logical
  grammars have a balanced type, then so has its conclusion.
\end{lemma}

\paragraph{Proof} This proof is most easily seen using proof
nets. Suppose the premisses of a rule are balanced. Then the premisses
have only a single possible proof net. Now if two atomic formulas are
equal in such a proof net, they must be the conclusions of a single
axiom link. ... TO BE CONTINUED ... NOTE: the key property is that
unified by MGU corresponds to a cut-link (after cut-elimination of the
logical links)

\qed
}

\begin{corollary} Given a principal type derived by
 the rules of hybrid type-logical grammar shown above, we can compute
 the corresponding lambda term up to
 $\beta\eta$ equivalence.
\end{corollary}

\paragraph{Proof}

Since the principal types computed are balanced by Lemma~\ref{lemma:balanced}, by the Coherence theorem (Theorem~\ref{thm:coherence}),
we can compute the corresponding
lambda term up to $\beta\eta$ equivalence. An easy way to do so is to
construct the proof net corresponding to the principal type (which is
unique because of balance) and to compute its lambda term; this
lambda term is the unique beta-normal eta-long term corresponding to
the principal type. \qed

\subsection{Example}
\label{sec:hex}

As an example of how to compute the principal derivation corresponding
to a hybrid derivation, we look at the following hybrid derivation.

{\label{hexproof}
$$
\infer[/I]{\lambda w. (e\, w):s/(np\bs s)}{\infer[|E]{\lambda
    v. (e\, (y\,v)):s}{\infer[|I]{\lambda x\lambda z.(x\, (y\,
      z)):s|np}{\infer[\bs E]{\lambda z.(x\, (y\, z)):s}{[x:np] & [y:np\bs
          s]}} & \lambda P\lambda v. ((P\, e)\, v):s|(s|np) }}
$$
}

\editout{
$$
\infer[???]{e^{1\ra 0} \vdash (\lambda z. e\, z)^{1\ra 0}}{
\infer[\ra E]{e^{1\ra 0}, y^{C\ra 1}\vdash (\lambda z. e\, (y\, z))^{C\ra 0}}{
\infer[\ra I]{y^{C\ra D} \vdash (\lambda x. \lambda z. x\, (y\,
  z))^{(D\ra B) \ra C\ra B}}{
\infer[\ra I]{x^{D\ra B}, y^{C\ra D} \vdash (\lambda z. x\, (y\,
  z))^{C\ra B}}{
\infer[\ra E]{x^{D\ra B}, y^{C\ra D}, z^C \vdash (x\, (y\, z))^B}{x^{A\ra B} \vdash x^{A\ra B} &
\infer[\ra E]{y^{C\ra D}, z^C \vdash y\, z:D}{y^{C\ra D} \vdash y^{C\ra
  D} & z^E \vdash z^E}}}}
&
e^{1\ra 0} \vdash (\lambda P. P\, e)^{((1\ra 0)\ra H\ra G) \ra (H\ra G)}}}
$$

The following works (for the principal type) but is not restrictive
enough (ie.\ overgenerates just like lambda grammars). Require
$C=\textit{max}/\textit{min}$ of the constants in the antecedent?
It seems better to use a test then reduce strategy, that is, test
first that the introduced resource is leftmost/rightmost then perform
the ``standard'' reduction shown below.
}

The corresponding principal derivation looks as follows (for reasons
of vertical space, the lexical entry for $e^{1\ra 0}$ has not been
eta-expanded to $\lambda P \lambda w. (P\, e)\, w$ as it should to
obtain the given principal type instead of $((1\ra 0)\ra G) \ra G$;
though either type will end up being instantiated to the same result type, the eta-expanded principal type $((1\ra 0)\ra H\ra G) \ra
H\ra G$ has the important advantage that it can be obtained without
instantiating type variables to complex types; similarly, $x^{A\ra B}$
and $y^{C\ra D}$ appear in eta-short form).

\medskip
\scalebox{.75}{$$
\infer{e^{1\ra 0} \vdash (\lambda z. e\, z)^{1\ra 0}}{
\infer[\!\ra\!I]{e^{1\ra 0} \vdash  (\lambda y. \lambda z. e\, (y\,
  z))^{(C\ra 1) \ra C\ra 0}}{
\infer[\!\ra\!E]{e^{1\ra 0}, y^{C\ra 1}\vdash (\lambda z. e\, (y\, z))^{C\ra 0}}{
\infer[\!\ra\!I]{y^{C\ra D} \vdash (\lambda x. \lambda z. x\, (y\,
  z))^{(D\ra B) \ra C\ra B}}{
\infer[\!\ra\!I]{x^{D\ra B}, y^{C\ra D} \vdash (\lambda z. x\, (y\,
  z))^{C\ra B}}{
\infer[\!\ra\!E]{x^{D\ra B}, y^{C\ra D}, z^C \vdash (x\, (y\, z))^B}{\infer{x^{A\ra B} \vdash x^{A\ra B}}{} &
\infer[\!\ra\!E]{y^{C\ra D}, z^C \vdash (y\, z)^D}{\infer{y^{C\ra D} \vdash y^{C\ra
  D}}{} & \infer{z^E \vdash z^E}{}}}}}
&
\!\!\!\!\!\!\!\!\!e^{1\ra 0} \vdash (\lambda P. P\, e)^{((1\ra 0)\ra H\ra G) \ra H\ra
  G}}}
& \infer[\!\ra\!I]{\vdash (\lambda v.v)^{J\ra J}}{\infer{v^J \vdash v^J}{}}}
$$}
\medskip

The $\bs E$ and the $/I$ rules correspond to three rules each in this
principal derivation (the derivation of $\lambda z. x\, (y\, z)$ for
$\bs E$ and the part of the derivation from $\lambda z. e\, (y\, z)$
to $\lambda z. e\, z$ for $/I$, this last rule satisfies the constraint
for the application of the rule, with $y$ appearing at the last position)

In principle, the computation of the principal type can fail because of the
constants (even though there might be a proof using
variables). However, this failure would mean the final term fails to
respect the word order of the input string. Principal types using
distinct variables for string positions would seem a useful tool for computing all possible word
orders for a given set of lexical entries, though.

\subsection{Semantics}
\label{sec:sem}

One of the attractive points of categorial grammars is that we have a
very simple and elegant syntax-semantics interface by means of the
Curry-Howard isomorphism between intuitionistic proofs and lambda
terms (or, in our case between linear intuitionistic proofs and linear
lambda terms). By interpreting the logical connectives for the implications ``$/$'', ``$\bs$'',
``$|$'' and ``$\multimap$'' as the type constructor ``$\ra$'' --- the
\emph{formulas as types} interpretation ---  our
derivations in the Lambek calculus, in lambda grammars, in hybrid
type-logical grammars and in first-order linear logic (where we treat
the quantifier as being semantically inert, that is, quantifier rules
are ``invisible'' to the meaning) correspond to $\lambda$-terms ---
the \emph{proofs as terms} interpretation. Using the Curry-Howard
isomorphism, we
can obtain semantics in the tradition of Montague simply by giving
lexical substitutions in the lexicon, using essentially the rules of
Table~\ref{tab:typing} (though we typically use the Church-style
typing) to assign a derivational meaning to a proof. 

The semantic version of the proof from the previous section looks as follows.

$$
\infer[/I]{\lambda Q. \forall z. (Q\, z):s/(np\bs s)}{\infer[|E]{\forall z. (Q\, z):s}{\infer[|I]{\lambda x. (Q\, x):s|np}{\infer[\bs E]{(Q\, x):s}{[x:np] & [Q:np\bs
          s]}} & \lambda P.\forall z. (P\,z):s|(s|np) }}
$$

Though syntactically, the Lambek elimination rule corresponds to
function composition (concatenation), \emph{semantically} it
corresponds to simple application and the introduction rule to
abstraction. Given the standard Montegovian semantics for ``everyone'' as
$\lambda P.\forall z. (P\, z)$ (the set of properties $P$ such that
all $z$ have this property), the previous proof actually produces an
equivalent term as the semantics for $s/(np\bs s)$, so the generalized
quantifier can function as a Lambek calculus subject quantifier while
keeping the same semantics.

%This section has been necessarily brief, but the basic idea is very
%simple. 
More detail about the syntax-semantics interface in categorial
grammars can be found in \cite{M95,mr12lcg}.

%\begin{array}{ccc}
%\infer[]
%\end{array}

\section{Equivalence}
\label{sec:equiv}

For the main result, we only need to show that a hybrid principal type
proof corresponds to a MILL1 proof, since we can reconstruct the
lambda term from the principal type.

The basic idea which makes the correspondence work is that there is a
1-1 mapping between the atomic terms of a predicate in MILL1 and the
principal type which is assigned to the corresponding term in a hybrid derivation.
So from the term assigned to a hybrid derivation, we compute the
principal type using the principal type algorithm (PTA) and this gives us
the first-order variables and from the first-order variables of a
MILL1 derivation we obtain the principal type and a hybrid lambda term
thanks to the coherence theorem, as shown schematically below.

\begin{tikzpicture}
\node (lt) {Hybrid lambda term};
\node (pt) [right=3em of lt] {Principal type};
\node (fo) [right=3em of pt] {First-order variables};
\draw[<->] (fo) -- (pt);
\path[->] ([yshift=+2pt] lt.east) edge node[above] {PTA} ([yshift=+2pt] pt.west);
\path[<-] ([yshift=-2pt] lt.east) edge node[below] {Coherence} ([yshift=-2pt] pt.west);
\end{tikzpicture}

\subsection{String positions, types and formulas}
\label{sec:string}

We need an auxiliary function $f$ (for \emph{flatten}) which reduces a complex type to a list
of atomic types. Following \citeasnoun{kanazawa11pg}, we compute this list by first taking the yield
of the type tree and then reversing this list, which is convenient for induction
since it has $f(\beta\ra \alpha) = f(\alpha)\concat f(\beta)$ ( ``$\concat$''
denotes list concatenation, $[ A ]$ the singleton list containing
element $A$ and $[ A_1,\ldots,A_n ]$ the $n$-element list with $i$th
element $A_i$). 
%However, any canonical, reversible way of
%$transforming complex types to lists of atoms
%works equally well. 

\begin{definition}\label{def:flatten} Let $\alpha$ be a type, the list $f(\alpha)$ is
  defined as follows.
\begin{align*}
f(A) &= [ A ] \quad \text{when $A$ atomic} \\
f(\beta\ra \alpha) &= f(\alpha)\concat f(\beta)
\end{align*}
\end{definition}

%We will use $f^{-1}$ for the inverse function from
%a list of atomic types to complex types.

For example, we have the following.

$$
\begin{array}{l}
f((B\ra 2) \ra (1\ra A) \ra B\ra A) \\
= f((1\ra A) \ra B\ra A) \concat f(B\ra 2) \\
= f(B\ra A) \concat f(1\ra A) \concat f(B \ra 2) \\
= [ A, B, A, 1, 2, B ]
\end{array}
$$

\begin{definition}\label{def:transl} Let $A$ be a formula in Hybrid Type-Logical
  Grammar, $\alpha$ its principal type and $L = f(\alpha)$ the
  flattened list of atomic types obtained from $\alpha$ according to Definition~\ref{def:flatten}. The translation of $A$ into
  first-order linear logic is defined as follows.
%We have the following translation from formulas in hybrid type-logical
%grammars, together with their polarity and (the flattened list
%corresponding to) their
%principal type, to formulas in first-order linear logic.
%
$$
\begin{array}{ll}
\leftimpl p \rightimpl^{[C_1,\ldots,C_n]} & = p(C_1,\ldots,C_n) \\
\leftimpl (A|B) \rightimpl^{f(\beta\ra\alpha)} &= \leftimpl B \rightimpl^{f(\beta)} \multimap \leftimpl A
\rightimpl^{f(\alpha)} \\ 
\leftimpl (A/B) \rightimpl^{[ C,D ]} &= \forall x. \leftimpl B \rightimpl^{[ D,x ]} \multimap \leftimpl A
\rightimpl^{[ C,x]} \\
\leftimpl (B\bs A) \rightimpl^{[ C,D ]} &= \forall x. \leftimpl B \rightimpl^{[ x,C ]} \multimap \leftimpl A
\rightimpl^{[ x,D]} \\
\end{array}
$$

We can obtain a closed formula by universally quantifying over all
variables in the list of arguments replacing all of them with
quantified variables using the universal closure operation (Definition~\ref{def:cl}).
$$
\| A \|_c^{L}  = \textit{Cl}(\| A \|^L)
$$
\end{definition}

\begin{proposition}\label{prop:fvt} Let $A$ be a formula in first-order linear logic
  and $H$ a formula in hybrid type-logical grammar and $A \equiv  \| H \|^{f(\alpha)}$. The free
  meta-variables of $A$ are exactly the type variables of $\alpha$
  (and of $f(\alpha)$).
\end{proposition}

\paragraph{Proof} Immediate by induction on $H$ using the
translation. All new variables introduced during the translation are
bound. \qed

\begin{lemma}\label{lem:mgu} Let $A_1$ and $A_2$ be first-order linear logic formulas
  obtained by the translation function from Hybrid Type-Logical
  Grammar formulas $H_1$ and $H_2$ with
  $\gamma_1$ and $\gamma_2$ as their respective principal types. In
  other words, $A_1 \equiv \| H_1 \|^{f(\gamma_1)}$ and $A_2 \equiv \| H_2 \|^{f(\gamma_2)}$.

$A_1$ unifies with $A_2$ with MGU $\sbst$ if and only if $H_1 \equiv
H_2$ and $\gamma_1$ unifies with $\gamma_2$ with this same MGU $\sbst$.
\end{lemma}

\paragraph{Proof} Suppose $A_1$ and
$A_2$ unify with MGU $\sbst$. We must show that $H_1 \equiv H_2$ and
that $\sbst$ is an MGU for $\gamma_1$ and $\gamma_2$. Showing $H_1 \equiv H_2$
is an easy induction (exploiting the fact that $A|B$ does not have a
quantifier prefix and therefore cannot unify with a Lambek connective
and that $A/B$ and $B\bs A$ cannot unify with each other because of the
condition preventing accidental capture of variables). Given that
$H_1$ and $H_2$ are identical, we know that $A_1$ and $A_2$ differ
only in the free variables (the bound variables are equivalent up to
renaming) and that the free variables for $A_1$ and $A_2$ are exactly
the type variables of $\gamma_1$ and $\gamma_2$ (by
Proposition~\ref{prop:fvt}). Therefore any substitution that makes
$A_1$ and $A_2$ equal (up to renaming of bound variables) makes
$\gamma_1$ and $\gamma_2$ equal.

For the other direction, suppose that $H_1 \equiv H_2$ and that
$\sbst$ is the MGU of $\gamma_1$ and $\gamma_2$. Since $\sbst$ is a MGU
$\sbst(\gamma_1) \equiv \sbst(\gamma_2)$ and therefore given that the translation
function uses identical hybrid formulas and identical principal types
we have that
$A_1 \equiv A_2$.
\qed

It is insightful to compare the translation of $(np\bs s)/np$ (with
principal type $2\ra 1$) to that of $(s|np)|np$ with principal type $(B\ra
2) \ra (1\ra A) \ra B\ra A$. Though the two end results are formulas
which are equivalent to each other (after universal closure of the meta-variables), there is a difference in the string position list for
the non-atomic subformulas: the Lambek formula only ever has a pair of
string positions, whereas the linear formula starts with a full list
of string positions which decreases at each step. In other words, for the Lambek formula, we compute the string positions step-by-step
whereas the lambda grammar version of the same formula precomputes all
string positions then divides them among the subformulas. 
%The positive
%translation of the $(s|np)|np$ is the same as the negative one. On the
%other hand, the positive translation for $(np\bs s)/np$ is show below
%on the left. 
$$
\begin{array}{l}
\| (np\bs s)/np \|^{[1,2]} %&  \| (np\bs s)/np \|^{-,[1,2]}   \\
\\
= \forall y. \| np \|^{[2,y]} \multimap \| np \bs s \|^{[1,y]} %&
                                %= \| np \|^{+,[2,B]} \multimap \| np
                                %\bs s \|^{-,[1,B]} \\
\\
= \forall y.  np(2,y) \multimap \| np \bs s \|^{[1,y]}  %& = np(2,B) \multimap \| np \bs s \|^{-,[1,B]} \\
\\ = \forall y.  np(2,y) \multimap \forall x. \| np \|^{[x,1]} \multimap \| s \|^{[x,y]} %& = np(2,B) \multimap \| np \|^{+,[A,1]} \multimap \| s \|^{-,[A,B]} \\
\\ = \forall y.  np(2,y) \multimap \forall x. [ np(x,1) \multimap
s(x,y) ] %& = np(2,B) \multimap np(A,1) \multimap s(A,B) \\
\end{array}
$$

$$
\begin{array}{l}
\| (s|np)|np \|^{[ A, B, A, 1, 2, B ]} \\
=  \| np \|^{[2,B]} \multimap \| s|np \|^{[ A, B, A, 1]} \\
= np(2,B) \multimap \| s|np \|^{[ A, B, A, 1]} \\
= np(2,B) \multimap \| np \|^{[A,1]} \multimap \| s \|^{[ A, B]} \\
= np(2,B) \multimap np(A,1) \multimap s(A,B) \\
\\
\| (s|np)|np \|_c^{[ A, B, A, 1, 2, B ]} \\
= \forall x. \forall y. \{ np(2,B) \multimap np(A,1) \multimap s(A,B)\}[A:=x,B:=y] \\
= \forall x. \forall y. [np(2,y) \multimap np(x,1) \multimap s(x,y)] \\
\end{array}
$$

Remember that sequents in hybrid type-logical grammar are of the form $x_1^{\alpha_1}:A_1, \ldots,
x_n^{\alpha_n}:A_n  \vdash M^{\beta}:B$ with
$M$ a linear lambda term containing exactly the free variables $x_1,
\ldots, x_n$ and that the principal type of $\lambda x_1,
\ldots x_n . M$ is balanced and of the form $\alpha_1 \rightarrow \ldots
\rightarrow a_n
\rightarrow \beta$. For the translation, we separate lexical axioms
from other axioms: lexical axioms correspond to closed formulas,
whereas the other axioms typically have free variables. With this in mind, we translate sequents as $\leftexpl A_1
\rightexpl^{f(\alpha_1)}, \ldots, \leftexpl A_n
\rightexpl^{f(\alpha_n)} \vdash \leftimpl B
\rightimpl^{f(\beta)}$, where the translation $\| . \|_c$ is used for
hypotheses which start at a lexicon rule and $\| . \|$ for hypothesis
which start at the axiom rule. For the right-hand side $B$, we use the
universal closure of all free variables in $f(\beta)$ \emph{minus} the
free variables on the left hand side of the sequent (the only free
variables are those used in the translation of hypothesis rules), this
is the universal closure of $B$ modulo $\Gamma$ of Definition~\ref{def:cl}.
%is, of course, required to make the application of the $\forall I$
%rule valid.

In order not to overburden our notation,
when the types are understood from the context, we
will often abbreviate this translation as $\leftexpl \Gamma\rightexpl
\vdash \leftimpl B \rightimpl$ (or even as $\Gamma\vdash  \| B \|$,
leaving the translation of $\Gamma$ implicit). As a special case of this translation, the sequent $w_1^{1\ra 0}:A_1, \ldots
w_n^{n\ra n-1}:A_n \vdash M^{n\ra 0}:B$, which is the endsequent
corresponding to a sentence in a hybrid type-logical grammars, is translated as $\leftexpl
A_1 \rightexpl_c^{[0,1]},\ldots, \leftexpl A_n \rightexpl_c^{[n-1,n]}
\vdash \leftimpl B \rightimpl_c^{[0,n]}$. 
%When the types are clear
%from the context, we will simply write $\leftexpl \Gamma \rightexpl
%\vdash \leftimpl B \rightimpl$ for the translation of $x_1^{\alpha_1}:A_1,\ldots
%x_n^{\alpha_n}:A_n \vdash M^{\beta}:B$, instead of the more cumbersome $\leftexpl A_1
%\rightexpl^{-,f(\alpha_1)},\ldots \leftexpl A_n
%\rightexpl^{-,f(\alpha_n)} \vdash \leftimpl B \rightimpl^{+,f(\beta)}$.

\editout{
$$
\infer*[\bo\forall I\bc^*]{}{\infer*[\bo\multimap
  I\bc^*]{}{\infer*[\bo\multimap I/\forall
    I\bc^*]{}{\infer[\shfocus]{}{\infer[\bo\forall E/\multimap
        E\bc]{}{\infer*{}{\infer[\forall E/\multimap E]{}{}} & }}}}}
$$
}

\paragraph{Example: gapping} To give an example, the gapping lexical entry for ``and''  of
\cite{kl12gap} looks as follows in our notation.
$$
((s|tv)|(s|tv))|(s|tv): \lambda \textit{STV2} \lambda \textit{STV1} \lambda \textit{TV} \lambda z.
(\textit{STV1}\ \textit{TV}) (\textit{and}\ (\textit{STV2}\ \lambda x.x))
$$

\noindent where $tv$ is short for $(np\bs s)/np$. The principal type
for this lambda term would be the following (the corresponding
formulas have been annotated above for ease of comparison).
$$
((\overset{tv}{\overbrace{E\ra E}})\ra \overset{s}{\overbrace{D\ra 4}}) \ra ((\overset{tv}{\overbrace{B\ra A}}) \ra \overset{s}{\overbrace{3\ra C}}) \ra (\overset{tv}{\overbrace{B\ra A}}) \ra \overset{s}{\overbrace{D\ra C}}
$$

If $tv$ were an atomic formula, the first-order linear logic formula
would look as shown below on the first line, the complete formula (for
the positive translation) is
shown just below it.
\begin{gather*}
(tv(E,E) \multimap s(4,D)) \multimap (tv(A,B) \multimap s(C,3))
\multimap tv(A,B) \multimap s(C,D)  \equiv \\ 
(\forall v. [ np(v,E) \multimap \forall w. [ np(E,w) \multimap s(v,w)] ] \multimap s(4,D)) 
\multimap \\  (\forall x' [ np(x,A) \multimap \forall y' [ np(B,y')
\multimap s(x',y') ] ]\multimap s(C,3)) \multimap \\
\qquad\qquad\qquad \forall x. [ np(x,A) \multimap \forall y.  [
np(B,y) \multimap s(x,y) ] ]
\multimap s(C,D)
\end{gather*}

Though the formula above looks intimidating (even before universal closure), it is easy to verify that it is \emph{equivalent} (up
to variable names) to the first-order linear logic formula which
corresponds to the analysis of gapping for the Displacement calculus
from Section~3.2.6 of \cite{mvf11displacement},
using the translation given in \cite{moot13lambek}.

\subsection{Proof-theoretic properties of the translation into MILL1}

\begin{figure}
\begin{center}
\begin{tikzpicture}
\node (lsp) at (7.6,4.2) {Subproofs $\Delta_i \vdashpos \| \mathcal{F}_1 \|$};
\node (hsp) at (6.5,5.25) {Subproofs $\Gamma_j \vdashpos \| \mathcal{F}_2 \|$};
%\fill[black!10] (0,3.25) rectangle (6.2,4.15);
\fill[black!10] (0,1.85) rectangle (6.1,2.75);
\draw[fill=black!10,black!10] (0,3.25) -- (6.1,3.25) -- (5.2,4.15) -- (0,4.15) -- (0,3.25);
\draw[dotted] (6.1,3.35) -- (6.4,3.65);
\draw[dotted] (5.7,3.75) -- (6.0,4.05);
\draw[dotted] (5.3,4.15) -- (5.6,4.45);
%\draw[dotted] (4.9,4.45) -- (5.2,4.75);
\draw[dotted] (4.95,4.5) -- (5.25,4.8);
\draw[dotted] (4.65,4.8) -- (4.95,5.1);
%\draw[dotted] (4.35,5.1) -- (4.65,5.4);
\draw[dotted] (4.4,5.05) -- (4.7,5.35);
%\draw (0,3.25) -- (6.2,3.25) -- (6.2,4.15) -- (0,4.15) -- (0,3.25);
\node (bottom) at (6,-0.2) {7};
\node (ii) at (6,0.8) {6};
\node (lli) at (6,1.8) {5};
\node (sf) at (5.75,3) {$\shfocus$\ \ 4};
\node (sft) at (6,3.2) {};
\node (sfb) at (6,2.8) {};
\node (le) at (5,4.2) {3};
\node (lbe) at (4,5.2) {2};
\node (fe) at (4,6.2) {1};
\path (bottom) edge node[left] {$[\forall I]^*$} (ii);
\path (ii) edge node[left] {$[\multimap I]^*$} (lli);
\path (lli) edge node [left] {$[\multimap I/\forall I]^*$} (sfb);
\path (sft) edge node [left] {$\ \ [\forall E/\multimap E]^*$} (le);
\path (le) edge node [left] {$\ \ [\multimap E]^*$} (lbe);
\path (lbe) edge node [left] {$[\forall E]^*$} (fe);
\node[text width=7.5em] (lambdai) at (1,0.8) {\hfill $\lambda$ grammar I $\left\{ \rule{0pt}{2.6em}\right.$};
\node[text width=7.5em] (lambeki) at (1,2.3) {\hfill Lambek I $\left\{ \rule{0pt}{1.4em}\right.$};
\node[text width=7.5em]  (lambeke) at (1,3.7) {\hfill   Lambek E $\left\{ \rule{0pt}{1.4em}\right.$};
\node[text width=7.5em] (lambdai) at (1,5.1) {\hfill $\lambda$ grammar E $\left\{
    \rule{0pt}{2.6em}\right.$};
\node[text width=7.5em] (shfocus) at (1,3){\hfill focus shift $\{\,$};
\node[text width=7.5em] (lexicon) at (1,6.2){\hfill lexicon/axiom $\{\,$};
\end{tikzpicture}
\end{center}
\caption{Schematic form of the main track of a translated hybrid
  sequent.}
\label{fig:transl}
\end{figure}
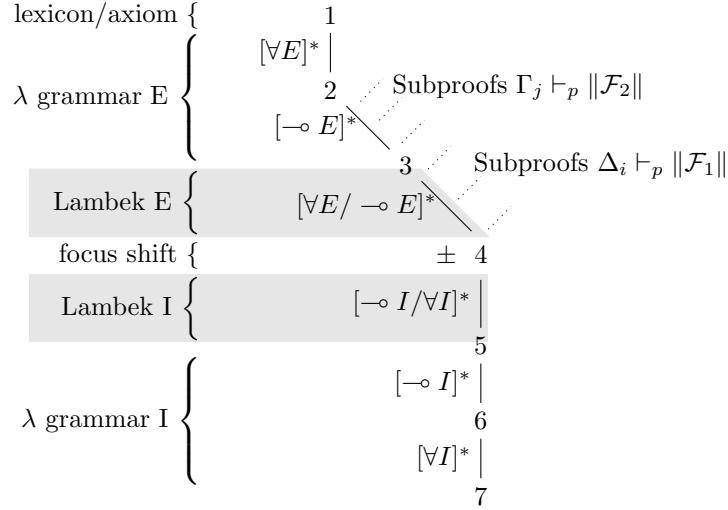

Before proving the main theorem, stating that for every hybrid proof
there is a first-order linear logic proof of its translation, we will
spend some time on the structure of normal/focused natural deduction
proofs and the consequences of the translation function.
Given that in hybrid type-logical grammars, the
lambda-grammar connective ``$|$'' always outscopes the Lambek
connectives ``$/$'' and ``$\bs$'', proofs using
the translated formulas into focused first-order linear logic look
schematically as shown in Figure~\ref{fig:transl}.

The figure shows the main track of a proof, which starts either with a
hypothesis/axioms, then has an elimination part, followed by a focus
shift followed by an introduction part ending in the conclusion of the
proof --- this is just the definition of a main track
(Definition~\ref{def:track}). The definition of formulas guarantees that the elimination part
starts with any number of $[| E]$ rules (possibly zero, like all other
parts, as indicated by the $*$ superscript in the figure) followed by
any combination of $[/ E]$ and $[\bs E]$ rules. The order is inverse
in the introduction part of the track, with $[/ I]$ and $[\bs I]$
preceding $[| I]$. For the translation of these rules into first-order
linear logic, the quantifiers corresponding to the 
rules for the lambda grammar connective ``$|$''  are obtained by
universal closure, so if they are present, it must be as a prefix at
the beginning of the proof or as a postfix at the end of the proof
 --- the subpaths labeled (1)-(2) and (6)-(7) in
 Figure~\ref{fig:transl} --- and the Lambek connectives correspond to
 a combination of a $\forall$ and a $\multimap$ rule upon translation.

\begin{proposition}\label{prop:forall} \emph{a.}\ The main track of a
  focused proof of a translated hybrid sequent looks as shown in Figure~\ref{fig:transl}.

\emph{b.}\ The subproofs $\Gamma_j \vdashpos \| \mathcal{F}_2
  \|$ contain the sequence of proof steps in (1)-(6), that is they do
  not end with any $\forall I$ rules corresponding to a hybrid
  connective.

\emph{c.}\ The subproofs in $\Delta_i \vdashpos \| \mathcal{F}_1 \|$ contain the
sequence of proof steps in (1)-(5), that is they do not end with any
lambda grammar introduction rules. 
\end{proposition}

\paragraph{Proof} These are immediate consequences of the translation function and the
structure of normal proofs.

\emph{a.}\ follows from the way the translation function is defined
and the standard structure of a main track.

\emph{b.}\ since normal proofs satisfy the subformula property and since hybrid connectives
are translated into prenex formulas, we do
not produce subformulas of the form $(\forall x_1,\ldots, x_n. [ A
\multimap B ]) \multimap C$ (for $n\geq 1$).

\emph{c.}\ would contradict the
definition of hybrid formulas, since it would have a Lambek connective
outscope a lambda grammar connective. \qed

An immediate corollary of Proposition~\ref{prop:forall} is that $\forall I$ rules corresponding to lambda
grammar connectives occur only at the end of the main track of a
proof, just like $\forall E$ rules corresponding to lambda grammar
connective occur only at the start of any track in which they occur.

\begin{definition} We say a first-order linear logic proof obtained by
  translating a hybrid proof is in \emph{quantifier-reduced form},
  when all $\forall E$ and $\forall I$ rules obtained by universal closure of
  lambda-grammar connectives have been removed from the proof.

  More precisely, the translation is kept as before with the following
  two exceptions:
\begin{itemize}
\item the Lexicon rule is translated as $\| A  \|_c^{f(\alpha)}
\vdash \| A \|^{f(\alpha)}$ (with the closure operation applied only to
the  translation of the antecedent)
\item the endsequent is translated as
$\Gamma \vdash \| C \|^{f(\beta)}$ (without the usual closure modulo
$\Gamma$).
\end{itemize}
\end{definition}

\begin{proposition} A sequent $\Gamma \vdash A$ produced by the
  translation function is derivable if and only if its
  quantifier-reduced form is.
\end{proposition}

\paragraph{Proof} Immediate by Proposition~\ref{prop:cl}. \qed

Quantifier-reduced form is a way of ``compiling'' away the predictable
prefixes of $\forall E$ rules (for each of the lexical leaves of the
proof) and the equally predictable postfix of
$\forall I$ rules introduced by the universal closure operation. 
This
simplifies the structure of the proof, as is clear from
Proposition~\ref{prop:qr} below and from
Figure~\ref{fig:transl} --- we keep only the subpath (2)-(6). It also
simplifies the correctness proof of the translation in the following
sections, since we avoid having to start each inductive step by a
number of $\forall E$ rules and end it with a number of $\forall I$ rules.

The quantifier-reduced
form of a proof is sensitive to the way we have obtained the formula:
the Lexicon rule for the Lambek formula $np\bs s$ has quantifier-reduced form $\forall
x. np(x,1)\multimap s(x,2)\vdashneg \forall
y. np(y,1)\multimap s(y,2)$ whereas Lexicon rule for the formula $s|np$ with principal
type $(1\ra A) \ra (2\ra A)$, which would normally be assigned the
same axiom, has quantifier-reduced form $\forall
x. np(x,1)\multimap s(x,2)\vdashneg np(A,1)\multimap s(A,2)$ (which we
can obtain from the previous sequent by a single application of
$\forall E$).
%axioms using only $\forall E$ and are therefore derived rules
%(derived \emph{axioms} even). In the context of a quantifier-reduced
%proof, we will still consider these derived
%rules instances of the \emph{Lexicon} rule. 
%From a quantifier-reduced
%proof, we can get back to the original proof, by spelling out the
%proof of these derived instances of the lexicon rule and by performing
%universal closure on the conclusion of the proof.

\begin{proposition}\label{prop:qr} Let $\delta$ be first-order linear logic
  proof in long normal form which has the translation of a hybrid sequent as its conclusion.
  All
  occurrences of $\forall E$ and $\forall I$ of the quantifier-reduced
  from $\delta'$ of $\delta$ occur respectively
  in the following contexts.

$$
\begin{array}{ccc}
\infer[\multimap E]{\Gamma,\Delta \vdashneg B}{\infer[\forall
  E]{\Gamma\vdashneg A\multimap B}{\Gamma\vdashneg \forall x. [ A
    \multimap B]} &\Delta \vdashpos A}
&&
\infer[\forall I]{\Gamma\vdashpos \forall x. [A \multimap
  B]}{\infer[\multimap I]{\Gamma\vdashpos A\multimap
    B}{\Gamma,A\vdashpos B}}
\end{array}
$$
\end{proposition}

\paragraph{Proof} Given Proposition~\ref{prop:forall} and the fact
that $\delta'$ is quantifier-reduced, all quantifiers occur in
(sub)formulas of the form $\forall x. [A \multimap B]$, which
corresponds to the translation of a Lambek formula. Given that
$\delta$ is in long normal form, meaning that the focus shift rule is
applied only to atomic formulas, so is its quantifier-reduced form $\delta'$.

Look at an arbitrary application of the $\forall E$ rule. We show it
must be part of a subproof of the form shown above on the left. After application of the
$\forall E$ rule, we have the sequent $\Gamma\vdashneg A\multimap
B$. The focus shift rule cannot apply, since $A\multimap B$ is not
atomic and $\delta'$ is in long normal form. Therefore, by inspection
of the available rules $\multimap E$ is the
only rule available and we are in the case shown above.

The case for the $\forall I$ rule is similar. To obtain a formula $\Gamma\vdashpos A\multimap
B$ as the premiss of the $\forall I$ rule, focus shift is excluded
because we have a complex formula. The only available alternative
removes the main connective as shown above on the right. \qed

 %We obtain the quantifier reduced form of $\delta$ by
%\emph{not} doing the universal closure for the right-hand side of a
%lexical axiom, creating 

%By Proposition~\ref{prop:cl}, we have $\Gamma \vdash A$ iff $\Gamma
%\vdash \textit{Cl}_{\Gamma}(A)$. Since the relevant cases here are the
%lexicon rules and the conclusion of the proof, $\Gamma$ is closed by
%construction, so $\textit{Cl}_{\Gamma}(A) \equiv \textit{Cl}(A)$.
%More precisely, in the focused
%calculus, we have $\Gamma \vdashneg \textit{Cl}_{\Gamma}(A)$ implies
%$\Gamma \vdashneg A$ (using one $\forall E$ rule for each eliminated
%quantifier) and $\Gamma \vdashpos A$ implies $\Gamma \vdashpos
%\textit{Cl}_{\Gamma}(A)$ (using one $\forall I$ rule for each
%introduced quantifier).

\subsection{Hybrid proof to MILL1 proof}

After this long setup, everything is in place to prove the main
theorem. Thanks to the way we have defined our basic notions
and translations, the proof is rather simple. We show that under the
given translation, the proof rules of hybrid type-logical grammar are
derived rules of MILL1. In the next section, we show the converse:
that MILL1 proofs using formulas obtained from the translation
correspond to proofs in hybrid type-logical grammars. 

The proof is actually stronger: we
show that proofs in the two systems generate the same
\emph{semantics}. This is easily verified since, as discussed in
Section~\ref{sec:sem},
the elimination (resp.\ introduction) rules for $/$, $\bs$ and $|$ correspond to the
elimination (resp.\ introduction) rule  for $\multimap$. The
elimination rules (for $/$, $\bs$, $|$ and $\multimap$) correspond to application and the introduction rule
correspond to abstraction. The quantifier $\forall$ is treated as semantically inert.

\begin{lemma}\label{lem:htom} Let $\delta$ be a hybrid proof of $\Gamma \vdash A$,
  then there is an MILL1 proof $\delta^*$ of its
  translation $\leftexpl \Gamma \rightexpl \vdash \leftimpl A \rightimpl$.
\end{lemma}

\paragraph{Proof} We produce a unfocused proof with
unification (that is, we do not distinguish between $\vdashpos$ and
$\vdashneg$). If desired, we can transform the proof obtained by this
lemma into a focused proof by Proposition~\ref{prop:focus}). We also produce a proof in quantifier-reduced
form.

Since the lexicon/axioms rules are in beta-normal eta-long form by
definition, we know from Lemma~3.22 of \citeasnoun{kanazawa11pg} that substitution is only of type
variables/atoms for type variables and never of a complex type for a
type variable, so the arity of our predicate symbols in first-order
linear logic is fixed.

Induction on the depth $d$ of the proof.

If $d=1$ we either have an axiom rule or a lexical hypothesis. In both
cases, we have a sequent $x:A \vdash M^{\alpha}:A$, with $\alpha$ the
principal type of $M$ and with $x$ of type
$\alpha$ in the axiom case and of type $i \ra i -1$ (for the $i$th
word) in the lexicon case.
We
translate the axiom by   $\| A \|^{f(\alpha)}
\vdash \| A \|^{f(\alpha)}$ (letting the free variables of $f(\alpha)$
become free meta-variables)
and the lexical hypothesis by
the axiom $\| A \|_c^{f(\alpha)}
\vdash \| A \|^{f(\alpha)}$, replacing the meta-variables in
$f(\alpha)$ on the left with
variables and quantifying over them, making the formula on the
left-hand side of the turnstile closed. Since we produce a proof in
quantifier-reduced form, we do not perform the closure on the
right-hand side of the turnstile (or, if you prefer, we perform the
closure but immediately follow it by $\forall E$ rules for all
quantifiers introduced by the closure operation).

If $d >1$, induction hypothesis gives us proofs of the premisses of
the rule and we proceed by case analysis on the last rule in the
hybrid proof. %Since we
%produce proofs in quantifier reduced form.
%To be fully
%precise, each inductive case starts with a set of $\forall E$ steps which
%replaces the quantifier prefix with meta-variables and ends with a universal closure step, which applies
%all possible (non-vacuous) $\forall I$ steps.

\begin{description}
\item{$[\bs E]$} By induction hypothesis, we have a proof $\delta_1$ of
  $\Gamma \vdash \|B \|^{[C,D]}$ and a proof $\delta_2$ of $\Delta\vdash \|
  B\bs A \|^{[F,E]}$. In addition, we know by induction hypothesis that
  a MGU $\sbst$ of $\langle \Gamma;D\rangle$ and $\langle\Delta;F\rangle$ exists. Therefore, we can construct a proof of
  the conclusion of the $/E$ rule as follows. Since $G$ is fresh,
  unifying it with $C$ is possible and produces a new substitution $\sbst'$.

$$
\infer[\multimap E]{\sbst'(\Gamma),\sbst'(\Delta) \vdash \| A \|^{[\sbst'(C),\sbst'(E)]}}{
  \infer*[\delta_1]{\Gamma \vdash \| B \|^{[C,D]}}{}
&
 %\infer[G=C,F=D]{\Delta \vdash \| B \|^{-,[C,D]} \multimap \| A
  % \|^{+,[C,E]}}{
  \infer[\forall E]{\Delta \vdash \| B \|^{[G,F]} \multimap \|
    A \|^{[G,E]}}{\infer[=_{\textit{def}}]{\Delta \vdash \forall x. \| B \|^{[x,F]} \multimap \|
    A \|^{[x,E]}}{\infer*[\delta_2]{\Delta \vdash \| B\bs A \|^{[F,E]}}{}}}
%}
}
$$

%Since $G$ is a newly introduced variable, we add the unification of
%$C$ and $G$ to the MGU $\sbst$ to produce a new MGU $\sbst'$.

%Where $\sbst$ is the most general unifier of $\langle \Gamma; \| B \|^{+,[C,D]} \rangle$
%and $\langle \Delta; \| B \|^{-,[G,F]} \rangle$, unifying $C$ with $G$
%and $D$ with $F$ (note that quantifiers are treated as transparent and
%that bound variables are treated as constants when computing the MGU,
%so the only interesting variables are $C$, $D$, $F$ and $G$).

\item{$[/ E]$} Symmetric.

\item{$[| E]$} By induction hypothesis, we have a proof $\delta_1$ of
  $\Gamma \vdash \| A| B \|^{f(\beta\ra\alpha)}$ and a proof
  $\delta_2$ of $\Delta \vdash \| B \|^{f(\gamma)}$. We also know
  there is an MGU $\sbst$ of $\langle\Gamma;\beta\rangle$ and
  $\langle\Delta;\gamma\rangle$. Therefore, we can combine these two
  proofs using $\multimap E$ and (by Lemma~\ref{lem:mgu}) this same unification, as follows.

$$
\infer[\multimap E]{\sbst(\Gamma),\sbst(\Delta) \vdash \| A
  \|^{f(\sbst(\alpha))}}{
\infer[=_{\textit{def}}]{\Gamma \vdash \| B \|^{f(\beta)} \multimap \| A
  \|^{f(\alpha)}}{\infer*[\delta_1]{\Gamma \vdash \| A| B \|^{f(\beta\ra\alpha)}}{}} &
\infer*[\delta_2]{\Delta \vdash \| B \|^{f(\gamma)}}{}
}
$$

\item{$[\bs I]$} By induction hypothesis, we have a proof $\delta_1$
  of $\Gamma, \| B \|^{[D,C]} \vdash \| A \|^{[D,E]}$. In addition,
  since all principal types are balanced and the two occurrences of
  $D$ occur in the translations of $B$ and $A$ respectively, we know
  there are no occurrences of $D$ in $\Gamma$. Hence, after the
  $\multimap I$ rule, we satisfy the
  condition for the $\forall I$ rule and can extend the proof as follows.

$$
\infer[=_{\textit{def}}]{\Gamma \vdash \| B \bs A \|^{[C,E]}}{
\infer[\forall I]{\Gamma \vdash \forall x. \| B \|^{[x,C]} \multimap \| A \|^{[x,E]} }{
   \infer[\multimap I]{\Gamma \vdash \|B\|^{[D,C]}  \multimap \| A \|^{[D,E]} }{
        \infer*[\delta_1]{\Gamma, \| B \|^{[D,C]}  \vdash \| A \|^{[D,E]} }{}
   }
}}
$$

\item{$[/ I]$} Symmetric.

\item{$[| I]$} Induction hypothesis gives us a proof $\delta_1$ of
  $\Gamma, \| B \|^{f(\beta)}\vdash \| A \|^{f(\alpha)}$, which we can
  extend as follows.

$$
\infer[=_{\textit{def}}]{\Gamma \vdash \| A | B
  \|^{f(\beta\ra\alpha)}}{
  \infer[\multimap I]{\Gamma \vdash \| B \|^{f(\beta)} \multimap \|A
    \|^f(\alpha)}{
      \infer*[\delta_1]{\Gamma, \| B \|^{f(\beta)}\vdash \| A \|^{f(\alpha)}}{}
  }
}
$$ 

\qed

\end{description}

\subsection{MILL1 proof to hybrid proof}

\begin{lemma}\label{lem:mtoh}  Let $\delta$ be the MILL1 derivation of the translation
  of a hybrid sequent, that is, of
  $\leftexpl A_1 \rightexpl^{[0,1]},\ldots, \leftexpl A_n \rightexpl^{[n-1,n]} \vdash \leftimpl B \rightimpl^{[0,n]}$. Then there is a
  hybrid proof $\delta^*$ of $x_1^{1\ra 0}:A_1,\ldots,x_n^{n\ra
    n-1}:A_n \vdash M^{n\ra 0}:B$, where $M \equiv_{\beta\eta} \lambda
  z. (x_1\, \ldots (x_n\, z))$. 
\end{lemma}

\paragraph{Proof} The fact that $M \equiv_{\beta\eta} \lambda
  z. (x_1\, \ldots (x_n\, z))$ follows immediately from the balanced
  occurrences of the type constants $0, \ldots, n$.

Let $\delta$ be the focused MILL1 derivation of $\leftexpl \Gamma
\rightexpl \vdashpos \leftimpl A \rightimpl$, or, the case being, of $\leftexpl \Gamma
\rightexpl \vdashneg \leftimpl A \rightimpl$. We assume $\delta$ to be
in quantifier-reduced form. %We assume, without loss of generality, that it is
%a proof in MILL1 with implicit $\forall E$.

We proceed by induction on the depth $d$ of the proof. 
%Proofs have at
%least depth 2, since an axiom must be followed by a focus shift rule
%to produce a positive sequent.

If $d=1$, then there are two cases. 

\begin{description}
\item{\emph{Lexicon}} If the rule was a lexical
hypothesis, then it is a proof of $\| A_i \|^{[i-1,i]}$ for one of the
$A_i$ of the endsequent of the proof. By construction, we can recover the principal type $\alpha$ and (by
Coherence) a unique
$\beta$-normal $\eta$-long lambda term $M$ of type
$\alpha$. Therefore, we have a 
hybrid proof $x_i^{i-1\ra i}:A_i \vdash M^{\alpha}:A_i$, with $\alpha$
the principal type by construction.

\item{\emph{Axiom}} If the rule was an axiom then the formula $A$ does
  not appear in the endsequent. We again recover the principal type
  $\alpha$ and the
  (eta-expanded) lambda term $M$ from the translation function and we return the hybrid proof
  $x^{\alpha}:A \vdash M^{\alpha}: A$, with $M$ the eta-expansion of
  $x$ to produce a valid Axiom rule.
\end{description}

If $d>1$, then we proceed by case analysis of the last rule of the
proof.

\begin{description}
\item{$[\shfocus]$} Induction hypothesis gives us the proof
  corresponding to the negative premiss of the rule. We return the
  same proof.

\item{$[\forall E/{}\multimap E]$} For the combination of a $\forall
  E/\multimap E$ rule, there are two cases to consider, depending on
  whether the translated formula had $/$ or $\backslash$ as main
  connective. In case it was $/$, our translation unfolds as shown
  below. The MGU
  $\sbst$ unifies $D$ with $F$ (it doesn't matter here if the $\forall
  E$ step has been done separately: in that case $x$ is replaced by a
  fresh metavariable
  $G$ and the MGU unifies $G$ with $E$).

%\| (A/B) \|^{[ C,D ]} &= 

$$
\infer[\multimap E]{\sbst(\| \Gamma\|), \sbst(\|\Delta \|) \vdashneg \| A \|^{[
    \sbst(C),\sbst(E)]} }{\infer[\forall E]{\| \Gamma \| \vdashneg \| B
    \|^{[D,E]}\multimap \| A \|^{[C,E]}}{\infer[=_{\textit{def}}]{ \| \Gamma \| \vdashneg
    \forall x.  \| B \|^{[ D,x ]} \multimap \| A \|^{[
    C,x]}}{\| \Gamma \| \vdashneg \| A/B \|^{[C,D]}}} & \| \Delta \| \vdashpos \| B \|^{[F,E]}}
$$

Lemma~\ref{lem:mgu} guarantees that the two hybrid formulas $B$ are
 indeed identical and
induction hypothesis gives us a proof $\delta_1$ of $\Gamma \vdash
M^{D\ra C}:A/B$ and a proof $\delta_2$ of $\Delta \vdash N^{E\ra F}:B$, which we can combine
by the $/ E$ rule, using the same substitution $\sbst$, to produce a
proof of $\Gamma,\Delta \vdash A^{\sbst(E)\ra\sbst(C)}$ as required. 

$$
\infer[/E]{\sbst(\Gamma),\sbst(\Delta)\vdash
  (\lambda z.M (N\, z))^{\sbst(E)\ra\sbst(C)}:A}{\infer*[\delta_1]{\Gamma \vdash M^{D\ra
      C}:A/B}{} & \infer*[\delta_2]{\Delta \vdash N^{E\ra F}:B}{}}
$$

According to Lemma~\ref{lemma:pt}, we have also computed the corresponding
principal type $\sbst(E)\ra\sbst(C)$.

The case for $\bs$ is symmetric.

\item{$[\multimap E]$} In a quantifier reduced proof, a solitary
  $\multimap E$ (without preceding $\forall E$ producing
  the major premiss of the rule, which was treated in the previous case)
  originated from a formula $A|B$. We
are in the following case.

$$
\infer[\multimap E]{\sbst(\Gamma),\sbst(\Delta) \vdashneg \| A
  \|^{f(\sbst(\alpha))}}{
      \infer[\equiv_{\textit{def}}]{\Gamma \vdashneg \| B \|^{f(\beta)} \multimap \| A \|^{f(\alpha)}}{
            \Gamma \vdashneg \| A| B \|^{f(\beta\ra\alpha)}} &
            \Delta \vdashpos \| B \|^{f(\gamma)}}
$$

%We can reconstruct the types $\beta\ra \alpha$ and $\gamma$ from the
%string positions $f(\beta\ra\alpha)$ and $f(\gamma)$ respectively.
By
induction hypothesis there is a proof of $\delta_1$ of $\Gamma \vdash
A|B$ (where the term $M$ of $A|B$ has principal type $\beta\ra \alpha$) and a proof $\delta_2$ of $\Delta
\vdash B$ (where the term $N$ assigned to $B$ has principal type
$\gamma$). By Lemma~\ref{lem:mgu}, the two hybrid $B$ formulas are
identical and we can use the MGU $\sbst$ as the most general unifier
of $\gamma$ and $\beta$. We can therefore combine these proofs
using the $| E$ rule  and $\sbst$ as follows.

$$
\infer[|E]{\sbst(\Gamma),\sbst(\Delta)\vdash
  (M\, N)^{\sbst(\alpha)}:A}{\infer*[\delta_1]{\Gamma\vdash M^{\beta\ra\alpha}:A|B}{} & \infer*[\delta_2]{\Delta \vdash
N^{\gamma}:B}{}}
$$

Producing principal type $\sbst(\alpha)$ for this derivation.

\item{$[\multimap I/\forall I]$} If it results from a translation with
  a pair of string formulas, we treat the combination of the
  $\multimap I$ and a $\forall I$ rule as a single step. By Proposition~\ref{prop:qr}, we can do so without loss of
  generality. Such a combination can only result from the
  translation of a positive formula with main connective $/$ or $\I$. We treat only $/ I$; the
  case for $\bs I$ is symmetric.

$$
\infer[=_{\textit{def}}]{\Gamma \vdashpos \| A/ B
  \|^{[C,D]}}{\infer[\forall I]{\Gamma
    \vdashpos \forall x. \| B \|^{[D,x]} \multimap \| A \|^{[C,x]}}{
      \infer[\multimap I]{\Gamma\vdashpos \| B \|^{[D,E]} \multimap \| A
        \|^{[C,E]}}{
           \Gamma,\|B \|^{[D,E]} \vdashpos \| A \|^{[C,E]}}
       }
    }
$$

We can simply extend the proof $\delta_1$ from the induction
hypothesis as follows.

$$
\infer[/I]{\Gamma \vdash ((\lambda x.M)(\lambda z.z))^{D\ra C}:A/B}{\infer*[\delta_1]{\Gamma,
    x^{E\ra D}:B \vdash
  M^{E\ra C}:A}{}}
$$

\item{$[\multimap I]$} Finally, the case where the $\multimap I$ is
  not followed by a $\forall I$ corresponds to the $| I$ rule. We are
  in the following situation.
%TODO: VERIFY! We move from explicit quantification to implicit
%quantification here! Better to use implicit quantification everywhere?

$$
\infer[=_{\textit{def}}]{\Gamma \vdashpos \leftimpl B| A
  \rightimpl^{f(\beta\ra\alpha)}}{\infer[\multimap I]{\Gamma\vdashpos \leftimpl B
    \rightimpl^{f(\beta)} \multimap \leftimpl A \rightimpl^{f(\alpha)}}{
      \Gamma, \leftexpl B \rightexpl^{f(\beta)} \vdashpos \leftimpl A \rightimpl^{f(\alpha)}}
    }
$$

We can simply extend the proof $\delta_1$ of $\Gamma, x^{\beta}:B \vdash
M^{\alpha}:A$ given by the induction hypothesis as follows.

$$
\infer[| I]{\Gamma \vdash
  (\lambda x.M)^{\beta\ra\alpha}:A|B}{\infer*[\delta_1]{\Gamma ,x^{\beta}: B
    \vdash M^{\alpha}:A}{}}
$$
\end{description}
\qed

%From the shape of the rules, it is clear that the only way to obtain
%a conclusion of the form $\forall x. B \multimap A$

\subsection{Main Theorem}
\label{sec:mainthm}

\begin{theorem}\label{thm:main} Derivability of hybrid type-logical grammars and their
  translation into first-order linear logic coincides. Moreover,
  proofs in the two systems produce the same semantic lambda terms.
\end{theorem}

\paragraph{Proof} Immediate from Lemma~\ref{lem:htom} and
Lemma~\ref{lem:mtoh} and the observation that $/E$, $\bs E$ and $|E$,  like
$\multimap E$ to which they correspond by translation,
are all translated as application on the meaning level and similarly for the different introduction rules and abstraction.  \qed

Thanks to Theorem~\ref{thm:main}, we can use the well-understood proof
theory of first-order linear logic for parsing/theorem proving hybrid type-logical
grammars. Besides (focused) natural deduction and proof nets,
discussed in Section~\ref{sec:mill1}, the work on sequent proof search of
\citeasnoun{foll}, which includes a treatment of the additives, can
also directly be applied. These proof systems all have their strengths
and inconveniences, but, since they are all equivalent we can choose
the most appropriate tool for the job. For example, focused natural
deduction and proof nets
simplify the work of enumerating readings for a
given statement, and, as shown in Figure~\ref{fig:cycle}, proof nets provide an easy way to show
underivability of a statement.
In addition, the main theorem has the following immediate consequence.

\begin{corollary} Hybrid type-logical grammars are NP-complete
\end{corollary}

\paragraph{Proof} Hardness follows from the fact that hybrid
type-logical grammars contain the Lambek calculus (the implicational
fragment of the Lambek calculus was shown to be NP-complete by \citeasnoun{savateev}) --- or alternatively
from the fact that they contain lexicalized abstract categorial
grammars \cite{groote01acg}. Since first-order linear logic is NP-complete, by
Lemma~\ref{lem:htom} and the fact that the translation is linear in
the size of the formulas, hybrid type-logical grammars are in NP. \qed

To compare hybrid type-logical grammars with lambda grammars, we first
define an interesting subclass of hybrid type-logical grammars which we will show
to be equivalent to lambda grammars.

\begin{definition} A hybrid proof is \emph{strictly separated} iff for
  every $/I$ and $\backslash I$ rule, the
  subproof leading to the premiss of this introduction rule consists
  only of Lambek elimination rules and premisses $A\vdash A$ with $A$
  a Lambek formula (ie.\ a member of $\mathcal{F}_1$, containing only
  $/$, $\backslash$ and simple atomic formulas).
\end{definition}

We can enforce strict separation directly in the proof theory by splitting the $\vdash$ symbol into
$\vdash_L$ and $\vdash_{\lambda}$, subscripting by $\vdash_L$ the premisses and
conclusions of the $/ E$, $\backslash E$, $/ I$, $\backslash I$ and
axiom/hypothesis for Lambek formulas as $\vdash_L$, subscripting by
$\vdash_{\lambda}$ the $| E$, $| I$ and axiom/hypothesis for formulas
not in $\mathcal{F}_1$ and adding the inclusion rule.

$$
\infer[L,\lambda]{\Gamma \vdash_{\lambda} M^{E\rightarrow
    D}:B}{\Gamma\vdash_L M^{E\rightarrow D}:B}
$$

Not all proofs in hybrid type-logical grammars are strictly separated,
as shown by the example in Section~\ref{sec:hex} on page~\pageref{hexproof}, where the final $/I$
rule is preceded by both $|E$ and $|I$.

\begin{lemma}\label{lem:hybridlambda} Strictly separated hybrid type-logical grammars generate the same
  string languages and the same string-meaning relations as lambda grammars.
\end{lemma}

\paragraph{Proof (sketch)} The main idea from  \cite{busz96}, who uses a variant of
the proof from \cite{pentus,pentus97}, is that we can replace Lambek
calculus formulas by \emph{sets} of atomic formulas (CFG nonterminals)
which behave combinatorially like AB formulas --- the CFG nonterminals
are essentially the names for AB formulas --- in such a way that these sets
generate the same lambda term
semantics. Here, we do the same for all Lambek \emph{sub}-formulas of
a given hybrid type-logical grammar.
 
By the definition of strict separation, we know that all Lambek rules
occur in subproofs where these rules are not intermingled with the
lambda grammar rules. Hence, Buszkowski's construction translates
these proofs of $\Gamma \vdash_L B$ into proofs of $\Gamma' \vdash_L
B$ where only the $/E$ and $\backslash E$ rules are used. Then, by
treating all Lambek formulas as CFG nonterminals and all
instantiations of the $/E$ and $\backslash E$ rules in the grammar as
CFG rules. That is, the instantiation of the the $\backslash E$ rule for
specific formulas $A$ and $B$

$$
\infer[\backslash E]{B}{A & A\backslash B}
$$

\noindent becomes a non-logical rule 

$$
\infer{C}{D & E} 
$$

(or, if we prefer
to write it as a CFG rule: $D, E \longrightarrow C$), where $C$ is the non-terminal
corresponding to formula $B$, $D$ corresponds to formula $A$ and $E$
corresponds to the formula $A\backslash B$.

%TODO: make this part clearer and more precise.
\editout{
Let $\mathcal{F}_1$ be the set of Lambek calculus formulas occurring as
subformulas of the lexicon of a hybrid type-logical grammar, let
$\mathcal{P}$ be the formulas of  $\mathcal{F}_1$ which occur
positively (plus the atomic formula $s$) and  $\mathcal{N}$ the
formulas of  $\mathcal{F}_1$ which occur negatively. The sets
$\mathcal{N}$ and $\mathcal{P}$ exhaust the Lambek formulas which can
appear.  By application of
the main theorem of \cite{busz96}, we produce an AB grammar
$G$ such that whenever $A_1, \ldots,
A_n \vdash C$ is derivable in the Lambek calculus (for all $A_i \in
\mathcal{N}$ and $C \in \mathcal{P}$), then the corresponding AB
grammar has formulas $B_i$
for each $A_i$ and $D$ for $C$ such that
$B_1,\ldots,B_n \vdash D$ is derivable using elimination rules only
\emph{and} this derivation produces the same semantics as the original proof.} 
%Since
%$G$ is a context-free grammar, we can translate it into a lambda
%grammar generating the same language, while keeping the
%semantics. Finally, we know from \cite{yk05acg} that we can lexicalize
%the resulting lambda grammar and thereby produce a lexicalized lambda
%grammar generating the same string language and string-meaning relations.}
\qed

\begin{lemma}\label{lem:nplambda} Parsing lambda grammars which are
  the translation of strictly separated hybrid type-logical grammars
  is NP-complete.
\end{lemma}

\paragraph{Proof} The construction of Lemma~\ref{lem:hybridlambda}
generates, by means of the \citeasnoun{busz96} proof, many non-logical grammar
rules. Given that such a system may not be decidable,  we need to
be careful. However, by the construction of \cite{busz96}, all
non-lexicalized rules are of the form $D, E \longrightarrow F$ with $D$, $E$ and
$F$ atomic formulas. Moreover, these atomic formulas correspond to AB
formulas, such that either $D = A/B$, $E=B$ and $F=A$ or $D=B$,
$F=B\backslash A$ and $F=A$ (for some Lambek formulas $A$ and
$B$). Therefore, we can start our proof by computing the closure of
these AB subproofs in $O(n^3)$, then continue the normal
lambda grammar proof, which is NP-complete. \qed

It should be obvious from the proof sketch of Lemmas~\ref{lem:hybridlambda} and~\ref{lem:nplambda} that though strictly
separated hybrid
type-logical grammars generate the same string languages and
string-meaning pairs as lambda grammars, hybrid type-logical grammars
allow a \emph{much} more compact specification of such grammars since
we avoid a brute-force explosion of the size of the lexicon and of the number of lexical entries per
word. Though I don't believe that the NP-complete problems we
encounter in computational linguistics are necessarily intractable ---
\citeasnoun{efficienthpsg} show that some NP-complete problems in
computational linguistics can be solved \emph{much} more
efficiently than $O(n^6)$ problems --- having an exponential explosion
of grammar size \emph{followed} by an NP-complete problem is
profoundly worrying for those interested in actually parsing the formalism.
%There will also be a large amount of redundancy in the resulting
%grammar: a $tv$ gapping grammar will add rules like $tv, np
%\longrightarrow vp$ and $np, vp\longrightarrow s$, with $tv$ the
%atomic formula corresponding to $(np\backslash s)/np$ and $vp$ the
%atomic formula corresponding to
%$np\backslash s$.

It is unclear whether we can generalize the proof of Lemma~\ref{lem:hybridlambda} to dispense
with the strict separation requirement on hybrid grammars. Allowing
interleaving of the Lambek grammar and lambda grammar rules seems to
require a generalization of the results of \cite{busz96} to the hybrid
type-logical grammar case and, unless we change the proof of the
theorem considerably, this would require a type of
interpolation proof for hybrid type-logical grammars, which, as we will
see in Section~\ref{visual}, seems problematic for the lambda grammar
part of the system. For example, looking back to the proof in
Section~\ref{sec:hex}, it is unclear how to replace the final $/I$
rule by the elimination rule for either $/$ or $\backslash$, besides
adding $s/(np\backslash s)$ directly as an additional lexical entry
for the quantifier. 

Also, though
it is certainly a desirable property of the hybrid system to derive
$s|(s|np) \vdash s/(np\backslash s)$ (for the given lexical lambda
term), since it relates the generalized quantifier formulas to one of
its standard Lambek calculus formulas, it is unclear if we actually
need this type of derivation to give a natural account of the
linguistic data. So the following question remains open: are there any
examples of hybrid type-logical grammar analyses where there is no
corresponding lambda grammar analysis? Having to resort to lexical
duplication is already a problem, both from a conceptual point of view
and from the point of view of parsing, but are there cases where even
this doesn't suffice?

Though we will leave this question unresolved, we investigate the descriptive
inadequacy of lambda grammars in Section~\ref{inadeq}.

\section{Comparison}
\label{sec:comparison}

The proof nets discussed in Section~\ref{sec:pn} provide an insightful way to
compare the different calculi discussed in this article in terms of
their basic ``building blocks'', seen from the
point of view of first-order linear logic. 

We need to be careful, since this comparison only gives
\emph{necessary} conditions to be in a certain fragment of first-order
linear logic, and as such, we can use it only as a diagnostic for
showing that possibilities are \emph{absent} from a logic. We can
directly use the different translation functions to give sufficient
conditions.

The conditions on the variables in the different fragments are also
absent from the visual representation. Nevertheless, we will see that
this comparison is insightful.

%Lambek grammars: prefix and suffix plus inverses. CFG only.

%Lambda grammars: application with abstraction as inverse. Prefix and suffix 
%definable, their inverses not. Choice
%between well-nested or not (order of lambda-terms). 

%D-grammars: prefix, suffix (also as \emph{tuple} concatenation), infix, circumfix with inverses for
%each. Well-nested only.

%Hybrid TLG: application, prefix, suffix plus inverses. Choice between
%well-nested or not (order of lambda-terms).

\subsection{A visual comparison of the different calculi}
\label{visual}

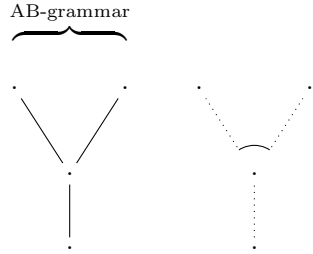
\begin{figure}
%Lambek grammars
%\medskip
\begin{center}
\begin{tikzpicture}
\node (forallnc) {$.$};
\node (forallnp) [above=2em of forallnc] {$.$};
\draw (forallnc) -- (forallnp);
\node (forallpc) [right=6em of forallnc] {$.$};
\node (forallpp) [above=2em of forallpc] {$.$};
\draw [dotted] (forallpc) -- (forallpp);
%
%\node (lollinc) [above=5em of forallnc] {$.$};
\node (tmplnl) [left=1.25em of forallnp] {};
\node (alollin) [above=2.5em of tmplnl] {$\ .\ $};
\draw (forallnp) -- (alollin);
\node (tmplnr) [right=1.25em of forallnp] {};
\node (blollin) [above=2.5em of tmplnr] {$\ .\ $};
\draw (forallnp) -- (blollin);
%
%\node (lollipc) [above=5em of forallpc] {$.$};
\node (tmplpl) [left=1.25em of forallpp] {};
\node (alollip) [above=2.5em of tmplpl] {$\ .\ $};
\node (tmplpr) [right=1.25em of forallpp] {};
\node (blollip) [above=2.5em of tmplpr] {$\ .\ $};
\begin{scope}
\begin{pgfinterruptboundingbox}
\path [clip] (forallpp.center) circle (2.5ex) [reverseclip];
\end{pgfinterruptboundingbox}
\draw [dotted] (forallpp.center) -- (blollip);
\draw [dotted] (forallpp.center) -- (alollip);
\end{scope}
\begin{scope}
\path [clip] (alollip) -- (forallpp.center) -- (blollip);
\draw (forallpp.center) circle (2.5ex);
\end{scope}
\node (tmp) [left=3.7em of alollip] {};
\node (lab)  [above=1em of tmp]
{$\overbrace{\rule{4.3em}{0pt}}^{\text{AB-grammar}} $};
\node (tmp) [left=0em of alollip] {};
%\node (ttl) [above=3.5em of tmp] {Lambek grammar};
\end{tikzpicture}
\end{center}
%\medskip
\caption{Lambek grammar}
\label{fig:llinks}
\end{figure}

Figure~\ref{fig:llinks} shows the Lambek calculus connectives as
links for first-order linear logic proof nets.
Curry's \citeyear{Curry61} criticism of the Lambek calculus connectives, seen from
the current perspective, is that they combine subcategorization information (functor-argument structure)
and string operations. Though from a modern proof-theoretical point
of view \cite{focus} it is perfectly valid to combine multiple
positive and multiple negative rules into a single rule, separating
the two gives more freedom (that is, it allows us to express more
relations between the string positions and go beyond simple
concatenation --- the prefix and postfix of the Lambek calculus).

\begin{figure}
%MILL1
%\medskip
\begin{center}
\begin{tikzpicture}
\node (forallnc) {$.$};
\node (forallnp) [above=2em of forallnc] {$.$};
\draw (forallnc) -- (forallnp);
\node (forallpc) [right=6em of forallnc] {$.$};
\node (forallpp) [above=2em of forallpc] {$.$};
\draw [dotted] (forallpc) -- (forallpp);
\node (lollinc) [above=5em of forallnc] {$.$};
\node (tmplnl) [left=1.25em of lollinc] {};
\node (alollin) [above=2.5em of tmplnl] {$\ .\ $};
\draw (lollinc) -- (alollin);
\node (tmplnr) [right=1.25em of lollinc] {};
\node (blollin) [above=2.5em of tmplnr] {$\ .\ $};
\draw (lollinc) -- (blollin);
\node (lollipc) [above=5em of forallpc] {$.$};
\node (tmplpl) [left=1.25em of lollipc] {};
\node (alollip) [above=2.5em of tmplpl] {$\ .\ $};
\node (tmplpr) [right=1.25em of lollipc] {};
\node (blollip) [above=2.5em of tmplpr] {$\ .\ $};
\begin{scope}
\begin{pgfinterruptboundingbox}
\path [clip] (lollipc.center) circle (2.5ex) [reverseclip];
\end{pgfinterruptboundingbox}
\draw [dotted] (lollipc.center) -- (blollip);
\draw [dotted] (lollipc.center) -- (alollip);
\end{scope}
\begin{scope}
\path [clip] (alollip) -- (lollipc.center) -- (blollip);
\draw (lollipc.center) circle (2.5ex);
\end{scope}
\node (tmp) [left=3.7em of alollip] {};
\node (lab)  [above=1em of tmp]
{$\overbrace{\rule{4.3em}{0pt}}^{\text{positive}} $};
\node (tmp) [right=0.9em of alollip] {};
\node (lab)  [above=1em of tmp]
{$\overbrace{\rule{4.3em}{0pt}}^{\text{negative}} $};
\node (tmp) [left=0em of alollip] {};
%\node (ttl) [above=3.5em of tmp] {MILL1};
\begin{pgfinterruptboundingbox}
\node (tmp) [left=6.5em of lollinc] {};
\node (lbr) [above=-1.0em of tmp] {$\text{\scriptsize
    functor/argument}\left\{ \rule{0pt}{2.1em}\right.$};
\node (tmpb) [below=4em of tmp] {};
\node (lbrb) [above=-1.8em of tmpb] {$\text{\scriptsize
    \ \ \ \ string positions}\left\{ \rule{0pt}{1.7em}\right.$};
\end{pgfinterruptboundingbox}
\end{tikzpicture}
\end{center}
%\medskip
\caption{MILL1}
\label{fig:lll}
\end{figure}

As shown in Figure~\ref{fig:lll}, the first-order linear logic solution decomposes the Lambek
connectives into separate subcategorization and string position
components. In a sense, this decomposition answers Curry's critique in
a very simple way.

\begin{figure}
%lambda grammars
%\medskip
\begin{center}
\begin{tikzpicture}
\node (forallnc) {$.$};
\node (forallnp) [above=2em of forallnc] {$.$};
\draw (forallnc) -- (forallnp);
\node (forallpc) [right=6em of forallnc] {};
%\node (forallpp) [above=2em of forallpc] {$.$};
%\draw [dotted] (forallpc) -- (forallpp);
%
\node (lollinc) [above=5em of forallnc] {$.$};
\node (tmplnl) [left=1.25em of lollinc] {};
\node (alollin) [above=2.5em of tmplnl] {$\ .\ $};
\draw (lollinc) -- (alollin);
\node (tmplnr) [right=1.25em of lollinc] {};
\node (blollin) [above=2.5em of tmplnr] {$\ .\ $};
\draw (lollinc) -- (blollin);
\node (lollipc) [above=5em of forallpc] {$.$};
\node (tmplpl) [left=1.25em of lollipc] {};
\node (alollip) [above=2.5em of tmplpl] {$\ .\ $};
\node (tmplpr) [right=1.25em of lollipc] {};
\node (blollip) [above=2.5em of tmplpr] {$\ .\ $};
\begin{scope}
\begin{pgfinterruptboundingbox}
\path [clip] (lollipc.center) circle (2.5ex) [reverseclip];
\end{pgfinterruptboundingbox}
\draw [dotted] (lollipc.center) -- (blollip);
\draw [dotted] (lollipc.center) -- (alollip);
\end{scope}
\begin{scope}
\path [clip] (alollip) -- (lollipc.center) -- (blollip);
\draw (lollipc.center) circle (2.5ex);
\end{scope}
\node (tmp) [left=3.55em of alollip] {};
\node (lab)  [above=1em of tmp]
{$\overbrace{\rule{4.3em}{0pt}}^{\text{2nd-order}\
    \lambda\text{-grammar}} $};
\node (tmp) [left=0em of alollip] {};
%\node (ttl) [above=3.5em of tmp] {$\lambda$-grammar};
\end{tikzpicture}
\end{center}
%\medskip
\caption{Lambda grammars}
\label{fig:lgl}
\end{figure}
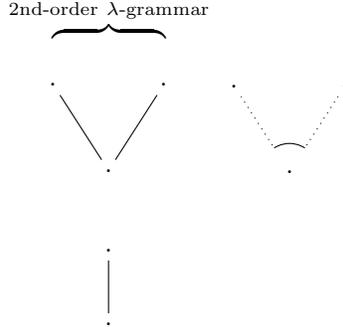

Curry's own solution is different and causes a loss of symmetry: as
Figure~\ref{fig:lgl} makes clear, the
positive universal link is missing! This 
loss of symmetry is easy to miss in a unification-based presentation of 
the logic where, in addition, the quantifiers occur only as an implicit prefix of
the formula. For a
logician/proof theorist, this is worrying since many classical results
and desirable properties of the system (restriction to atomic axioms,
cut elimination, interpolation\footnote{Interpolation, proved first
  for the Lambek calculus in \cite{Roorda} is a key component of the
  context-freeness proof for the Lambek calculus of
  \citeasnoun{pentus97} and is likely to play a similar role in proofs
  about the generative capacity of these alternative and extended systems.}) depend on this symmetry. However, it is also the cause of empirical inadequacy:  positive $A/B$ and $B\bs A$ can no longer
be represented, hence no satisfactory treatment of adverbs,
coordination, gapping etc.; we will elaborate this point in detail in
Section~\ref{inadeq}.

Another way to look at this is that lambda grammars require
all formulas to be expressed in prenex normal form --- something we
exploit in the translation function. However, since
we are using linear logic, not all formulas have a prenex normal form.
The following are all underivable (assuming no occurrences of $x$ in
$B$). Refer back to Figure~\ref{fig:cycle} to see why the first
statement is underivable.
\begin{align*}
(\forall x. A) \multimap B &\nvdash \exists x (A
\multimap B) \\
\exists x (A
\multimap B) & \nvdash (\forall x. A) \multimap B \\
B \multimap \exists x. A & \nvdash \exists
x. (B\multimap A) \\
\exists
x. (B\multimap A) & \nvdash B \multimap \exists x. A
\end{align*}

% === Moved to discussion
% Though Buszkowski's \citeyear{busz96} variant of Pentus' \citeyear{pentus} theorem shows that we
% do not need positive $A/B$ and $B\bs A$, eliminating these occurrences
% explodes the size of the lexicon. This means lambda grammars suffer not only
% from descriptive inadequacy (ie.\ an incapacity for expressing
% linguistic generalizations directly in the formalism, needing
% excessive duplication in the lexical entries), they are also
% ill-suited for parsing: they first
% have an exponential blowup of grammar size and \emph{then} are
% NP-complete, whereas Lambek grammars, MILL1 grammars, hybrid grammars
% and D grammars all avoid this exponential blowup (since they avoid
% this descriptive inadequacy problem) and  are just NP-complete.

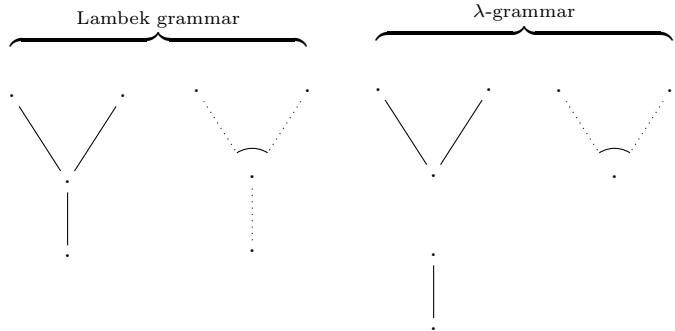
\begin{figure}
%Hybrid
%\medskip
\begin{center}
\begin{tikzpicture}
\node (forallnc) {$.$};
\node (forallnp) [above=2em of forallnc] {$.$};
\draw (forallnc) -- (forallnp);
\node (forallpc) [right=6em of forallnc] {};
%\node (forallpp) [above=2em of forallpc] {$.$};
%\draw [dotted] (forallpc) -- (forallpp);
%
\node (lollinc) [above=5em of forallnc] {$.$};
\node (tmplnl) [left=1.25em of lollinc] {};
\node (alollin) [above=2.5em of tmplnl] {$\ .\ $};
\draw (lollinc) -- (alollin);
\node (tmplnr) [right=1.25em of lollinc] {};
\node (blollin) [above=2.5em of tmplnr] {$\ .\ $};
\draw (lollinc) -- (blollin);
\node (l1) [above=1em of blollin] {$\ \ \qquad\overbrace{\rule{11.2em}{0pt}}^{\lambda\text{-grammar}}$};
\node (lollipc) [above=5em of forallpc] {$.$};
\node (tmplpl) [left=1.25em of lollipc] {};
\node (alollip) [above=2.5em of tmplpl] {$\ .\ $};
\node (tmplpr) [right=1.25em of lollipc] {};
\node (blollip) [above=2.5em of tmplpr] {$\ .\ $};
\begin{scope}
\begin{pgfinterruptboundingbox}
\path [clip] (lollipc.center) circle (2.5ex) [reverseclip];
\end{pgfinterruptboundingbox}
\draw [dotted] (lollipc.center) -- (blollip);
\draw [dotted] (lollipc.center) -- (alollip);
\end{scope}
\begin{scope}
\path [clip] (alollip) -- (lollipc.center) -- (blollip);
\draw (lollipc.center) circle (2.5ex);
\end{scope}
% rename and move forallnc
\node (tmp) [left=13em of forallnc] {};
\node (forallnc) [above=2em of tmp] {$.$};
\node (forallnp) [above=2em of forallnc] {$.$};
\draw (forallnc) -- (forallnp);
\node (tmplnl) [left=1.25em of forallnp] {};
\node (alollin) [above=2.5em of tmplnl] {$\ .\ $};
\draw (forallnp) -- (alollin);
\node (tmplnr) [right=1.25em of forallnp] {};
\node (blollin) [above=2.5em of tmplnr] {$\ .\ $};
\draw (forallnp) -- (blollin);
\node (tmp) [left=13em of forallpc] {};
% bottom of Lambek par
\node (forallppb) [above=5em of tmp] {$.$};
% bottom of complete Lambek slash
\node (forallpcb) [below=2em of forallppb] {$.$};
\node (tmplplb) [left=1.25em of forallppb] {};
\node (alollipb) [above=2.5em of tmplplb] {$\ .\ $};
\node (tmplprb) [right=1.25em of forallppb] {};
\node (blollipb) [above=2.5em of tmplprb] {$\ .\ $};
\begin{scope}
% reverseclip doesn't seem to work here, I'm note sure why...
\begin{pgfinterruptboundingbox}
\path[invclip] (forallppb.center) circle (2.5ex);
%\path [clip] (forallppb.center) circle (2.5ex) [reverseclip];
\end{pgfinterruptboundingbox}
%\clip (tmplplb) rectangle (blollipb);
%\clip (forallppb.center) circle (2.5ex);
%\clip (forallppb.center) rectangle (alollipb);
%\clip (forallppb.center) rectangle (blollipb);
\draw [dotted] (forallppb.center) -- (blollipb);
\draw [dotted] (forallppb.center) -- (alollipb);
%\draw [dotted] (blollipb) -- (forallppb.center);
%\draw [dotted] (alollipb)  -- (forallppb.center);
\end{scope}
%\draw (tmplplb) rectangle (blollipb);
%\draw (tmplblp) rectangle (blollipb);
% circle segment connecting par line segments
\begin{scope}
\path [clip] (alollipb) -- (forallppb.center) -- (blollipb);
\draw (forallppb.center) circle (2.5ex);
\end{scope};
\draw[dotted] (forallpcb) -- (forallppb);
\node (l2) [above=1em of blollin] {$\ \
  \qquad\overbrace{\rule{11.2em}{0pt}}^{\text{Lambek grammar}}$};
\node (tmp) [left=7em of alollip] {};
%\node (ttl) [above=3.5em of tmp] {Hybrid grammar};
\end{tikzpicture}
\end{center}
%\medskip
\caption{Hybrid grammar}
\label{fig:hgl}
\end{figure}

The hybrid solution to this problem is shown in Figure~\ref{fig:hgl}: reintroduce the positive Lambek connectives
directly. There are now two ways of coding the negative Lambek
connectives. The resulting system is also greater than the sum of its parts, since
gapping, which has a satisfactory neither in Lambek grammars nor in
lambda grammars, can be elegantly treated in hybrid categorial grammar
\cite{kl12gap,kl13emp}.

Symmetry is still lost\footnote{Neither full logical symmetry nor
  having the Lambek calculus as a subsystems is of course necessary to
  have an empirically valid formal system, as shown, for example by
  CCG \cite{steedman}. However it calls for further investigation as to
  what exactly is absent from the system and if this absence is
  important from a descriptive point of view. For lambda grammars, we will do this in detail in Section~\ref{inadeq}.}, but empirically the system seems
comparable to the Displacement calculus \cite{mvf11displacement}: the Displacement calculus has
the full symmetry absent from hybrid type-logical grammars. In spite
of this,
% but hybrid
%type-logical grammars (unlike the Displacement calculus) can represent
%non-well-nested MCFGs, though it is debatable whether this distinction has
%an empirical bite. 
as we have seen at the end of
Section~\ref{sec:string}, in many cases, the analyses proposed for the two
formalisms basically agree, as is made especially clear by their
translation into MILL1.

The differences between the two systems seems to be that hybrid
type-logical grammars can, like lambda grammars, generate
non-well-nested string languages and that Displacement grammars (seen
from the point of view of hybrid type-logical grammars) allow
the Lambek connectives to outscope the discontinuous connectives.
Further analysis is necessary to decide which of these two systems has
the better empirical coverage.

%D grammars, binary

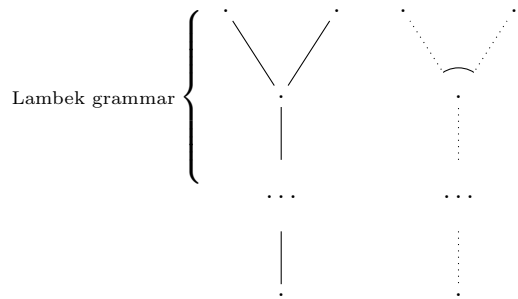
\begin{figure}
%\medskip
\begin{center}
\begin{tikzpicture}
\node (forallnc) {};
\node (forallnp) [above=2em of forallnc] {$.$};
\draw (forallnc) -- (forallnp);
\begin{pgfinterruptboundingbox}
\node (lbr) [left=2em of forallnp] {$\text{\scriptsize Lambek grammar}\left\{ \rule{0pt}{3.6em}\right.$};
\end{pgfinterruptboundingbox}
\node (bottn) [below=4em of forallnp] {};
\node (botn) [below=2em of bottn] {$.$};
\draw (bottn) -- (botn);
\node (dots) [above=.2em of bottn] {$\ldots$};
\node (forallpc) [right=6em of forallnc] {};
\node (forallpp) [above=2em of forallpc] {$.$};
\draw [dotted] (forallpc) -- (forallpp);
\node (bottp) [below=4em of forallpp] {};
\node (botp) [below=2em of bottp] {$.$};
\draw[dotted] (bottp) -- (botp);
\node (dots) [above=.2em of bottp] {$\ldots$};
%\node (lollinc) [above=5em of forallnc] {$.$};
\node (tmplnl) [left=1.25em of forallnp] {};
\node (alollin) [above=2.5em of tmplnl] {$\ .\ $};
\draw (forallnp) -- (alollin);
\node (tmplnr) [right=1.25em of forallnp] {};
\node (blollin) [above=2.5em of tmplnr] {$\ .\ $};
\draw (forallnp) -- (blollin);
%
%\node (lollipc) [above=5em of forallpc] {$.$};
\node (tmplpl) [left=1.25em of forallpp] {};
\node (alollip) [above=2.5em of tmplpl] {$\ .\ $};
\node (tmplpr) [right=1.25em of forallpp] {};
\node (blollip) [above=2.5em of tmplpr] {$\ .\ $};
\begin{scope}
\begin{pgfinterruptboundingbox}
\path [clip] (forallpp.center) circle (2.5ex) [reverseclip];
\end{pgfinterruptboundingbox}
\draw [dotted] (forallpp.center) -- (blollip);
\draw [dotted] (forallpp.center) -- (alollip);
\end{scope}
\begin{scope}
\path [clip] (alollip) -- (forallpp.center) -- (blollip);
\draw (forallpp.center) circle (2.5ex);
\end{scope}
\node (tmp) [left=0.0em of alollip] {};
\end{tikzpicture}
\end{center}
%\medskip
\caption{D grammars, binary}\
\label{fig:dl}
\end{figure}

D grammars \cite{mvf11displacement} have a different perspective,
which is shown in Figure~\ref{fig:dl}. Functor argument structure
and string positions are still joined, but a greater number of
combinations are possible (from 0 to $n$ quantifiers, for a small
value of $n$ determined by the grammar). Lambek grammars are now the restriction to
a single quantifier for each binary connective.

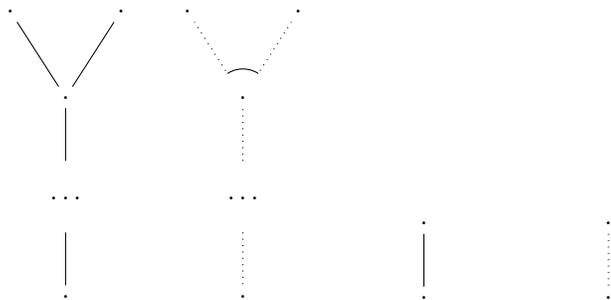
\begin{figure}
%D grammars, including synthetic connectives.
%\medskip
\begin{center}
\begin{tikzpicture}
\node (forallnc) {};
\node (forallnp) [above=2em of forallnc] {$.$};
\draw (forallnc) -- (forallnp);
\node (bottn) [below=4em of forallnp] {};
\node (botn) [below=2em of bottn] {$.$};
\draw (bottn) -- (botn);
\node (dots) [above=.2em of bottn] {$\ldots$};
\node (bottp) [below=4em of forallpp] {};
\node (botp) [below=2em of bottp] {$.$};
\draw[dotted] (bottp) -- (botp);
\node (dots) [above=.2em of bottp] {$\ldots$};
\node (forallpc) [right=6em of forallnc] {};
\node (forallpp) [above=2em of forallpc] {$.$};
\draw [dotted] (forallpc) -- (forallpp);
%
%\node (lollinc) [above=5em of forallnc] {$.$};
\node (tmplnl) [left=1.25em of forallnp] {};
\node (alollin) [above=2.5em of tmplnl] {$\ .\ $};
\draw (forallnp) -- (alollin);
\node (tmplnr) [right=1.25em of forallnp] {};
\node (blollin) [above=2.5em of tmplnr] {$\ .\ $};
\draw (forallnp) -- (blollin);
%
%\node (lollipc) [above=5em of forallpc] {$.$};
\node (tmplpl) [left=1.25em of forallpp] {};
\node (alollip) [above=2.5em of tmplpl] {$\ .\ $};
\node (tmplpr) [right=1.25em of forallpp] {};
\node (blollip) [above=2.5em of tmplpr] {$\ .\ $};
\begin{scope}
\begin{pgfinterruptboundingbox}
\path [clip] (forallpp.center) circle (2.5ex) [reverseclip];
\end{pgfinterruptboundingbox}
\draw [dotted] (forallpp.center) -- (blollip);
\draw [dotted] (forallpp.center) -- (alollip);
\end{scope}
\begin{scope}
\path [clip] (alollip) -- (forallpp.center) -- (blollip);
\draw (forallpp.center) circle (2.5ex);
\end{scope}
\node (tt) [right=6em of bottp] {$.$};
\node (tb) [below=2em of tt] {$.$};
\draw (tt) -- (tb);
\node (pt) [right=6em of tt] {$.$};
\node (pb) [below=2em of pt] {$.$};
\draw[dotted] (pt) -- (pb);
\node (tmp) [right=5.0em of alollip] {};
%\node (et) [above=1em of tmp] {D grammars};
\end{tikzpicture}
\end{center}
\caption{D grammars}
\label{fig:dlu}
\end{figure}
%\medskip

D grammars enriched with bridge, left projection and right projection,
shown in Figure~\ref{fig:dlu},
permit combinations of string position/subcategorization which
are not of the same polarity. These uses are rather restricted
compared to the visually similar quantifier link of first-order
linear logic:
essentially, they enable us to require that a pair of positions spans the
empty string.

Summing up, first-order linear logic decomposes the connectives of
different grammatical frameworks --- the Lambek calculus, lambda
grammars, Hybrid Type-Logical Grammars and the Displacement calculus
--- in a natural way into its four types of links. This visual
comparison both highlights the differences between this calculi and
opens the way for a more detailed comparison of the descriptive
limitations of one calculus compared to another.

Given that it is a decomposition of connectives, the MILL1 translation
is slightly bigger in terms of the total number of connectives in the
lexical entries. However, the basic operation are simple and
well-understood and the first-order variables actually function as
powerful constraints during proof search. Thanks to the embedding
results of this paper and of \cite{moot13lambek}, we can import the large range of linguistic
phenomena treated by Displacement grammars and Hybrid Type-Logical
Grammar directly into MILL1.

From the point of view of first-order linear logic, the connectives of
the other calculi are synthetic connectives, combined connectives of
the same polarity. We can mix and match these synthetic connectives as we see
fit. We can also exploit the symmetry of
first-order linear logic and use lambda grammar lexical entries as
arguments, restoring the symmetry of lambda grammars (and of Hybrid
Type-Logical Grammars). In addition, we can add the product $\otimes$ and
quantifier $\exists$ to our calculus essentially for free. Moreover,
as discussed in \cite{mill1,moot13lambek} we can use the quantifiers
of first-order linear logic to give an account of agreement and
island constraints as well. So we can improve upon Displacement
grammar analyses by adding agreement and island constraints and
improve upon Hybrid Type-Logical Grammar analyses by adding symmetry,
agreement and island constraints, all with the same logical primitives.

\section{Descriptive Inadequacy of Lambda Grammars}
\label{inadeq}

As already alluded to in Section~\ref{visual}, the asymmetry of
lambda grammars is the cause of descriptive inadequacy. Researcher in
lambda grammars have been aware of problems with coordination at least
since \citeasnoun{muskens01lfg}, who briefly mentions an apparent
incompatibility between lambda grammars
and the categorial grammar treatment of coordination, but the problem
can be traced back to \cite{Curry61} where the analysis of the coordination ``both
\ldots and \ldots'' in \S 5-6 is problematic. Kubota \& Levine \citeyear{kl13coord,kl13emp} show how
catastrophic the predictions of lambda grammars are; we will repeat
their observations below while adding several additional troublesome cases. This
problem has been little noted and little discussed\footnote{At least in the
lambda grammar and abstract categorial grammar literature, the problem
is discussed in the context of linear grammar in \cite{worth14coord}.}. %\marginpar{Is there
%  discussion at all besides M and K\&L?}. 
Indeed, one can find several
claims in the literature which deny there is a problem: Muskens claims
elsewhere \cite{muskens03lambda} that ``Since word order is now completely encoded in the
phrase structure term, there is no longer any need for a
directionality of the calculus'' and that ``The availability of
syntactic $\lambda$-terms reins in the overgeneration of the
traditional undirected calculi.''. However, as we will show below,
using lambda terms to limit the overgeneration of undirected calculi
is only partially successful and it is exactly for this reason that  
a satisfactory treatment of coordination has remained elusive. Worse, the problem of overgeneration is not
limited to coordination, but a problem with \emph{any} higher-order type of the
Lambek calculus. The standard higher-order lambda grammar treatments
for generalized quantifiers and for non-peripheral are the only cases we know of where
lambda grammars make the right predictions. But even here, the
lambda grammar analysis does not generalize: generalized quantifiers
can be see as instances of Moortgat's \citeyear{m96q} $q(A,B,C)$
operator, and the lambda grammar treatment only works when $B$ is atomic and therefore for
quantifiers, of type $q(np,s,s)$, but not
for reflexives, of type $q(np,np\bs s,np\bs s)$.
For non-peripheral extraction, the lambda grammar analysis again
presupposes the extracted element is an atomic formula and therefore
the treatment does not generalize to gapping (for more on gapping see Section~\ref{sec:problems}).

To give an idea of how widespread and serious the problems are, the
following is a non-exhaustive list of problems for lambda grammars.

\ex. \label{sent:adverb} John deliberately hit Mary. (adverbs)

\ex. John bought a sandwich and ran to the train. (VP coordination)

\ex. \label{sent:fish} John caught and ate a fish. (TV coordination)

\ex. \label{sent:coord} John likes both black and gray t-shirts. (adjective coordination, after Curry, 1961)

\ex. \label{sent:rnr}  John loves but Mary hates Noam. (right-node raising)

\ex. John bought himself a present. (reflexives)

\ex. John studies logic and Charles, phonetics. (gapping)

\ex. John left before Mary did. (ellipsis)

\ex. \label{sent:end} John ate more donuts than Mary bought bagels. (comparative sub-deletion)

These problems range from the mundane to the more involved, but the
important point is that, taken together, these problems occur
very frequently and that \emph{all} cases listed above have a simple and elegant
treatment in the Displacement calculus \cite{mvf11displacement}, in Hybrid
type-logical grammars \cite{kl12gap,kl13dgap} and in multimodal
type-logical grammars \cite{cgellipsis,KurtoMM}. Sentence~\ref{sent:adverb} to
\ref{sent:rnr} are simply and correctly handled by Lambek grammars
 and Sentence~\ref{sent:adverb} to \ref{sent:coord} even by AB grammars.

Let me be precise about what I mean by descriptive inadequacy in this
context, since some authors use the term with a slightly different
meaning. A theory suffers from descriptive inadequacy if it fails to
capture linguistic generalizations and instead has to resort to
enumerating the linguistic data. In a lexicalized formalism like
categorial grammars, this means we want to avoid multiplying the
number of lexical
entries for the words in our grammar as much as
possible\footnote{Maybe a more reasonable measure would prefer the \emph{sum} of
  the size for all entries assigned to a word to be as small as possible, since a single entry
  $A_1 \oplus \ldots \oplus A_n$ is not really simpler that $n$ distinct
  entries $A_1,\ldots,A_n$. 

It should also be noted as the size of our grammar
  increases (in terms of the number of words and constructions it is
  able to handle), so does the size of our lexicon. So this is a relative
  measure rather than an absolute one.}. So in the context of the examples
above, we would like Sentence~\ref{sent:adverb} to use the same
lexical entries as the sentence ``John hit Mary'', with the lexical
assignment to ``deliberately'' being to only addition and we would
like Sentence~\ref{sent:fish} to use the same lexical
entries as the sentences ``John caught a fish'' and ``John ate a
fish'', with the lexical assignment to ``and''  being the only
difference. When I say that lambda grammars suffer from descriptive
inadequacy, this does not mean that they are fundamentally unable to
handle Sentences~\ref{sent:adverb} to \ref{sent:end}, since Lemma~\ref{lem:hybridlambda}
guarantees that they can (given that the phenomena listed above all
have strictly separated hybrid proofs). I mean that they cannot treat the sentences
above without introducing otherwise unmotivated additional lexical
entries --- in fact, not without an exponential blowup of the size of the
lexicon, as is clear from Lemma~\ref{lem:hybridlambda}.

In Section~\ref{solutions} we will discuss the consequences of these problems in detail,
as well as some possible modifications to
lambda grammars which may solve these problems, chiefly among those are
extensions to hybrid type-logical grammar and to first-order linear logic.

\subsection{Inhabitation machines}

To show the main results, we need some additional notions of the typed
lambda calculus. An inhabitation machine (see \cite{barendregt13types}) is a type of grammar which,
given a type, enumerates all possible terms of this type. Their use
for categorial grammars has been pioneered by \citeasnoun{benthem}.

From page 33 of \cite{barendregt13types}, the following two-level grammar
(defined on type-context pairs) enumerates all closed
inhabitants in beta-normal eta-long form of a given type.

$\Gamma$ is a context, $\Gamma,x^{\alpha}$ denotes $\Gamma\cup \{
  x^{\alpha} \}$ (where $x$ is distinct from the terms in $\Gamma$, so the result is again a
valid context), $A$ is an atomic type, $\alpha$, $\beta$ are
arbitrary types and $\vec{\alpha} \ra \beta$ is short for $\alpha_1 \ra
\ldots \ra \alpha_n \ra \beta$.

$$
\begin{array}{rl}
L(A;\Gamma) &\Longrightarrow xL(\alpha_1;\Gamma)\ldots L(\alpha_n;\Gamma)
\text{\qquad\qquad if}\ x:\vec{\alpha}\ra A \in \Gamma \\
L(\alpha\ra\beta;\Gamma) &\Longrightarrow \lambda x^{\alpha}. L(\beta;\Gamma,x^{\alpha})
\end{array}
$$

The lambda grammar case is considerably more restricted: the lexical
lambda terms must be linear and contain, for a given
word $w$ with corresponding variable $w_s$, a single occurrence of $w_s$. That is, we start with
$\Gamma = \{ w_s^{\sigma\ra \sigma} \}$ and for the application rule, we partition
$\Gamma - \{ x^{\vec{\alpha} \ra A} \} $ into jointly exhaustive, pairwise disjoint subsets
and divide these over the different subterms. In addition, we want our
lexical term to produce the correct word order and to be compatible
with the syntactic lambda grammar derivation.

\subsection{Problems for lambda grammars}
\label{sec:problems}

In this next section, we will show several problematic cases for lambda
grammars, using inhabitation
machines to exhaust all possible solutions and find all of them inadequate.

\subsubsection*{Adverbs}
As a first problem for lambda grammars,  the Lambek calculus formula of an
adverb such as ``deliberately'', as it occurs in a sentence like
``Eduardo deliberately fell'', is $(np\backslash s)/(np
\backslash s)$ --- it modifies a verb having taken all arguments
except its subject and this verb phrase is on the immediate right of
the adverb. If we translate this formula to a first-order formula and
move (where possible) the quantifiers to the prefix and eliminate
them, we obtain the formula $(\forall
c. np(c,2) \multimap s(c,D)) \multimap np(E,1) \multimap s(E,D)$ but
we cannot use the principal type $((2\ra c) \ra D \ra c)\ra (1\ra E) \ra D\ra E$
(with $c$ a fresh type constant) since it is uninhabited.

\begin{figure}
\begin{center}
\begin{tikzpicture}[node distance=1.5cm]
\node[draw] (top) {$((\str) \ra \str) \ra (\str) \ra \str$};
\node[draw, below of=top] (sigma) {$\sigma$};
\path[->] (top) edge node[right] {$\lambda
  \textit{VP}^{(\sigma\ra\sigma)\ra\sigma\ra\sigma}
  \textit{NP}^{\sigma\ra\sigma} z^{\sigma}$} (sigma);
\path[->] (sigma) edge [loop left] node[left] {$d$} () 
                             edge [loop right] node[right] {$\textit{NP}$} ();
\node[below of=sigma] (z) {$z$};
\node[left of=z] (vp) {$VP$};
\node[right of=z] (y) {$y$};
\path[->] (sigma) edge node {} (vp);
\path[->] ([xshift=-1pt] vp.north east) edge node {} ([xshift=-3pt] sigma.south);
%\path[->] ([xshift=3pt] vp) edge node {} ([xshift=3pt] sigma);
\path[->] (sigma) edge node {} (y);
\path[->] (sigma) edge node {} (z);
\node[draw,below of=vp] (sisi) {$\sigma\ra\sigma$};
\path[->] (vp) edge node {} (sisi);
%\path[->] (sisi.south west) edge node {$\lambda y^{\sigma}$}
%(sigma.north west);
%\draw[->]  (sisi.south west) to[out=45,in=115] (sigma.north west);
\draw[->]  (sisi.north west) to[out=145,in=115] (sigma.north west);
\node[left of=vp,node distance=1.15cm] {$\lambda y^{\sigma}$};
\end{tikzpicture}
\end{center}
\caption{Inhabitation machine for an adverb type.}
\label{fig:inhab}
\end{figure}
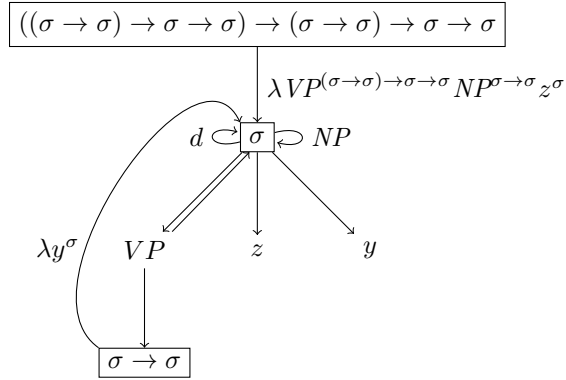

The lambda grammar syntactic type $(s|np) | (s|np)$ translates to the
prosodic type $((\str) \ra \str) \ra (\str) \ra \str$ and
produces the inhabitation machine shown in Figure~\ref{fig:inhab}. We use the variable $d$
(of type $\sigma\ra\sigma$)
to stand for the occurrence of the string ``deliberately''. We can see that the $\textit{VP}$ node in
the figure requires first an argument of type $\sigma\ra\sigma$ (the
downward arrow) then an argument of type $\sigma$ (the upward arrow)
to produce a term of type $\sigma$. Valid linear paths through the
machine must pass each term label exactly once, and must pass the
$\lambda y$-label (on the curved arrow upwards to $\sigma$) before the $y$ variable.

\begin{figure}
\begin{center}
\begin{tikzpicture}[node distance=1.5cm]
\node[draw] (top) {$((\str) \ra \str) \ra (\str) \ra \str$};
\node[draw, below of=top] (sigma) {$\sigma$};
\path[->] (top) edge node[right] {$\lambda
  \textit{VP}^{(\sigma\ra\sigma)\ra\sigma\ra\sigma}
  \textit{NP}^{\sigma\ra\sigma} z^{\sigma}$} (sigma);
\path[->] (sigma) edge [loop left] node[left] {$d$} () 
                             edge [loop right] node[right] {$\textit{NP}$} ();
\node[below of=sigma] (tmp) {};
\node[left of=tmp] (vp) {$VP$};
\node[right of=tmp] (z) {$z$};
\path[->] (sigma) edge node {} (vp);
\path[->] ([xshift=-1pt] vp.north east) edge node {} ([xshift=-3pt] sigma.south);
\path[->] (sigma) edge node {} (z);
\node[draw,below of=vp] (sisi) {$\sigma\ra\sigma$};
\path[->] (vp) edge node {} (sisi);
\node[draw,below of=sisi] (sbis) {$\sigma$};
\path[->] (sisi) edge node[right] {$\lambda y^{\sigma}$} (sbis);
\node[below of=sbis] (y) {$y$}; 
\path[->] (sbis) edge node {} (y);
\path[->] (sbis) edge [loop left] node[left] {$d$} () 
                             edge [loop right] node[right] {$\textit{NP}$} ();
\end{tikzpicture}
\end{center}
\caption{Simplified inhabitation machine for an adverb type.}
\label{fig:inhabb}
\end{figure}
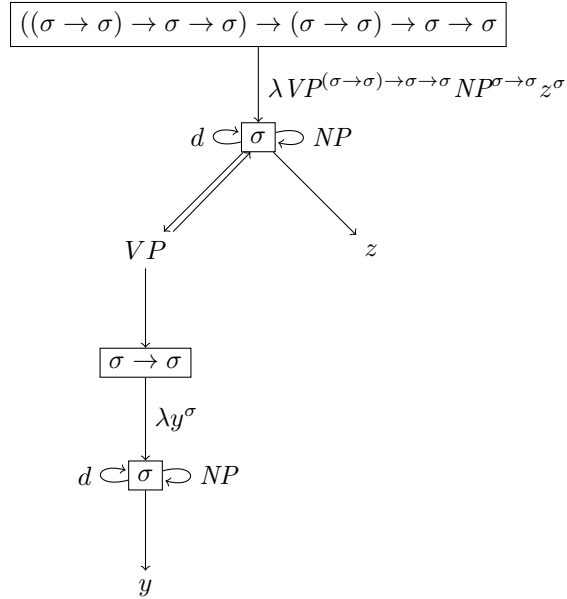

Figure~\ref{fig:inhabb} spits the $\sigma$ node in two, making the scope of
the $y$ variable clearer.

The word order of the sentence constrains the paths we can take. We
must take an $\textit{NP}$ arc before we take a $d$ arc, since
``deliberately'' occurs after the subjet noun phrase. So from the top
$\sigma$ node, we can only take three possible paths, as shown
below. For comparison, the uninhabited type corresponding most closely
to the first-order formula is shown as item~\ref{uninhabited}. We can
see that the three other types are obtained by replacing the $c$
constant by a $C$ variable and exchanging one of the occurrences of
$C$ with another atomic type in such a way that the resulting type is inhabited.

%Combinatorics (TODO: complete)

\begin{align}
%\lambda \textit{VP} \lambda \textit{NP} \ \lambda z. d (\textit{NP} ((\textit{VP}\ \lambda y.y)\ z)) \\
\label{delibone} \lambda \textit{VP}  \lambda \textit{NP} \lambda z. & \textit{NP}\ (d\
((\textit{VP}\ \lambda y.y)\ z)): \\ & ((C\ra C)\ra D\ra 2) \ra (1\ra E)
\ra D \ra E \nonumber \\
%\lambda \textit{VP} \lambda \textit{NP} \lambda z. (\textit{VP}\ \lambda y.y)\ (d (\textit{NP} z)) \\
\label{delibtwo} \lambda \textit{VP} \lambda \textit{NP} \lambda
z. &\textit{NP}\ ((\textit{VP}\ \lambda y.d\ y)\ z): \\
 & ((2\ra 1)\ra D\ra C) \ra (C\ra E)
\ra D \ra E \nonumber  \\
\label{delibthree} \lambda \textit{VP} \lambda \textit{NP} \lambda z. & ((\textit{VP}\
\lambda y.\textit{NP}\ (d\ y))\ z): \\
& ((2\ra C)\ra D\ra E) \ra (1\ra C)
\ra D \ra E \nonumber \\
\label{uninhabited} \textit{Uninhabited:} \\
& ((2\ra c) \ra D \ra c)\ra (1\ra E) \ra D\ra E \nonumber
\end{align}

% Lambda term~\ref{delibtwo} requires the syntactic $np$ argument of
% $np\ra s$ to span exactly the positions of the word ``deliberately'',
% which is not possible. Lambda term~\ref{delibthree} on the other hand
% requires that this same $np$ arguments ends at the rightmost position
% of the word ``deliberately''. Neither of these possibilities is
% syntactically consistent. Another way to see this inconsistency is by
% looking at the principal types, which are shown for the adverb
% occurring between positions 1 and 2. Looking at the simple sentence
% ``Eduardo deliberately fell'', with ``Eduardo'' assigned $np$ and
% principal type $1\ra 0$  and fell assigned $np\multimap s$ and
% principal type $(2\ra A) \ra (3\ra A)$.

We investigate the three possibilities in turn.
 
Lambda term~\ref{delibone} comes closest to the first-order linear logic formula, but it
is a lambda term modeled after those used for extraction and, as such, it takes
a sentence missing a noun phrase \emph{anywhere} as its argument, instead of
a verb phrase. Therefore, it
incorrectly predicts that the three following sentences are all grammatical. 

\ex. \label{wronga} John deliberately Mary hit.

\ex. \label{wrongb} John deliberately Mary insinuates likes
Susan.

\ex. \label{wrongc} John deliberately Mary hit the sister of.

Predicting that sentence~\ref{wronga} means
``It was deliberate on the part of John that Mary hit him'', with sentence~\ref{wrongb} meaning approximately ``John made Mary insinuate that he likes Susan''
and sentence~\ref{wrongc} meaning something like ``Mary hit the
sister of John and this was deliberate on the part of John''. It seems
very difficult to block this example without also blocking the noun
``boy which Mary likes the sister of'' (not super-natural, but we want
to allow these kinds of extractions which are essentially
indistinguishable from the current formula).

Lambda term~\ref{delibtwo} shifts from the extraction-like lambda term
and its corresponding overgeneration to a lambda term similar to those
used for in situ binding/quantifying in\footnote{As we have seen, a
  generalized quantifier like ``everyone'' is assigned the lambda term
  $\lambda P.P(e)$ with $e$ being the string constant corresponding to
  the word ``everyone''.},
where we require a sentence missing a noun phrase at the position of
``deliberately'' as argument. Though this analysis again allows us to
derive the correct word order, it also makes
the dubious claim that there is an $np$ constituent at the position of the
adverb. In addition, it overgenerates as follows.

\ex. \label{wrongd} Mary John hit deliberately.

\ex. \label{wronge} Mary the friend of deliberately left.

\ex. \label{wrongf} Mary John gave the friend of deliberately a book.

Though it is possible to argue that sentence~\ref{wrongd} is a sort of
topicalization (with stress on \emph{Mary}), it is problematic that this topicalization is
triggered by the adverb, since topicalization is independent of the
presence or absence of adverbs. Moreover, we generate the semantics ``It was deliberate on the part of Mary that
John hit her'' for sentence~\ref{wrongd}. We generate the semantics ``It was deliberate on
the part of Mary that her friend left'' for sentence~\ref{wronge} and
similarly ``Mary incited John to give her friend a book'' for sentence~\ref{wrongf}.

Finally, lambda term~\ref{delibthree} selects for a sentence missing a
noun phrase with the only condition that this noun phrase occurs
directly before the adverb. Here, we make the odd claim that the noun
phrase and the adverb \emph{together} span the position of an $np$:
that is, it claims that an adverb is a post-modifier of an $np$. In
addition, it is again an in situ binding/quantifying in analysis, but this time with
the complex string ``$np$ deliberately'' (where lambda
term~\ref{delibtwo} used an in situ binding analysis of just the word ``deliberately'').

\ex. \label{wrongg} John hit Mary deliberately.

\ex. \label{wrongh} The friend of Mary deliberately left.

\ex. \label{wrongi} The friend of Mary deliberately who lives in Paris
left.

Though sentences~\ref{wrongg} and \ref{wrongh} are syntactically
correct, the problem is that we generate the semantics ``It was deliberate on the part of Mary that
John hit her'' for sentence~\ref{wrongg} and a reading ``It was
deliberate on the part of Mary that her friend left'' for
sentence~\ref{wrongh} and \ref{wrongi}.

In sum, we cannot capture the essence of the Lambek calculus
formula $(np\backslash s)/(np
\backslash s)$ in lambda grammars. Other adverb formulas --- $(np\backslash s)\backslash (np
\backslash s)$ (an adverb occurring after the verb phrase) and $(n/n)/(n
/ n)$ (for adverbs such as ``very''), etc.\ --- suffer from the same
problem. The best approximations 
%of these formulas
that we can obtain all suffer from overgeneration because
non-com\-mu\-ta\-ti\-vi\-ty is insufficiently enforced.

There is, of course, a solution which replaces the complex $np\backslash s$
argument by a new atomic formula, say $vp$ and then, for all
lexical items of the form $((np\backslash s)/A_n)\ldots /A_1$, adds an
additional formula $(vp/A_n)\ldots / A_1$. This would essentially double the
number of lexical formulas for verbs, adverbs and prepositions --- syntactic
categories which already have a high number of lexical formulas ---
for just a single type of problematic example... More such examples
will follow.
%resulting ultimately in an exponential explosion of the
%lexicon.

We
will discuss this potential solution in a bit more detail in
Section~\ref{solutions}, but it should already be clear that this is
not a particularly attractive option, since it is a prototypical
example of descriptive inadequacy, the reasons for doubling
the lexicon are purely theory-internal: no other categorial grammar,
not even AB grammars, have this type of overgeneration for the simple
cases we've shown.

\subsubsection*{Coordination}

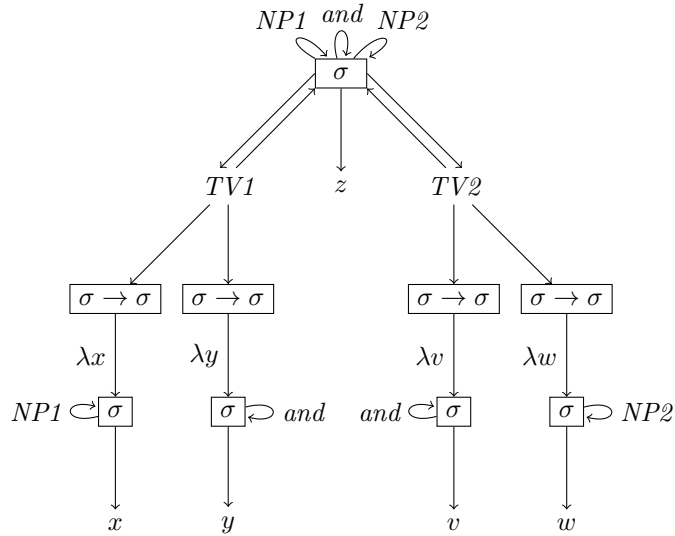
\begin{figure}
\begin{center}
\begin{tikzpicture}[node distance=1.5cm]
\node[draw] (SIGMA) {$\ \sigma\ $};
\path[->] (SIGMA) edge [loop above] node[above] {$\textit{and}$} ()
                             edge [in=30,out=50,loop] node[above] {$\ \ \ \ \textit{NP2}$} ()
                             edge [in=130,out=150,loop] node[above]
                             {$\textit{NP1}\ \ \ \ $}  ();
\node (Z) [below of=SIGMA] {$z$};
\node (TV1) [left of=Z] {$\textit{TV1}$};
\node (TV2) [right of=Z] {$\textit{TV2}$};
\draw[->] (SIGMA.west) -- ([xshift=-3pt] TV1.north);
\draw[->] ([xshift=3pt] TV1.north) --  (SIGMA.south west);
\draw[->] (SIGMA.east) -- ([xshift=3pt] TV2.north);
\draw[->] ([xshift=-3pt] TV2.north) -- (SIGMA.south east);
\draw[->] (SIGMA.south) -- (Z.north);
\node[draw] (OBJ1) [below of=TV1] {$\sigma\ra\sigma$};
\node[draw] (SUBJ1) [left of=OBJ1] {$\sigma\ra\sigma$};
\node[draw] (SUBJ11) [below of=SUBJ1] {$\sigma$};
\node[draw] (OBJ11) [below of=OBJ1] {$\sigma$};
\node (X) [below of=SUBJ11] {$x$};
\node (Y) [below of=OBJ11] {$y$};
\path[->] (SUBJ11) edge [loop left] node[left] {$\textit{NP1}$} ();
\path[->] (OBJ11) edge [loop right] node[right] {$\textit{and}$} ();
\path[->] (SUBJ1) edge node[left] {$\lambda x$} (SUBJ11);
\path[->] (OBJ1) edge node[left] {$\lambda y$} (OBJ11);
%\draw[->] (OBJ11) -- (TV2);
\draw[->] (TV1) -- (OBJ1);
\draw[->] (TV1) -- (SUBJ1);
\draw[->] (OBJ11) -- (Y);
\draw[->] (SUBJ11) -- (X);
\node[draw] (SUBJ2) [below of=TV2] {$\sigma\ra\sigma$};
\node[draw] (OBJ2) [right of=SUBJ2] {$\sigma\ra\sigma$};
\node[draw] (SUBJ22) [below of=SUBJ2] {$\sigma$};
\node[draw] (OBJ22) [below of=OBJ2] {$\sigma$};
\node (V) [below of=SUBJ22] {$v$};
\node (W) [below of=OBJ22] {$w$};
%\draw[->] (SUBJ22) -- (TV1);
\draw[->] (TV2) -- (OBJ2);
\draw[->] (TV2) -- (SUBJ2);
\draw[->] (OBJ22) -- (W);
\draw[->] (SUBJ22) -- (V);
\path[->] (SUBJ2) edge node[left] {$\lambda v$} (SUBJ22);
\path[->] (OBJ2) edge node[left] {$\lambda w$} (OBJ22);
\path[->] (SUBJ22) edge [loop left] node[left] {$\textit{and}$} ();
\path[->] (OBJ22) edge [loop right] node[right] {$\textit{NP2}$} ();
\end{tikzpicture}
\end{center}
\caption{Simplified inhabitation machine for transitive verb
  conjunction}
\label{fig:inhabtv}
\end{figure}

As noted by Kubota \& Levine \citeyear{kl13coord,kl13emp}, we can play a similar game for ``John
caught and ate a fish'', which looks as shown in Figure~\ref{fig:inhabtv}; for the sake of
space, we do not show the prefix $\lambda \textit{TV2}. \lambda \textit{TV1}. \lambda \textit{NP2}.
\lambda \textit{NP1}. \lambda z$, where $\textit{TV2}$ is the
transitive verb to the right of ``and'' (``ate'' in the current
example), $\textit{TV1}$ is the transitive verb to the left of it
(``caught''), $\textit{NP1}$ is the subject, $\textit{NP2}$ is the
object and $z$ is the end of the complete string.

Remark that ``and'' takes all constituents as argument: the two
transitive verbs, the subject noun phrase and the object noun phrase,
so it would seem that we \emph{should} be able to generate the right string.

As before, we have split the $\sigma$ and $\sigma\ra\sigma$ nodes for
readability; the actual graph merges all $\sigma$ and all
$\sigma\ra\sigma$ nodes. The implausible analyses with $\textit{TV1}$
and subject of $\textit{TV2}$ and with $\textit{TV2}$ as object of
$\textit{TV1}$ are not shown in the figure, but they fail for the same
reasons discussed below.

The graph of Figure~\ref{fig:inhabtv} shows that the $\textit{TV1}$ node takes first its subjet (down and to the left
of it), then its object (directly below) and finally an argument of
type $\sigma$ (the upward arrow back to $\sigma$) and similarly for
$\textit{TV2}$. The $\textit{TV1}$ node (optionally) takes
$\textit{NP1}$ as its subjet and $\textit{TV2}$ (optionally) takes
$\textit{NP2}$ as its object.

Two combinations are fairly limited: the second argument of
$\textit{TV1}$ is either $\textit{NP1}$ or the empty string and the
first argument of $\textit{TV2}$ is either $\textit{NP2}$ or the empty
string. However, if the lexical entry contains the subterm
$((\textit{TV1}\, M)\, \textit{NP1})$ (for some $M$ at the place of
the object), then we are
essentially using a quantifying-in analysis for the subject: $(\textit{TV1}\, M)$ is a
sentence missing a noun phrase \emph{anywhere} and applying this term
to an argument puts this argument back at the place of the missing
noun phrase. Consequently, it would allow the derivation of ``caught
John and ate a fish''. Similar overgeneration occurs for ``ate'' and
``a fish'' if we use the quantifying-in combination $(\textit{TV2}\,
\textit{NP2})$ for the object.

If we want to avoid both types of overgeneration (subject
quantifying-in and object quantifying in), 
the only remaining analysis
%\editout{\footnote{Some (implausible) alternative analyses exist: for example, the
% object of $\textit{TV1}$ can have either ``and'' or $\textit{TV2}$ as
% head term (similarly, the subject of $\textit{TV2}$ can have either
% ``and'' or $\textit{TV1}$ as head term). All these alternatives
% suffer from overgeneration, as we have seen with lambda
% term~\ref{delibtwo} on page~\pageref{delibtwo}.}}
consists of choosing $\lambda x.x$,
$\lambda y.y$, $\lambda v.v$ and $\lambda w.w$  as arguments for the
two transitive verbs.\footnote{This solution still overgenerates
  because it equates transitive verb with ``sentence missing two noun
  phrases'' and therefore incorrectly predicts that ``John likes $[]_{np}$ 's
  friend from $[]_{np}$'' can felicitously fill this role as follows.

\ex. Mary went to and John likes 's friend from Paris.

Meaning ``Mary went to Paris and John likes Mary's friend from there''.}
 This solution is shown in full below.
\begin{align*}
\lambda \textit{TV2}. &\lambda \textit{TV1}.  \lambda \textit{NP2}.
\lambda \textit{NP1}. \lambda z. \\ &\textit{NP1}\, ((\textit{TV1}\, \lambda
x. x\, \lambda y.y) (\textit{and}\, ((\textit{TV2}\, \lambda
v. v\, \lambda w.w) (\textit{NP2}\, z))))
\end{align*}

% TODO: transform into HTLG proof
\editout{
$$
\infer{s}{\infer{np}{\textit{John}} &
\infer{np\ra s}{
\infer{np\ra np\ra s}{
               \infer{np\ra np\ra s}{\textit{caught}} &
               \infer{a2}{\infer{a1}{\textit{and}} & \infer{np\ra
                   np\ra s}{\textit{ate}}}}
& \infer{np}{\textit{a fish}}}}
$$
}

\begin{figure}
\begin{center}
\begin{sideways}
$$
\infer{\overset{((((\textit{and}\,
a)\, c)\, f)\, j)^{5\ra 0}}{s}}{\infer{np^{1\ra 0}}{\textit{John}} &
\infer{\overset{(((\textit{and}\,
a)\, c)\, f)^{(1\ra
                     K)\ra 5\ra K}}{s|np}}{
\infer{\overset{((\textit{and}\,
a)\, c)^{(L\ra 4)\ra (1\ra
                     K)\ra L\ra K}}{(s|np)|np}}{
               \infer{\overset{c^{(B\ra 2)\ra (1\ra A)\ra B\ra A}}{(s|np)|np}}{\textit{caught}} &
               \infer{\overset{(\textit{and}\, a)^{((E\ra E)\ra(F\ra F)\ra
                     2\ra I)\ra(L\ra
                     4)\ra(I\ra K)\ra L\ra K}}{((s|np)|np)\,|\,((s|np)|np)}}{\infer{\overset{\textit{and}^{((G\ra G)\ra(H\ra H)\ra J\ra 3)\ra((E\ra E)\ra(F\ra F)\ra
                     2\ra I)\ra(L\ra
                     J)\ra(I\ra K)\ra L\ra K}}{(((s|np)|np)\,|\,((s|np)|np))\,|\,((s|np)|np)}}{\textit{and}}
                 & \infer{\overset{a^{(D\ra 4)\ra(3\ra C)\ra D\ra C}}{(s|np)|np}}{\textit{ate}}}}
& \infer{\overset{f^{5\ra 4}}{np}}{\textit{a fish}}}}
$$
\end{sideways}
\end{center}
\caption{Proof of ``John caught and ate a fish'' (simplified)}
\label{fig:fish}
\end{figure}

As we can see from the proof in Figure~\ref{fig:fish}, this lexical type allows us to derive
``John caught and ate a fish'' with the correct semantics. The proof
has been slightly simplified by using distinct variables for the words
instead of complex lambda terms (ie.\ we have not done lexical
substitution). This has the advantage that we can use the resulting
lambda term for computing the semantics as well, for which we use the
following (standard) semantic substitutions\footnote{To keep this example simple, we have
  treated ``a fish" as an individual constant instead of a quantified
  noun phrase, since quantification is irrelevant for this
  example.}. We can obtain the prosodic lambda terms from the
principal types and the string positions (eg.\ $(2\ra 1) \vdash (B\ra 2)
\ra (1\ra A) \ra B\ra A$ for ``caught'', which is the standard
transitive verb principal type we have seen before). %We have also
%performed the substitutions of the $\ra E$ rule directly. 
The semantic terms below are all standard.
\begin{align*}
\textit{and} &= \lambda \textit{TV1} \lambda \textit{TV2}\lambda
y \lambda x.  ((\textit {TV1}\, y)\, x) \wedge ((\textit{TV2}\, y)\, x) \\
j &= \textit{john'} \\
f &= \textit{a\_fish'} \\
c &= \textit{caught'} \\
a &= \textit{ate'} 
\end{align*}

Unfortunately, this analysis of ``and'' also make the
(rather catastrophic) prediction that ``John caught and ate a fish'' has a
second reading which can be paraphrased as ``John caught a fish and a
fish ate John''. This reading is easy to miss when we look only at
eta-short proofs, since the key point of this second derivation
involves switching the two arguments of the transitive verb, as
shown in Figure~\ref{fig:fishb}.\footnote{The term $\lambda f. \lambda x.\lambda y. ((f\, y)\,
  x)$ which switches subject and object is of course the \textbf{C} combinator we have already seen in
  Section~\ref{sec:exc}. It commutes the two
  arguments of a function $f$, and the proof shown in
  Figure~\ref{fig:fishb} has a subproof which computes
  $\textbf{C}a\equiv \lambda x.\lambda y. (a\, y)\,x$.} The crux
of this second proof is that swapping the two arguments of ``ate'' is
a purely local operation which has no visible effects on the word
order: the only difference between the proof in Figure~\ref{fig:fish}
and the proof in
Figure~\ref{fig:fishb} is in the subproof with undischarged hypothesis
``ate'' (with term $a$ resp.\ $\lambda x. \lambda y. (a\, y)\,x$).
\editout{
$$
\infer[\multimap I_1]{\lambda x. \lambda y. ((a\, y)\, x) :np \multimap np \multimap s}{\infer[\multimap
  I_2]{\lambda y. ((a\, y)\, x): np\multimap s}{\infer[\multimap E]{((a\, y)\, x):s}{[x:np]^1 & \infer[\multimap
      E]{(a\, y):np \multimap s}{[y:np]^2 & a:np \multimap np \multimap s }}}}
$$

eta-expanded

$$
((\textit{and}\,
\lambda y. \lambda x. ((a\, y)\, x)\, c)\, f)\, j
$$
}
\editout{
$$
\infer{s}{\infer{np}{\textit{John}} &
\infer{np\ra s}{
\infer{np\ra np\ra s}{
               \infer{np\ra np\ra s}{\textit{caught}} &
               \infer{a2}{\infer{a1}{\textit{and}} & \infer[I_2]{np\ra
                   np\ra s}{\infer[I_1]{np\ra s}{\infer{s}{\infer{[np]^1}{x} &\infer{np\ra
                         s}{\infer{np\ra
                   np\ra s}{\textit{ate}} & \infer{[np]^2}{y}}}}}}}
& \infer{np}{\textit{a fish}}}}
$$

$$
\infer{s}{\infer{np}{\textit{John}} &
\infer{np\ra s}{
\infer{np\ra np\ra s}{
               \infer{np\ra np\ra s}{\textit{caught}} &
               \infer{a2}{\infer{a1}{\textit{and}} & \infer[I_1]{np\ra
                   np\ra s}{\infer[I_2]{np\ra s}{\infer{s}{\infer{[np]^1}{x} &\infer{np\ra
                         s}{\infer{np\ra
                   np\ra s}{\textit{ate}} & \infer{[np]^2}{y}}}}}}}
& \infer{np}{\textit{a fish}}}}
$$
}

\begin{figure}
\begin{center}
\begin{sideways}
$$
\infer{\overset{((((\textit{and}\,
a)\, c)\, f)\, j)^{5\ra 0}}{s}}{\infer{np^{1\ra 0}}{\textit{John}} &
\infer{\overset{(((\textit{and}\,
a)\, c)\, f)^{(1\ra
                     K)\ra 5\ra K}}{s|np}}{
\infer{\overset{((\textit{and}\,
a)\, c)^{(L\ra 4)\ra (1\ra
                     K)\ra L\ra K}}{(s|np)|np}}{
               \infer{\overset{c^{(B\ra 2)\ra (1\ra A)\ra B\ra A}}{(s|np)|np}}{\textit{caught}} &
               \infer{\overset{(\textit{and}\, a)^{((E\ra E)\ra(F\ra F)\ra
                     2\ra I)\ra(L\ra
                     4)\ra(I\ra K)\ra L\ra K}}{((s|np)|np)\,|\,((s|np)|np)}}{\infer{\overset{\textit{and}^{((G\ra G)\ra(H\ra H)\ra J\ra 3)\ra((E\ra E)\ra(F\ra F)\ra
                     2\ra I)\ra(L\ra
                     J)\ra(I\ra K)\ra L\ra K}}{(((s|np)|np)\,|\,((s|np)|np))\,|\,((s|np)|np)}}{\textit{and}}
                 & \infer{\overset{(\lambda x. \lambda y. ((a\,y)\,x))^{(3\ra C)\ra(D\ra
                       4)\ra D\ra C}}{(s|np)|np}}{\infer{\overset{(\lambda y. ((a\,y)\,x))^{(D\ra
                       4)\ra D\ra C}}{s|np}}{\infer{\overset{((a\,y)\,x)^{D\ra C}}{s}}{\infer{\overset{(a\,y)^{(3\ra C)\ra D\ra C}}{s|np}}{\infer{\overset{a^{(D\ra 4)\ra(3\ra C)\ra D\ra
                       C}}{(s|np)|np}}{\textit{ate}} &
                 \overset{y^{D\ra 4}}{np} } & \overset{x^{3\ra C}}{np}}}}
}}
& \infer{\overset{f^{5\ra 4}}{np}}{\textit{a fish}}}}
$$
\end{sideways}
\end{center}
\caption{Proof of ``John caught and ate a fish'' with semantics ``John
  caught a fish and a fish ate John'' (simplified).}
\label{fig:fishb}
\end{figure}

As shown in the figure, the second proof computes the following ``deep structure''.
$$(((\textit{and}\,
(\textbf{C} a))\, c)\, f)\, j = (((\textit{and}\,
\lambda x. \lambda y. ((a\, y)\, x)\, c)\, f)\, j$$

In a similar way, we can obtain a third and a fourth reading,
corresponding the string ``John caught and ate a fish'' but to the meanings ``A fish caught John and John ate a
fish'' and ``A fish caught and ate John'' respectively, as follows.
\begin{align*}
(((\textit{and}\,
a)\, (\textbf{C} c))\, f)\, j &= (((\textit{and}\, a)\, \lambda v.\lambda w. (c\, w)\,
v)\, f)\, j\\
 (((\textit{and}\,
(\textbf{C} a))\, (\textbf{C} c))\, f)\, j &= (((\textit{and}\,
\lambda x. \lambda y. ((a\, y)\, x)\, \lambda v.\lambda w. (c\, w)\,
v)\, f)\, j
\end{align*}

The problem is that though we would \emph{want} the two $(s|np)|np$
arguments of ``and'' to be transitive verbs, they \emph{mean} ``a
sentence missing two $\textit{np}$ arguments anywhere'', which is what
causes the problems with commutativity.

We can again remedy this by adding new lexical entries, for example
choosing $tv$ for the two transitive verbs and $(tv\backslash ((np\backslash
s)/np))/tv$ for the conjunction, but this would mean adding several
other lexical entries to analyse sentences like ``John has understood and
will probably implement Dijkstra's algorithm'', which are handled by
the Lambek calculus analysis --- since ``has understood'' and ``will
probably implement'' can both be analysed as $(np\backslash s)/np$ ---
but not by the new atomic $tv$ analysis. So adding lexical entries is
not only inelegant and an admittance of descriptive inadequacy, but
such additions can cascade throughout the grammar.

I would seem that another simple potential solution would be to add
case to lambda grammars. While adding case to
first-order linear logic is something we can do essentially for free
using extra arguments, adding case to lambda grammars at least
complicates either the grammars or the types. In addition, though case
 would exclude the subject-object
swaps we have seen in this section, is is easy to see this would not be a real
solution, because the sentences in \ref{subjex} below are all sentences
missing a subject/nominative $np$, those in \ref{objex} sentences
missing an object/accusative $np$
and those in \ref{tvex} sentences missing both a subject and an
object (for clarity, the missing subjects and objects have been shown
as $[]_s$ and $[]_o$ respectively). So while adding case excludes some bad derivations, we would still
predict sentences like ``*Sue likes Mary and John saw the man whom
likes'' is grammatical (with meaning ``Sue likes Mary and John saw the man whom Sue
likes.''), that `` *John saw the friend of who lives in Paris and
Ted likes Sue'' is grammatical and means ``John saw the friend of Sue who lives in Paris
and Ted likes Sue'' and that ``*Sue John believes avoids but Ted saw
whom kissed Peter'' is grammatical and means ``John believes Sue avoids
Peter but Ted saw Peter whom Sue kissed''.

\ex.\label{subjex}  
\a. $[]_{s}$ likes Mary.
\b. John believes $[]_s$ left.
\b. John saw the man whom $[]_s$ likes.

\ex.\label{objex} \a. John likes $[]_o$.
\b.  John saw the friend of $[]_o$ who lives in Paris.
\b. Ted gave $[]_o$ flowers.

\ex.\label{tvex} \a. $[]_s$ gave $[]_o$ flowers.
\b. John believes $[]_s$ avoids $[]_o$.
\b. John saw $[]_o$ whom $[]_s$ kissed.

While it would certainly be possible to appeal to island constraints
or other independently motivated mechanisms to exclude coordination of
the phrases listed above, it seems that use case for this purpose is inherently on
the wrong track: it uses lambda terms to encode word order for
negative implications and a cascade of stop-gap solutions to
constrain word order for positive implications. As the examples above
make clear, for coordination, we
don't coordinate (partial) constituents which have the same case
marking, but rather those which have the same \emph{structure} and
Lambek calculus formulas, such as $(np\bs s)/np$ for transitive verb conjunction, are a good proxy for this notion of the same structure.

Though we have given an in-depth analysis only of transitive verb
conjunction, other conjunctions of complex types (adjectives,
intransitive verbs, etc.) suffer similar problems. In all cases, the
lambda grammar analysis is between a rock and a hard place, suffering
either from overgeneration (as the adverb case) or from bizarre
readings (as in the transitive verb conjunction case).

\subsubsection*{Gapping}
As a last problem case, the standard (multimodal) categorial grammar analysis of gapping
\cite{cgellipsis}, of which we have seen the hybrid version in
Section~\ref{sec:string}, does not fare any better when we try to
translate it into lambda grammar. The analysis of a sentence like

\ex.  John studies logic and Charles phonetics.

\noindent would assign ``and'' the formula.\editout{
$$((np\multimap np \multimap s)
\multimap s) \multimap ((np\multimap np \multimap s)
\multimap s) \multimap (np\multimap np \multimap s)
\multimap s$$
}
$$
((s|((s|np)|np))\,|\,(s|((s|np)|np)))\,|\,(s|((s|np)|np))
$$

The idea behind the analysis of \citeasnoun{cgellipsis} is that
``and'' takes first two sentences missing a transitive verb as its
arguments, then a transitive verb to produce a sentence by placing the
transitive verb back to its normal place in the first argument (which
is the sentences to its left missing a transitive verb) and using the
empty string instead of the transitive verb in the second sentence. In
short, it uses a quantifying-in analysis for the transitive verb in
the sentence to the left and an extraction analysis for the
``missing'' transitive  in the sentence to the right. What is nice
about this analysis, is that we use a normal coordination formula
for ``and'', an instance of the schema $(X|X)|X$ with (in this case)
$X = (s|np)|np$. Since the analysis of \citeasnoun{cgellipsis} uses a combination of quantifying in and
extraction, it is tempting to think that the lambda grammar analysis
is unproblematic. However, they key point of the analysis is that we
need both extraction and in situ binding for a complex formula, a
transitive verb, though unlike for the coordination case it occurs in
a negative position in the gapping coordination type. Let's
investigate the possibilities.

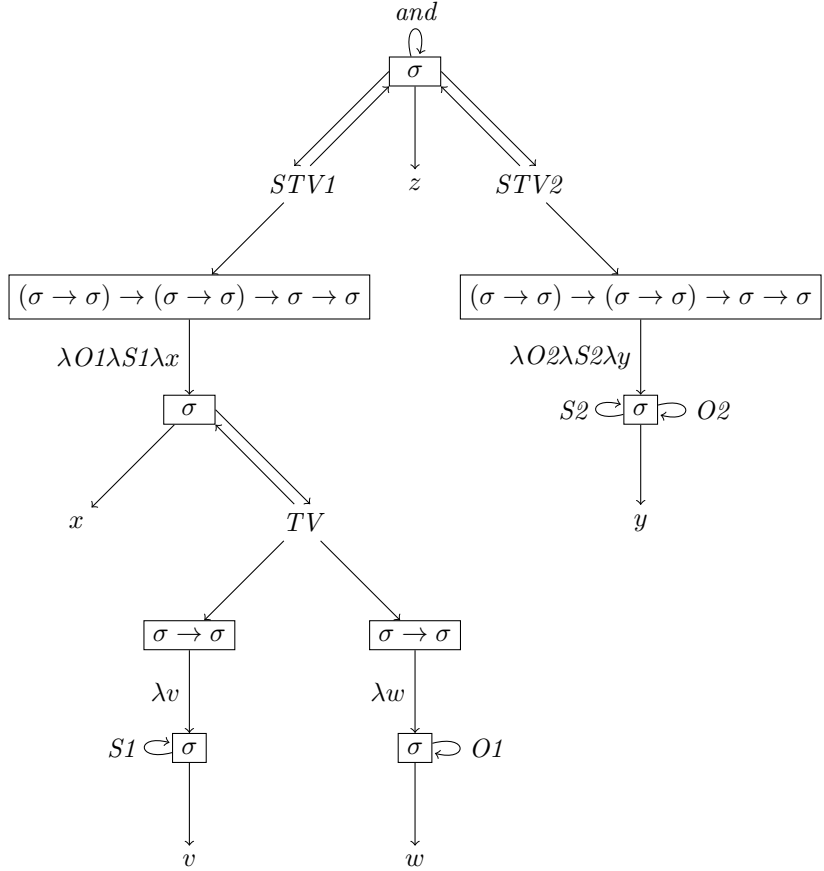
\begin{figure}
\begin{center}
\begin{tikzpicture}[node distance=1.5cm]
\node[draw] (SIGMA) {$\ \sigma\ $};
\path[->] (SIGMA) edge [loop above] node[above] {$\textit{and}$} ();
\node (Z) [below of=SIGMA] {$z$};
\node (STV1) [left of=Z] {$\textit{STV1}$};
\node (dmy1) [below of=STV1] {};
\node[draw] (SSS) [left of=dmy1] {$(\sigma\ra\sigma)\ra(\sigma\ra\sigma)\ra\sigma\ra\sigma$};
\node[draw] (ZZZ) [below of=SSS] {$\ \sigma\ $};
\node (dmy3) [below of=ZZZ] {};
\node (XXX) [left of=dmy3] {$x$};
\node (TV1) [right of=dmy3] {$\textit{TV}$};
\node (TV2) [right of=Z] {$\textit{STV2}$};
\path[->] (SSS) edge node[left] {$\lambda \textit{O1} \lambda \textit{S1} \lambda x$} (ZZZ);
%\path[->] (ZZZ) edge %%%%%
\draw[->] (SIGMA.west) -- ([xshift=-3pt] STV1.north);
\draw[->] ([xshift=3pt] STV1.north) --  (SIGMA.south west);
\draw[->] (SIGMA.east) -- ([xshift=3pt] TV2.north);
\draw[->] ([xshift=-3pt] TV2.north) -- (SIGMA.south east);
\draw[->] (SIGMA.south) -- (Z.north);
\node (dmy4) [below of=TV1] {};
\node[draw] (OBJ1) [right of=dmy4] {$\sigma\ra\sigma$};
\node[draw] (SUBJ1) [left of=dmy4] {$\sigma\ra\sigma$};
\node[draw] (SUBJ11) [below of=SUBJ1] {$\sigma$};
\node[draw] (OBJ11) [below of=OBJ1] {$\sigma$};
\node (X) [below of=SUBJ11] {$v$};
\node (Y) [below of=OBJ11] {$w$};
% S1 and O1 are not really possible here
%\path[->] (ZZZ) edge [loop left] node[left] {$\textit{S1}$} ();
%\path[->] (ZZZ) edge [loop right] node[right] {$\textit{O1}$} ();
%
\draw[->] (ZZZ.east) -- ([xshift=3pt] TV1.north);
\draw[->] ([xshift=-3pt] TV1.north) -- (ZZZ.south east);
\path[->] (SUBJ11) edge [loop left] node[left] {$\textit{S1}$} ();
\path[->] (OBJ11) edge [loop right] node[right] {$\textit{O1}$} ();
%\path[->] (OBJ11) edge [loop right] node[right] {$\textit{and}$} ();
\path[->] (SUBJ1) edge node[left] {$\lambda v$} (SUBJ11);
\path[->] (OBJ1) edge node[left] {$\lambda w$} (OBJ11);
%\draw[->] (OBJ11) -- (TV2);
\draw[->] (STV1) -- (SSS);
\draw[->] (ZZZ) --(XXX);
%\draw[->] (ZZZ) -- (TV1);
%
\draw[->] (TV1) -- (OBJ1);
\draw[->] (TV1) -- (SUBJ1);
\draw[->] (OBJ11) -- (Y);
\draw[->] (SUBJ11) -- (X);
\node (dmy2) [below of=TV2] {};
\node[draw] (SUBJ2) [right of=dmy2] {$(\sigma\ra\sigma)\ra(\sigma\ra\sigma)\ra\sigma\ra\sigma$};
%\node[draw] (OBJ2) [right of=SUBJ2] {$\sigma\ra\sigma$};
\node[draw] (SUBJ22) [below of=SUBJ2] {$\sigma$};
%\node[draw] (OBJ22) [below of=OBJ2] {$\sigma$};
\node (V) [below of=SUBJ22] {$y$};
%\node (W) [below of=OBJ22] {$w$};
%\draw[->] (SUBJ22) -- (TV1);
%\draw[->] (TV2) -- (OBJ2);
\draw[->] (TV2) -- (SUBJ2);
%\draw[->] (OBJ22) -- (W);
\draw[->] (SUBJ22) -- (V);
\path[->] (SUBJ2) edge node[left] {$\lambda \textit{O2} \lambda \textit{S2} \lambda y$} (SUBJ22);
%\path[->] (OBJ2) edge node[left] {$\lambda w$} (OBJ22);
\path[->] (SUBJ22) edge [loop left] node[left] {$\textit{S2}$} ();
\path[->] (SUBJ22) edge [loop right] node[right] {$\textit{O2}$} ();
\end{tikzpicture}
\end{center}
\caption{Simplified inhabitation machine for gapping.}
\label{fig:inhabg}
\end{figure}

The formula for transitive verb gapping produces the (simplified and reduced) inhabitation machine
shown in Figure~\ref{fig:inhabg}.

As before, the prefix $\lambda \textit{STV2}. \lambda \textit{STV1}.
\lambda \textit{TV}. \lambda z.$ has been remove from the figure;
$\textit{STV1}$ denotes the sentence missing a transitive verb to the
left of ``and'' (``John logic'' in our case) and $\textit{STV2}$
denotes the sentence missing a transitive verb to the right of ``and''
(``Charles phonetics'' in our case), $\textit{TV}$ the transitive verb
(here: ``studies'') and $z$ the end of the string.

We can obtain the full combinatorics by identifying all nodes with the
same type. The current reduced graph emphasizes the reasonable
lambda terms: for example, $\textit{TV}$ can only be an argument of
$\textit{STV1}$, corresponding to the the quantifying-in analysis
producing the desired word order ``John
studies logic'', similarly, the first argument of the transitive verb
has been restricted to the subject and the second argument to the object.
In fact, getting the word order and semantics right leaves a unique
lambda term --- this is just the term from \cite{bourreau13ellipse},
where the subterm $(\textit{STV1}\, \textit{TV})$ has been eta-expanded.\footnote{The eta-short term looks as follows.
\begin{align*}
\lambda \textit{STV2}. \lambda \textit{STV1}.
\lambda \textit{TV}. \lambda z. (&(\textit{STV1}\,\textit{TV})\\
& (\textit{and}\ (\textit{STV2}\, \lambda \textit{O2} \lambda \textit{S2} \lambda
y. \textit{S2}\, (\textit{O2}\, y))\, z))
\end{align*}}

\begin{align*}
\lambda \textit{STV2}. \lambda \textit{STV1}.
\lambda \textit{TV}. \lambda z. (&(\textit{STV1}\,\lambda \textit{O1}
\lambda \textit{S1} \lambda x.(((\textit{TV}\, \lambda w.\textit{O1}\,
w)\, \lambda v.\textit{S1}\, v))\, x)\\
& (\textit{and}\ (\textit{STV2}\, \lambda \textit{O2} \lambda \textit{S2} \lambda
y. \textit{S2}\, (\textit{O2}\, y))\, z))
\end{align*}

The principal type of this term is.
\begin{align*}
(((L\ra K) \ra (K \ra M) &\ra L \ra M) \ra J \ra 4) \ra \\
(((D\ra C) \ra (F\ra E) &\ra H \ra G) \ra 3 \ra I) \ra \\
((D \ra C) \ra (F \ra E) &\ra H \ra G) \ra J \ra I
\end{align*}
Given this lambda term and principal type, the we can derive the
correct word order and semantics as shown in Figure~\ref{fig:jsl}. 
We have again abbreviated the proof, using the following abbreviations
for readability.
\begin{align*}
X &= s|((s|np)|np) \\
\alpha &= ((L\ra K)\ra (K\ra M) \ra L\ra M)\ra J\ra 4 \\
\beta &= (D\ra C)\ra (F\ra E)\ra H\ra G\\
%\beta &= ((D\ra C)\ra (F\ra E) \ra H\ra G)\ra 3\ra I \\
%\gamma &= ((D\ra C) \ra (F\ra E)\ra H\ra G)\ra J\ra I
\end{align*}
We have also performed the substitutions necessary for the $\ra E$
rules directly on the hypotheses of the proof. We obtain the semantics by
substituting the following terms for the constants in the lambda term
computed for this proof.
\begin{align*}
\textit{and} &= \lambda \textit{STV1} \lambda \textit{STV2} \lambda
\textit{TV}. (\textit {STV1} \textit{TV}) \wedge (\textit{STV2}
  \textit{TV}) \\
j &= \textit{john'} \\
l &= \textit{logic'} \\
c &= \textit{charles'} \\
p &= \textit{phonetics'} \\
s & = \textit{studies'} 
\end{align*}
But again, there is an alternative
proof, shown in Figure~\ref{fig:jslb}. This proof swaps both the arguments of $P$ and the
two abstractions of $s$ (``studies''). The net result is that the left
conjunct stays as before, syntactically and semantically, since the
two swaps cancel out against each other. Now the
right conjunct has its arguments swapped in the semantics only, giving the absurd reading ``John
studies logic and phonetics studies Charles''.
\newcommand{\bigforma}{\overset{\textit{and}^{\alpha\ra (\beta\ra 3\ra
      I)\ra \beta\ra J\ra I}}{(X|X)|X}}
\newcommand{\bigformb}{\overset{(\textit{and}\, (\lambda
    Q. (Q\,p)\,c))^{(\beta \ra 3\ra I) \ra \beta \ra 6\ra I}}{X|X}}
\editout{
\begin{align*}
b1 &= ((np \ra np\ra s)\ra s) \ra ((np \ra np\ra s)\ra s) \ra (np \ra
np \ra s) \ra s \\
b2 &= ((np \ra np\ra s)\ra s) \ra (np \ra
np \ra s) \ra s\\
\end{align*}
$$
\infer{s}{
\infer{(np \ra np\ra s)\ra s}{
\infer[I_1]{(np \ra np \ra s)\ra s}{\infer{s}{\infer{np}{\textit{John}} &
    \infer{np\ra s}{\infer{[np \ra np\ra s]^1}{P} &
      \infer{np}{\textit{logic}}}}} &
\infer{\bigformb}{
\infer{\bigforma}{\textit{and}} &
\infer[I_2]{(np \ra np \ra s)\ra s}{\infer{s}{\infer{np}{\textit{Charles}} &
    \infer{np\ra s}{\infer{[np \ra np\ra s]^2}{Q} &
      \infer{np}{\textit{phonetics}}}}}}} &
\infer{np\ra np\ra s}{\textit{studies}}}
$$

$$
\infer{s}{
\infer{(np \ra np\ra s)\ra s}{
\infer[I_1]{(np \ra np \ra s)\ra s}{\infer{s}{\infer{np}{\textit{John}} &
    \infer{np\ra s}{\infer{[np \ra np\ra s]^1}{P} &
      \infer{np}{\textit{logic}}}}} &
\infer{\bigformb}{
\infer{\bigforma}{\textit{and}} &
\infer[I_2]{(np \ra np \ra s)\ra s}{\infer{s}{\infer{np}{\textit{Charles}} &
    \infer{np\ra s}{\infer{[np \ra np\ra s]^2}{Q} &
      \infer{np}{\textit{phonetics}}}}}}} &
\infer{np\ra np\ra s}{\textit{studies}}}
$$
}
\begin{figure}
\begin{center}
\begin{sideways}
$$
\infer{\overset{(((\textit{and}\, (\lambda
    Q. (Q\,p)\,c))\, (\lambda P. (P\,l)\,j))\, (\lambda yx. (s\,y)\,x))^{6\ra 0}}{\vphantom{|}s}}{
\infer{\overset{((\textit{and}\, (\lambda
    Q. (Q\,p)\,c))\, (\lambda P. (P\,l)\,j))^{((3\ra 2)\ra(1\ra 0) \ra
      3\ra I)\ra 6\ra I}}{s|((s|np)|np)}}{
\infer{\overset{(\lambda P. (P\,l)\,j)^{((3\ra 2)\ra(1\ra 0)\ra
      B'\ra A')\ra B'\ra
      A'}}{s|((s|np)|np)}}{\!\!\!\!\!\!\infer{\overset{((P\,l)\,j)^{B'\ra
        A'}}{\vphantom{|}s}\!\!\!\!\!\!}{\infer{\overset{j^{\vphantom{(P)^{1\ra
              0}}1\ra 0}}{\vphantom{|}np}}{\textit{John}} &
    \infer{\overset{(P\,l)^{(1\ra 0)\ra B'\ra
          A'}}{s|np}}{\overset{P^{(3\ra 2)\ra (1\ra 0)\ra B'\ra A'}}{(s|np)|np} &
      \infer{\overset{l^{3\ra 2}}{\vphantom{|}np}}{\textit{logic}}}}} &
\infer{\bigformb}{
\infer{\bigforma}{\textit{and}} &
\infer{\overset{(\lambda Q. (Q\,p)\,c)^{((6\ra 5)\ra(5\ra 4)\ra
      D'\ra C')\ra D'\ra C'}}{s|((s|np)|np)}}{\infer{\overset{((Q\,p)\,c)^{D'\ra C'}}{\vphantom{|}s}}{\infer{\overset{c^{5\ra 4}}{\vphantom{|}np}}{\textit{Charles}} &
    \infer{\overset{(Q\, p)^{(5\ra 4)\ra D'\ra C'}}{s|np}}{\overset{Q^{(6\ra 5)\ra (5\ra 4)\ra D'\ra C'}}{(s|np)|np} &
      \infer{\overset{p^{6\ra 5}}{np}}{\textit{phonetics}}}}}}} &
\infer{\overset{(\lambda yx. (s\,y)\,x)^{(B\ra
      2)\ra(1\ra A)\ra B\ra A}}{(s|np)|np}}{\infer{\overset{(\lambda
      x.(s\,y)\,x)^{(1\ra A)\ra B\ra A}}{s|np}}{\infer{\overset{((s\,y)\, x)^{B\ra A}}{\vphantom{|}s}}{\overset{x^{1\ra A}}{\vphantom{|}np} & \infer{\overset{(s\, y)^{(1\ra
            A)\ra B\ra A}}{s|np}}{\infer{\overset{s^{(B\ra 2)\ra (1\ra
              A)\ra B\ra A}}{(s|np)|np}}{\textit{studies}} &
        \overset{y^{B\ra 2}}{\vphantom{|}np} }}}}} 
$$
\end{sideways}
\end{center}
\caption{Proof of "John studies logic and Charles phonetics".}
\label{fig:jsl}
\end{figure}

\begin{figure}
\begin{center}
\begin{sideways}
$$
\infer{\overset{(((\textit{and}\, (\lambda
    Q. (Q\,p)\,c))\, (\lambda P. (P\,j)\,l))\, (\lambda xy. (s\,y)\,x))^{6\ra 0}}{s}}{
\infer{\overset{((\textit{and}\, (\lambda
    Q. (Q\,p)\,c))\, (\lambda P. (P\,j)\,l))^{((1\ra 0)\ra(3\ra 2) \ra
      3\ra I)\ra 6\ra I}}{s|((s|np)|np)}}{
\infer{\overset{(\lambda P. (P\,j)\,l)^{((1\ra 0)\ra(3\ra 2)\ra
      B'\ra A')\ra B'\ra
      A'}}{s|((s|np)|np)}}{
         \infer{\overset{((P\,j)\,l)^{B'\ra A'}}{s}}{
               \infer{\overset{(P\,j)^{(3\ra 2)\ra B'\ra A'}}{s|np}}{
                    \infer{\overset{j^{1\ra 0}}{np}}{\textit{John}} &
                    \overset{P^{(1\ra 0)\ra (3\ra 2)\ra B'\ra A'}}{(s|np)|np}}
          &  \infer{\overset{l^{3\ra 2}}{np}}{\textit{logic}}
               }
        } &
\infer{\bigformb}{
\infer{\bigforma}{\textit{and}} &
\infer{\overset{(\lambda Q. (Q\,p)\,c)^{((6\ra 5)\ra(5\ra 4)\ra
      D'\ra C')\ra D'\ra C'}}{s|((s|np)|np)}}{\infer{\overset{((Q\,p)\,c)^{D'\ra C'}}{s}}{\infer{\overset{c^{5\ra 4}}{np}}{\textit{Charles}} &
    \infer{\overset{(Q\, p)^{(5\ra 4)\ra D'\ra C'}}{s|np}}{\overset{Q^{(6\ra 5)\ra (5\ra 4)\ra D'\ra C'}}{(s|np)|np} &
      \infer{\overset{p^{6\ra 5}}{np}}{\textit{phonetics}}}}}}} &
\infer{\overset{(\lambda xy. (s\,y)\,x)^{(1\ra
      A)\ra(B\ra 2)\ra B\ra A}}{(s|np)|np}}{\infer{\overset{(\lambda
      y.(s\,y)\,x)^{(B\ra 2)\ra B\ra A}}{s|np}}{\infer{\overset{((s\,y)\, x)^{B\ra A}}{s}}{\overset{x^{1\ra A}}{np} & \infer{\overset{(s\, y)^{(1\ra
            A)\ra B\ra A}}{s|np}}{\infer{\overset{s^{(B\ra 2)\ra (1\ra
              A)\ra B\ra A}}{(s|np)|np}}{\textit{studies}} &
        \overset{y^{B\ra 2}}{np} }}}}}
$$
\end{sideways}
\end{center}
\caption{Proof of "John studies logic and Charles phonetics" with
  semantics ``John studies logic and phonetics studies Charles''.}
\label{fig:jslb}
\end{figure}

\editout{
$$
\infer{s}{
\infer{(np \ra np\ra s)\ra s}{
\infer[I_1]{(np \ra np \ra s)\ra s}{\infer{s}{\infer{np}{\textit{logic}} &
    \infer{np\ra s}{\infer{[np \ra np\ra s]^1}{P} &
      \infer{np}{\textit{John}}}}} &
\infer{\bigformb}{
\infer{\bigforma}{\textit{and}} &
\infer[I_2]{(np \ra np \ra s)\ra s}{\infer{s}{\infer{np}{\textit{Charles}} &
    \infer{np\ra s}{\infer{[np \ra np\ra s]^2}{Q} &
      \infer{np}{\textit{phonetics}}}}}}} &
\infer[I_3]{np\ra np\ra s}{\infer[I_4]{np\ra s}{\infer{s}{\infer{[np]^3}{x} & \infer{np\ra
      s}{\infer{np\ra np\ra s}{\textit{studies}} & \infer{[np]^4}{y}}}}}}
$$
}

Since this article is already rather long, we cannot treat the other problem cases mentioned at the start of
Section~\ref{inadeq}. However, the examples which have been treated in
detail serve as a blueprint to constructing similar problems for the
additional listed problem cases. 
%it is easily seen that we can exploit
%the limitations of lambda grammars in a way similar to the examples
%which have been treated in detail:
 In all cases, the fundamental asymmetry of 
lambda grammars means we have insufficient tools at our disposal to constrain the word order for positive
implications and that the best possible approximations are inadequate both
syntactically and semantically.

\subsection{Solutions for lambda grammars}
\label{solutions}

Given the descriptive challenges for lambda grammars, it seems natural to ask how
lambda grammars could evolve to rise to these challenges.

\begin{enumerate}
\item \emph{Stasis.} Keeping the formalism and the analyses as
  they are is a possible, if not a very attractive solution, since it
  would require us to significantly tone down the ambitions of the
  syntax-semantics interface of the formalism, thereby losing one of
  the attractive aspects of categorial grammars.
%  results in an explosion of the lexicon lexicon (essentially by

  We can also choose to embrace descriptive inadequacy and use the
  result from \cite{busz96}, which, as discussed in Section~\ref{sec:mainthm},
  translates 
  Lambek grammars into AB grammars while preserving the semantics, though at the price of an
  explosion in lexicon size (as Pentus', 1995 original proof) --- to obtain
  at least the most of the
  coverage of hybrid type-logical grammars directly within
  lambda grammars (though this presupposes strict separation, as required
  for the application of Lemma~\ref{lem:hybridlambda}). This would save lambda grammars empirically,
  incorporating the syntax-semantics interface of the hybrid system, but we
  would then have a combinatorial explosion followed by an
  NP-complete
  problem (according to Lemma~\ref{lem:nplambda}). Given that the original Pentus proof, with $O(|G | n^3)$ complexity for
  some colossal $| G |$, never resulted in fast, practical parsers for
  the Lambek calculus because of the grammar size constant, having
  a similar constant for an NP-complete problem does not bode well for
  parsing the resulting grammar.

 So it seems we have two unappealing options here: give up --- or
 significantly reduce the ambition of --- the syntax-semantics
 interface or give up actually parsing lambda grammars.

\item Change the \emph{terms} and/or their
  \emph{interpretation}. The lambda grammars discussed above produce
  strings. Several authors have looked at lambda grammars which
  generate different types of structures, such as trees \cite{muskens01lfg,groote02tag}. When we
  generate trees, we can add a separate yield algebra which tells us
  how to interpret the possible word orders generated by a given
  tree. This ``multiple transduction'' approach has several other
  instances and though it is conceivable that such multiple transductions may help alleviate some of the \emph{symptoms},
  it does not address their root cause, which is the asymmetry of the
  system. 

What we need is a system which can reject ``candidate derivations''
which have been computed at the previous level.
Such a solution can be found in \cite{muskens97comb}, who produces first-order logic
  formulas and adds a separate theorem-prover component (which is
  essentially a model-builder for multimodal categorial
  grammars). Muskens' solution would solve the problems with
  lambda grammars by essentially generating potential derivations and
  asking a multimodal grammar if these derivations are valid. However,
  this setup does not seem to have any benefits over a direct
  multimodal implementation and suffers from the same complexity
  problems as multimodal categorial grammars. \citeasnoun{pp10acg},
  discussed below since they change the types in addition to the terms, also fall into this
  category.

Another potential solution in this family would be to add
\emph{term equations} (and corresponding reductions) to the lambda
calculus. However, it is unclear what sort of form such a solution
would take.

As we have seen in the treatment of transitive verb conjunctions,
adding case offers (at best) a partial solution by attacking the
symptoms rather than the underlying cause of the problem. It should also be noted that
the only solution of this kind which has been worked out in any detail
uses dependent types and this complicates the \emph{types} as well as the terms
\cite{pp10acg,pompigne13}, which moves us to the next point.

\item Change the \emph{types}. Various authors \cite{maarek,pp10acg} have looked at extending the type theory of
  lambda grammars beyond the simply typed lambda calculus. Of these
  extensions, dependent types seem well-suited to the challenges posed
  in this paper, though they would need to be added on a much larger
  scale than previously assumed and they would complicate the
  type/term calculus and its mathematical properties considerably. 
%Since smart parsing algorithms for
% lambda grammars (de Groote 2007) already first-order linear logic as
%  their core engine, it seems much easier, if we desire to obtain
%  dependent types, to simply compute them using the Curry-Howard
%  isomorphism from the first-order linear logic proof.

An alternative solution, proposed in the context of linear grammars \cite{worth14coord} 
uses subtyping combined with restrictions on the form
of subterm cooccurrences. It is unclear to me at the moment whether this
type of treatment is equivalent to other proposals (eg.\ those of
hybrid type-logical grammars) and whether it corresponds to a natural
fragment of first-order linear logic.
\end{enumerate}

Is seems that the easiest way to fix the inadequacies of
lambda grammars would by to extend the system to hybrid type-logical
grammar: existing linguistic analyses in lambda grammars can be
preserved and/or corrected while the system keeps much of the flavor
of lambda grammars.

An alternative, especially for those convinced of the need to extend
lambda grammars to handle linguistic features and island constraints \cite{pp10acg,pompigne13}, is to move to
first-order linear logic, which also allows us to preserves the things
that work in lambda grammars but incorporate a simple treatment of
both features and island constraints without having to change the underlying logical theory. Those particularly attached to
dependent types can obtain them from MILL1 proofs by means of the
Curry-Howard isomorphism; first-order logic is a fairly weak fragment
of the lambda calculus with dependent types \cite{su06ch} so we need
to verify whether
it is expressive enough. Since smart parsing algorithms for lambda
grammars \cite{groote07parsing} already use  first-order (linear) logic to
drive proof search, this
solution stays close to the computational core of lambda grammars: it
remedies the severe problems but also allows us to include treatments
for which much more complicated analyses have been proposed. However,
it is not clear in such a setup what the typed lambda terms actually
contribute and we would have a much simpler system if we simply
removed the typed lambda terms from the surface structure component
and handle all of the surface structure in first-order linear logic.

\section{Conclusions}

In this paper, we have shown that Hybrid Type-Logical Grammars
\cite{kl13coord} (and by extension lambda grammars/abstract categorial
grammars) can be embedded in first-order linear logic by means of a
simple translation, formula to formula and proof to proof. This
provides cleaner proof-theoretic foundations for Hybrid Type-Logical
Grammars but also suggests new ways of parsing these grammars. As an
immediate corollary, we have also shown that Hybrid Type-Logical
Grammars are NP-complete (like lambda grammars and the Lambek
calculus).

We have also seen how this translation provides a new perspective of
the known (but often ignored) problems of lambda grammars with
coordination and shown that the lack of left-right symmetry (or, at
the very
least, the absence of a way to emulate the Lambek calculus introduction
rules) results in overgeneration and descriptive inadequacy problems
for a much larger class of cases than previously assumed.

Combined with the results from \cite{mill1} and \cite{moot13lambek},
this means that the Lambek calculus, the Displacement calculus, lambda
grammars and Hybrid Type-Logical Grammars can all be translated into
first-order linear logic by means of simple translations and that,
moreover, many of the analyses of linguistic phenomena in these different
systems converge upon translation into first-order linear logic. 

First-order linear logic can thus be seen as a way to decompose the
connectives of all these logics, separating the functor/argument
structure from the word order operations.

\section*{Acknowledgments}

This paper is deeply indebted to Yusuke Kubota and Robert Levine, whose ESSLLI 2013 course awoke my
curiosity both about the proof theoretic aspects of hybrid
type-logical grammar and about
the descriptive inadequacies of lambda grammars/abstract categorial
grammars --- the two principal themes of the current paper.

Early versions of these ideas were presented at the LIX Colloquium on
the Theory and Application of Formal Proofs (Palaiseau, November 2013), Computational
Linguistics in the Netherlands (Leiden, January 2014) and the Polymnie
workshop (Toulouse, March 2014). I would like all the people present
there for their questions and constructive comments, notably Crit Cremers, Philippe de Groote, Dominic Hughes
and
Dale Miller. 

Last, but certainly not least, I would like to thank Michael Moortgat,
Carl Pollard and Christian Retor\'{e} for their discussion about the themes of this paper.

All remaining errors are of course my own.

This work has
    benefitted from the generous
support of the French agency Agence Nationale de la Recherche as part of the project Polymnie
(ANR-12-CORD-0004).

\bibliographystyle{agsm}
\bibliography{moot}

%\pagebreak
%\tableofcontents

\end{document}